%% file: Main_TIFS_draft.tex
\documentclass[journal]{IEEEtran}
\newcommand{\STAB}[1]{\begin{tabular}{@{}c@{}}#1\end{tabular}}

\usepackage{array}
\usepackage[utf8]{inputenc}
\usepackage{amsmath, commath}
\usepackage{amsthm}
\usepackage{amsfonts}
\usepackage{amssymb}
\usepackage{bm}
\usepackage{bbm}
\usepackage{graphicx}
\usepackage{graphics}
\usepackage{xcolor}
\usepackage{float}
\usepackage{tikz}
\usetikzlibrary{shapes, snakes, patterns}

\usepackage{epstopdf}
\usepackage[export]{adjustbox}
\usepackage[list=true]{subcaption}
\usepackage{color}
\usepackage{algorithm}
\usepackage{algorithmic}

\newtheorem*{definition}{Definition}

\usepackage[justification=centering]{caption}
\usepackage{enumerate} 
\usepackage{cite}
\usepackage{suffix}
\usepackage{mathtools}
\usepackage{tabularx, booktabs}
\usepackage{multirow}
\usepackage{diagbox}
\usepackage{amsfonts}
\usepackage{amssymb}
\usepackage{textcomp}
\usepackage{dsfont} 
\usepackage{subcaption}
\usepackage{pgfplots,pgfplotstable}
\pgfplotsset{compat=1.3}
\usetikzlibrary{automata, positioning, arrows,shapes.multipart,fit}

\DeclarePairedDelimiter\ceil{\lceil}{\rceil}

\def\x{{\mathbf x}}

\def\h{{\mathbf h}}

\def\y{{\mathbf y}}

\def\n{{\mathbf n}}
\def\x{{\mathbf x}}
\def\0{{\mathbf 0}}

\def\i1{{\mathbf 1}}

\title{{Smart Channel State Information Pre-processing for Joint Authentication and Secret Key Distillation}}
\author{Muralikrishnan Srinivasan$^1$, Sotiris Skaperas$^2$,  Arsenia Chorti$^2$, Mahdi Shakiba Herfeh$^2$, Muhammad K. Shehzad$^3$, and Philippe Sehier$^3$\\
$^1$Chalmers University of Technology, Gothenburg, Sweden\\
$^2$ETIS UMR 8051 / CY Paris University, ENSEA, CNRS, 95000, Cergy FR\\
$^3$Nokia Bell-Labs, Route de Villejust,
91620 Nozay, France\\
mursri@chalmers.se, \{sotiris.skaperas, arsenia.chorti, mahdi.shakiba-herfeh\}@ensea.fr\\
\{{muhammad.shehzad, philippe.sehier\}@nokia.fr}}

\begin{document}

\maketitle
\begin{abstract}
While the literature on RF fingerprinting-based authentication and key distillation is vast, the two topics have customarily been studied separately. 
In this paper, starting from the observation that the wireless channel is a composite, deterministic / stochastic process, we propose a power domain decomposition that allows performing the two tasks simultaneously. We devise intelligent pre-processing schemes to decompose channel state information (CSI) observation vectors into ``predictable'' and ``unpredictable'' components. The former, primarily due to large-scale fading, can be used for node authentication through RF fingerprinting. The latter, primarily due to small-scale fading, could be used for semantically secure secret key generation (SKG). To perform the decomposition, we propose: (i) a fingerprint ``separability'' criterion, expressed through the maximisation of the total variation distance between the empirical fingerprint measures; (ii) a statistical independence metric for observations collected at different users, expressed through a normalised version of the $d$-dimensional Hilbert Schmidt independence criterion (dHSIC) test statistic. We propose both explicit implementations, using principal component analysis (PCA) and kernel PCA and black-box, unsupervised learning, using autoencoders. Our experiments on synthetic and real CSI datasets showcase that the incorporation of RF fingerprinting and SKG, with explicit security guarantees, is tangible in future generations of wireless.  


\end{abstract}
\section{Introduction}

Sixth generations (6G) systems will be required to meet diverse constraints in an integrated ground-air-space global network. In particular, meeting overly aggressive latency constraints and operating in massive connectivity regimes with low energy footprint and low computational effort while providing explicit security guarantees can be challenging \cite{zou2016survey}.
In addition, the extensive introduction of artificial intelligence (AI) and machine learning (ML) and the rapid advances in quantum computing are further developments that will increase the attack surface of 6G systems  \cite{mavroeidis2018impact, chen2016report}. 
More importantly, the massive deployment of low-end Internet of things (IoT) nodes, often produced following non-homogeneous production processes and with expected lifespans exceeding $10$ years, poses pressing questions concerning future security architectures.  


In this setting, quality of security (QoSec) is envisioned as a flexible security framework for future networks with diverse requirements. By mirroring the differentiated services (DiffServ) networking paradigm, various security levels could be conceptualized, moving away from static security controls, captured currently in zero-trust security architectures \cite{INGR21}. In parallel, the integration of communications and sensing, along with embedded (on-device) AI, can provide the foundations for building autonomous and adaptive security controls, orchestrated by a vertical security plane in coordination with a vertical semantic plane, dubbed as context-aware smart security \cite{Chorti22}, \cite{Andre22}.


It is in this framework that we envision the incorporation of physical layer security (PLS) schemes in 6G security protocols, introducing security controls at all layers for the first time \cite{INGR21}. This exciting prospect does not come, however, without challenges. Despite intense research interest in PLS for more than two decades, its incorporation in actual security products remains largely elusive, with a few exceptions in RF fingerprinting (physec.de) and multi-factor authentication (silencelaboratories.com). The hurdles to be overcome concern primarily two key issues: the difficulty in providing explicit security guarantees with PLS irrespective of the link quality and the potential degradation in terms of achievable rates. For incorporating PLS in 6G, both of these issues need to be addressed.

With this in mind, in the first part of this work, we aim to provide a draft roadmap for the inclusion of PLS in 6G. Our goal is to pave the way for developing practical PLS schemes with explicit security guarantees, taking into account inconsistencies between idealized channel models -- used in information-theoretic proofs -- and actual transmission conditions. To this end, we draw inspiration from the methodology used by the cryptographic community. We revisit how idealized semantic security primitives, such as pseudorandom number generators and functions, are used to build real cryptosystems. We use these ideas as guidelines for the proposed roadmap.
Furthermore, we explicitly propose using hybrid PLS - crypto schemes, an approach that allows addressing jointly two key issues. First, to relax the pressure for high secrecy transmission rates, and, second, to shield PLS schemes from active attacks, e.g., by ensuring the integrity of exchanged messages or side information with the use of message authentication codes and cryptographic hashing \cite{Mitev20}. 

The second part of this article showcases the proposed roadmap through technical results. In more detail, we devise smart pre-processing algorithms of observed channel state information (CSI) vectors for RF fingerprinting and secret key generation (SKG). The separability of the RF signatures (fingerprints) and the unpredictability of the shared randomness in SKG are put forward as the design criteria. Overall, we look at the wireless channel as a composite process, with both deterministic and stochastic components \cite{goldsmith2005wireless}. 
Inspired by Wold's decomposition theorem \cite{Anderson}, we propose to express the observed CSI as the \textit{sum} of a predictable (deterministic) component -- primarily due to large scale fading -- and an unpredictable (stochastic) component -- primarily due to small scale fading. 

The premise of the proposed pre-processing is that large-scale fading (path-loss and shadowing) depends heavily on the location and can be helpful for authentication purposes \cite{mitev2020multi}. On the other hand, small-scale fading can constitute an entropy source for SKG with early results presented in \cite{srinivasan2021CSI} and \cite{icc}. We study a power-domain decomposition of the CSI in the two components by using unsupervised learning approaches (principal component analysis (PCA), kernel-PCA (KPCA), and autoencoders (AE)). The proposed decomposition is designed to meet the following criteria: (i) maximum separability of RF fingerprints, captured through the total variation distance (TVD) between the empirical measures; (ii) minimum dependence between the sources of shared randomness at different nodes, measured through a normalized version of the $d$-dimensional Hilbert Schmidt
independence criterion (dHSIC) \cite{dhsic} test statistic. We validate the proposed pre-processing both on synthetic - using the Quadriga models \cite{jaeckel2014quadriga} - and real CSI datasets \cite{Nokia_dataset}.  

The rest of the paper is organized as follows. Section II describes the proposed roadmap for incorporating PLS in practical 6G systems. Section III reviews the background concepts for RF fingerprinting based node authentication and  SKG and introduces the metrics proposed as design criteria. Section IV presents the proposed approaches for disentangling predictable CSI components from unpredictable ones using PCA, KPCA, and two different AE. Section V includes numerical results and Section VI presents the discussion and concluding remarks.

\section{Proposed Roadmap for Incorporating PLS in 6G}
This section aims to provide a methodology that will overcome existing hurdles in translating PLS theory to practical systems. In this direction, we begin the discussion from the fact that the study of reliable communications involves a typical bottom-up approach; building on idealized source and channel models, powerful theorems on the fundamental limits of achievable rates have been built. Such results guide the design of actual wireless systems operating over actual transmission channels, given that in practice (a few) connection outages and errors are tolerated. 
However, the requirements are different in information security, in which the guarantees to be met need to be tight and explicit. 

In cryptographic proofs, such explicit guarantees are expressed through the concept of the  "adversarial advantage'', which needs to be negligible (i.e., strictly upper-bounded in polynomial time, reflecting limitations in computational resources) \cite{Boneh}. Related proofs of semantic security (chosen-plaintext attack -- CPA security, chosen ciphertext attack -- CCA security, etc.) are built around \textit{games} played between legitimate and adversarial parties, when the latter possess specific capabilities, ranging from launching passive attacks (eavesdropping) to launching active attacks (tampering, man-in-the-middle, spoofing). In the cryptographic proofs, several idealized mathematical constructions are used, such as pseudorandom number generators (PRG), pseudorandom functions (PRF) and permutations (PRP), and random oracles (RO), as abstractions of the fundamental cryptographic primitives \cite{Boneh}. 

Consequently, practical cryptosystems are built to \textit{resemble}, as closely as possible, the above-mentioned idealized mathematical structures. In particular, secure stream ciphers should resemble PRGs, while block ciphers should be built to resemble PRFs or PRPs. As an example of the latter, non-linearities are a core element of their design, aiming to induce maximal confusion \cite{Shannon}, e.g., with the use of S-boxes. Their design typically involves selected functions from the class of bent functions that are subsequently modified to withstand linear and differential cryptanalysis \cite{s-box}. As practical crypto primitives are designed to resemble the respective idealized mathematical structures, practical cryptography lingers towards semantic security. 

Turning our attention to PLS, information-theoretic proofs guaranteeing perfect secrecy or distributed key distillation without leakage are similarly built using idealized channel models. Importantly, it has been shown that it is possible to prove semantic security for wiretap channels \cite{Tessaro}, while recently, fundamental results have been published regarding the finite blocklength \cite{Poor} along with related approaches for analysis \cite{Zoran} and code design \cite{SSP21}, \cite{ITW21}. However, it remains challenging to translate the theoretical results to practical systems for various reasons, including that actual transmission channels might not match idealized channel models. As an example, although the impact of correlations on the secrecy capacity has been extensively explored \cite{CorrFading}, \cite{CorrMIMO} \cite{CorrNakagami}, when accounting for more complex phenomena, captured typically through compound or arbitrarily varying channel models, translating channel secrecy capacity expressions  \cite{Notzel, Mansour} to practical code designs becomes tedious. 
More importantly, the adequacy of practical channel models -- proposed by 3GPP -- for information-theoretic security is still unclear. We note in passing that, for mmWave and sub-THz bands, existing 3GPP models have been shown to have limitations \cite{Rappaport1,Rappaport2}. We, therefore, need to shed light on how  PLS can provide explicit security guarantees in non-idealized transmission conditions. 

A first step to resolve related inconsistencies that will be tangible in 6G systems is the site-specific, online learning and characterization of the channel \cite{Rangar}, which can allow constructing secrecy maps \cite{Zoran} from actual channel estimates. It should be noted, however, that there are inherent limitations in the online evaluation of secrecy outages directly from field data, e.g., during the network operation, as the dataset size limits the precision in the estimation of mutual information from measured CSI vectors \cite{MI}.

To bypass this issue, we propose a further step -  
to engineer the transmission, given the channel model learned, so that the end-to-end transmission resembles transmission over an idealized channel, for which the security guarantees are explicit. A general approach in this direction could exploit 
dimensionality reduction as a generic tool to carve out a memoryless / block fading channel from an actual, site-specific channel with correlations across different dimensions. In subsequent sections, we will provide relevant results for the case of SKG. 

A final point that should be addressed concerns the potential impact of proposed pre-processing approaches on the achievable secrecy or secret key rates. Indeed, in numerous measurement campaigns for SKG \cite{Paul1}, or secrecy outage probabilities in sub THz bands \cite{Ma}, it has been shown that in real-world scenarios, high entropy (for the extraction of keys) or a consistent signal to noise ratio (SNR) advantage (for wiretap coding) cannot always be guaranteed, e.g., for target rates of $0.5$ b/s/Hz. Due to this negative result, there has been considerable skepticism concerning the feasibility of practical PLS systems. In order to address this issue, we first need to identify the target secrecy or secret key rate. We propose, at present, to use PLS mainly for two security operations: (i) for the distillation (using SKG) or distribution (in wiretap channels) of secret keys in hybrid PLS - cryptosystems \cite{Mitev20, mitev2020authenticated}; (ii) for localization or RF fingerprint-based authentication \cite{Eduard}, \cite{Wafa}. 

Consequently, the target rate would be dictated by cryptographic considerations (mode of operation) or networking parameters (e.g., re-keying due to handovers in high mobility environments). An example of the former is that in the record protocol of the transport layer security (TLS) protocol version 1.3, the suggested cipher suite is the \verb|TLS_AES_256_GCM_SHA384 [GCM]|, requiring essentially 96 bytes of key material (32-byte keys, 48 byte MAC secrets, and 2 x 8 byte IVs), while the largest record that can be transmitted amounts to $2^{14}$ bytes \cite{RFC}. Assuming PLS is used to generate or distribute these keys, then key rates (over multiple subcarriers and antennas) as low as $0.006$ bits/sec could be comfortably targeted\footnote{ Target key rates can be much lower if we consider key re-use over multiple records or zero-round-trip time operation, that can incorporate SKG \cite{Mitev20}.}. This toy example showcases that in hybrid PLS - cryptosystems, the target rates can be very low, alleviating concerns related to diminishing throughput.

To summarise, we put across three key points that, in our opinion, can facilitate the incorporation of PLS in 6G:
\begin{enumerate}

    \item Continuous, online learning of the channel that is site-specific, context-aware \cite{Chorti22}, performed by authenticated entities such as trusted base stations during the actual operation of the network. 
\item Transmission and channel engineering of end-to-end PLS links to provide explicit security guarantees, using dimensionality reduction as a generic tool.
\item PLS to be used for the distillation and distribution of authentication or symmetric keys in hybrid PLS - crypto systems. Target key rates can be determined by the cryptographic suite of upper-layer protocols or network properties (such as handovers).
\end{enumerate}

In the rest of this paper, we will showcase some elements of the proposed roadmap for RF fingerprinting and secret key distillation. We will first begin reviewing related background concepts in Section III while also presenting the proposed metrics and design criteria.

\section{Background Concepts and Proposed Metrics for RF Fingerprinting and SKG}
This section reviews background concepts for RF fingerprinting based node authentication,  SKG and related proposed metrics.
\subsection{Localization / RF fingerprinting based node authentication}
In PLS, authentication requires a verifiable source of uniqueness, related, for example, to node positioning and fingerprinting. The RF fingerprints used for authentication must be statistically separable for each location (although not necessarily decorrelated). Also, it is beneficial if the fingerprints vary only slowly \cite{MahdiBookChapter};
please refer to \cite{wang2016physical, li2017fingerprints, fang2018learning} for some representative examples exploiting different types of channel parameters. Our contribution in this work is to introduce a probability distribution distance metric as an explicit design criterion of pre-processing to extract maximally separable fingerprints from CSI vectors.
To measure the separability of RF fingerprints, e.g., between nodes at neighbouring locations, we use the TVD, which measures the L1 distance between the empirical measures.
\begin{definition}
The TVD between two probability distributions $\mu$ and $\nu$ on a countable configuration space $\mathcal{X}^N$ is defined by \cite{levinmarkov}
\begin{equation}
   TVD= \norm{\mu - \nu}_{TV} = \frac{1}{2}\sum_{x \in \mathcal{X}^N} \abs{\mu(x) - \nu(x)}. \nonumber
\end{equation}
\end{definition}
As the TVD increases, the measures are more distinguishable and vice-versa.
 
\subsection{Secret key generation (SKG)}
SKG from wireless fading coefficients exploits three facts: (i) channel reciprocity between two nodes, referred to as Alice and Bob, during the channel coherence time; (ii) spatial independence (typically measured through decorrelation), in theory at distances of the same order of magnitude as the wavelength; and (iii) temporal variation, e.g., due to node mobility \cite{zhang2020new}. Based on Jakes' model, the channel will be uncorrelated when a third party is located half-wavelength away \cite{goldsmith2005wireless}. However, experimental results show that a half-wavelength distance spatial decorrelation is valid only in rich scattering environments \cite{edman2016security,zenger2016passive, dautov2019effects, ji2020vulnerabilities}, while related attacks using ray-tracing have been successfully demonstrated \cite{Paul2}. Albeit, it is not uncommon in the existing literature that SKG is performed without systematically removing predictable components of the wireless channel coefficients \cite{li2018high, peng2017secret, xi2016instant} and without accounting for them explicitly in the privacy amplification step.

In this work, we focus on spatial correlations at attackers (referred to as Eves) in the vicinity of legitimate users. Without any pre-processing of the observed CSI, an attacker can, in principle, distil highly correlated sequences and thus substantially decrease the effective size of the keyspace\footnote{Although any correlations in the time, frequency, space or antenna domains between the reconciled sequences at Alice, Bob and potential Eves should be explicitly taken into account in the evaluation of the {conditional} min-entropy to estimate the target privacy amplification rate, such measures are omitted in many published works. }. These correlations can indeed be removed through privacy amplification. However, this approach (i.e., without any pre-processing of the channel coefficients) entails the use of much larger quantizers and heavier hashing, potentially leading to less energy-efficient solutions.


Another gain of the proposed CSI pre-processing is that it can facilitate relating SKG to the concept of a cryptographic PRG at the advantage distillation phase \cite{Rock}.
\begin{definition}
Pseudorandomness of generator $G$ \cite{Rock}: the ensemble $\{G(U_n)\}_{n\in\bm{N}}$ is pseudorandom, iff for any probabilistic polynomial-time algorithm $A$, for any positive
polynomial $p$, and for all sufficiently large $n$,
\begin{equation}
|Pr(A(G(U_n); 1^{l(n)})) = 1) - Pr(A(U_{l(n)}; 1^{l(n)}) = 1)| <
\frac{1}{p(n)}. \nonumber
\end{equation}\end{definition}
In the definition of the PRG given above, the probabilistic polynomial-time algorithm $A$ can be seen as a polynomial-time \textit{statistical test}. If an algorithm $A$ exists such that the above condition does not hold, we say the generator $G$ fails the test $A$. From a pseudorandom generator, we expect an observer who knows $i$ bits of the
output to be unable to predict the $(i + 1)$-th bit with a probability non-negligibly greater than $\frac{1}{2}$. This property is called (next-bit) unpredictability. 

Various polynomial time
statistical tests of randomness have been introduced and form the NIST suite. There also exist non-polynomial time tests, the most famous among them being the spectral test, which measures the correlation between overlapping $n$-tuples in the output of a PRG. However, these tests do not account for dependencies and cross-correlations between legitimate and adversarial nodes that can be present in the case of SKG. To address this vital issue, in the following, we alternatively propose a novel metric \textit{to infer} predictability  /unpredictability, by measuring spatial dependencies between observed CSI vectors\footnote{The proposed methodology can be straightforwardly extended to incorporate time, frequency and antenna domains.}.


\subsubsection{Proposed statistical independence metric}

In the PLS literature, to the best of our knowledge without exception, the Pearson cross-correlation coefficient (CC) has been used as an indirect measure of unpredictability. 
However,  decorrelation does not suffice to prove unpredictability as there might be non-linear dependencies and / or the underlying distributions might not be Gaussian. On the contrary, independence of the observations at legitimate and adversarial entities suffices to infer unpredictability and allows alignment of the design of SKG with the definition of a PRG, right at the advantage distillation step;  
hence we explicitly introduce spatial independence of SKG seeds as a proposed design criterion for semantically secure key distillation from wireless fading coefficients.

In this direction, we propose to measure dependencies between observed CSI vectors through the use of the test statistic of a recently introduced kernel-based statistical test of independence, dubbed as the $d$-parameter Hilbert-Schmidt independence criterion ($d$-HSIC) \cite{dhsic}. The test applies a positive-definite kernel on $N$-dimensional random variables (RVs) and maps its distribution into the reproducing kernel Hilbert space to determine whether the $N$-dimensions are independent. 
More precisely, let $ \tilde{\bm{H}}=\left( \tilde{\bm{h}}^{1},\cdots, \tilde{\bm{h}}^{N}\right)$ be an $M \times N$ matrix of the observation vectors $\tilde{\bm{h}}^{i} =[\tilde{h}^{i}_{1},\cdots,\tilde{h}^{i}_M]^T$ for $i\in \{1,\cdots,N \}$. The null hypothesis indicates that the vectors $ \tilde{\bm{h}}^{i}$, $i\in \{1,\cdots,N\}$ are mutually independent and therefore their joint probability density function can be expressed as the product of the marginals,
$H_0: {F_{ \tilde{\bm{h}}^1,\cdots, \tilde{\bm{h}}^N}}=F_{ \tilde{\bm{h}}^1}{\cdots}F_{ \tilde{\bm{h}}^N}$, whereas the alternative, ${H_A}: (\text{not }H_0)$,
denotes that $ \tilde{\bm{H}}$ consists of at least two dependent vectors. An estimator (test statistic) $dHSIC$ of the statistical functional was defined as follows \cite[Def 2.6]{dhsic},
\begin{equation}\label{dhsicest}
\small{
    \begin{aligned}
        dHSIC( \tilde{\bm{H}}) &= \frac{1}{M^2}{\sum_{i,j=1}^{M} \prod_{l=1}^{N}\left(\mathbf{1}_{M\times{M}}{\circ}K_{ij}^{l} \right)}  + {\frac{1}{M^{2N}}}{\prod_{l=1}^{N}\sum_{i,j=1}^{M} K_{ij}^{l}} \\
    &- {\frac{2}{M^{N+1}}}{\sum_{i,j=1}^{M}}\prod_{l=1}^{N}\left(\mathbf{1}_{M\times{1}}\circ K_{ij}^{l} \right),
    \end{aligned}
    }
\end{equation}
where the operator $\circ$ denotes the Hadamard product and $\mathbf{1}_{M\times{M}}$ is an $M \times M$ matrix of ones. Also, $\mathbf{K}^{l}=\left({\mathbf{K}^{l}_{ij}}\right)=\left(k^{l}(x_i,x_j)\right)\in\mathbb{R}^{M\times{M}}$ is the Gram matrix of the positive semi-definite Gaussian kernel $k^{l}$, defined $\forall{x_i,x_j}\in\mathbb{R}$ by, $k^{l}=\exp\left(-\frac{\norm{x_i-x_j}^2}{\sigma^2}\right),$
with bandwidth $\sigma=\sqrt{\frac{\text{med}\left({\norm{x_i-x_j}^2} \right)}{2}}$ and  $\text{med}(\cdot)$ is the median heuristic. 

According to \cite[Theorem 3.1]{dhsic}, with respect to the hypothesis test at hand, the critical value (for a specific significance level $\alpha$) can be obtained as below, 
\begin{equation}\nonumber
CV_{\alpha} =
\left[\mathbf{D}^{dHSIC'}\right]_{\ceil*{(B+1)(1-\alpha)}+\sum_{i=1}^{B}{  \mathds{1}_{\{dHSIC'( \tilde{\bm{H}})=dHSIC'({ \tilde{\bm{H}}}_i)\}} }},
\end{equation}
where the vector $\mathbf{D}^{dHSIC'}$ contains the $B$ Monte-Carlo realisations of $dHSIC'{({ \tilde{\bm{H}}}})$ in an increasing order; the re-sampling function $dHSIC'\left({ \tilde{\bm{H}}}\right)$, ${ \tilde{\bm{H}}}=\left(r_1( \tilde{\bm{h}}_{1}),\cdots,r_N( \tilde{\bm{h}}_{M})\right)$ is constructed by $r_1, \cdots, r_M$ random re-samplings without replacement. The operators $\ceil*{.}$ and $\left[.\right]_{j}$ denote the ceiling function and the $j$-th element of a vector respectively, and  $\mathbbm{1}_{\{\cdot\}}$ is the indicator function.   

Based on the $dHSIC$ test statistic and the corresponding critical value, we propose here a normalised metric for measuring the level of dependence, expressed as   
 \begin{equation}
\overline{\Delta}=
\frac{dHSIC(\tilde{\bm{H}})}{CV_{\alpha}}\mathbbm{1}_{\{dHSIC( \tilde{\bm{H}})>{CV_{\alpha}}\}}.
 \end{equation}
$\overline{\Delta}$ is close to unity in case of low dependence between the variables and grows without bound in case of high dependence. 

Finally, we note that during the pre-processing, lowering dependencies between observations at different spatial locations should be balanced with preserving, as much as possible, the reciprocity between the observations of Alice and Bob, discussed next.
\subsubsection{Reciprocity and mismatch probability}
On one hand, we introduce pre-processing to lower dependencies between legitimate and adversarial observations. On the other hand, this pre-processing will inevitably impact the reciprocity of the SKG seeds. In this work
we use a one-bit quantizer about the median point along the time dimension to check the reciprocity of the SKG seeds in the uplink (Alice to Bob) and the downlink (Bob to Alice).
The mismatch probability (MP) between Alice and Bob is given by the ratio of the number of bits in disagreement to the total number of bits. Mismatches can be corrected during the information reconciliation step; however, the higher the MP, the lower the rate of the reconciliation decoder, while reconciliation becomes impractical beyond a certain point for short blocklengths.

\section{Proposed Power Domain Pre-processing}


In this section, we present the proposed approaches for disentangling predictable CSI components (primarily due to large-scale fading) from unpredictable ones (primarily due to small-scale fading) using PCA, KPCA and two different autoencoders (AE).
As argued in the previous section, we will use the TVD to measure the statistical separability of RF fingerprints and propose to evaluate spatial dependencies of SKG seeds using the introduced metric $\overline{\Delta}$ \cite{dhsic}. To validate the proposed methodology, we perform experiments on synthetic and real datasets, described first in the following.

\subsection{Datasets}
We have performed evaluations using Quadriga synthetic data and experimental data collected from Nokia \cite{Nokia_Measurement}. 

\subsubsection{Quadriga synthetic dataset}
\begin{figure}[t] 
\centering
    \includegraphics[width=0.4\textwidth]{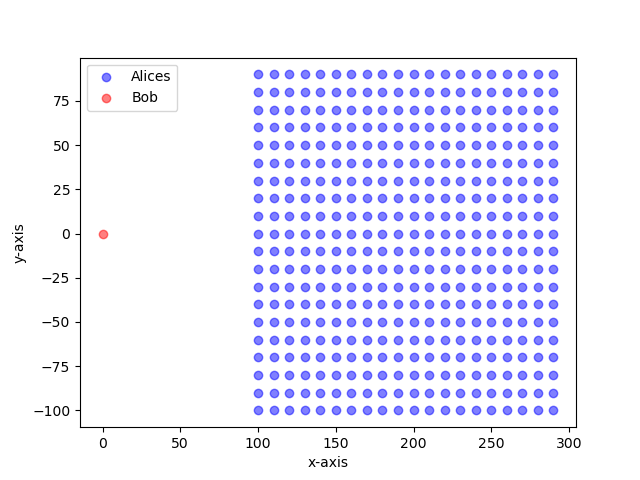}
    \caption{Node positions in the Quadriga synthetic dataset}\label{fig:quadrig}
    \end{figure}

We considered single-antenna legitimate nodes, referred to as Alices and a base station referred to as Bob; Alices' spatial locations are denoted by $\{\x_n\}_{n=1}^{N}$ $n= 1,\hdots,N$, where $\{x_n\}_{n=1}^{N} \in \mathbb{R}^L$ and $L$ denotes the spatial dimensions considered (typically $L=2$). We obtain the channel response at $N= 400$ equi-distant ($1$~m) spatial locations within a square area on the ground, between $x=100$ and $x=290$ and $y=-100$ and $y=90$ and a base station located at $(x,y,z) = (0,0,10)$ using the "Berlin-UMa-NLOS" configuration in Quadriga channel models  \cite{jaeckel2014quadriga}, as depicted in Fig. \ref{fig:quadrig}. We assume that for any specific Alice, all other Alices can act as attackers (Eves).
    
     This configuration's terrestrial Urban Macrocell parameters are extracted from measurements in Berlin, Germany. To create temporal variations in the channel, the Alices are assumed to move at a low speed of $0.5$~m/s. The number of CSI snapshots per Alice is set to $M=256$, while the carrier frequency is set to $2.68$~GHz.

Let the channel function mapping the spatial locations to the $M\times{1}$ CSI vectors $\{\h_n\}_{n=1}^{N}$ be denoted by $\mathcal{H} : \mathbb{R}^L \to \mathbb{R}^M$, where $M$ is the number of snapshots in the time domain (after concatenation of the real and imaginary components into a single column vector). The CSI observations at
Alice and Bob after the exchange of pilot signals can be modelled as
\begin{equation}{\label{observations}}
    \y_{nu} = \h_n s +\n_{nu}, \: n=1, \hdots, N \text{, } u\in\{a, b\},
\end{equation}
where the index $a$ denotes an Alice, $b$ denotes Bob; $\n_{na}$ and $\n_{nb}$ are circularly symmetric Gaussian noise variables and the pilot symbols $s$ are drawn from the binary phase-shift keying (BPSK) constellation \cite{chorti2017optimal}. The zero-force CSI estimates at Alice and Bob, respectively, are denoted by $\h_{na}= \y_{na}$ and  $\h_{nb}= \y_{nb}$ for $ n=1, \hdots, N$. 

\subsubsection{Nokia experimental dataset}

\begin{figure*}
    \centering
    \includegraphics[width=18cm, height=6cm]{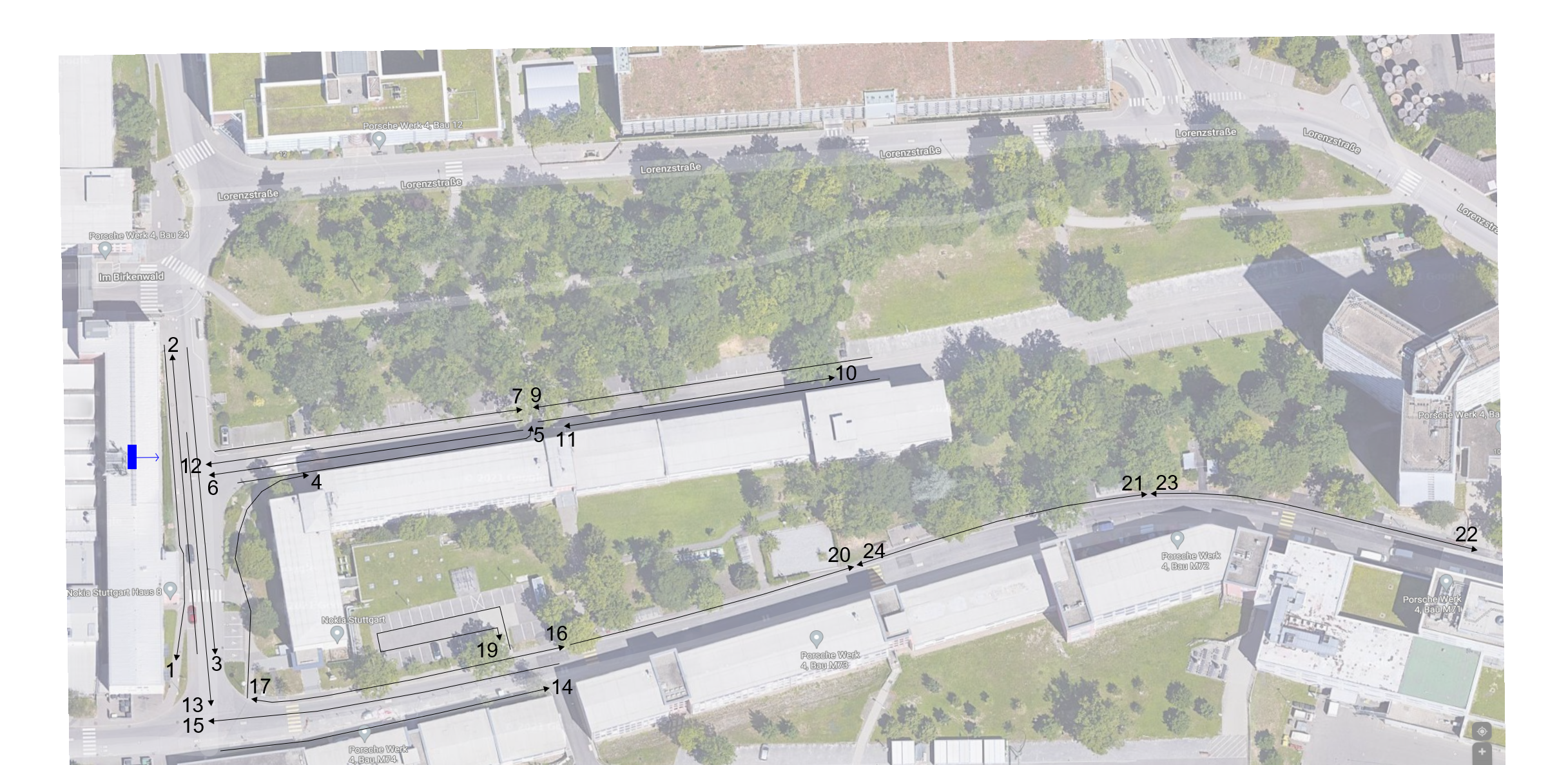}
    \caption{\textcolor{black}{Pictorial representation of the Nokia campus in Stuttgart, Germany, where the dataset was recorded. The blue bar on the left side corresponds to the location of the massive MIMO transmit antenna, placed on a rooftop, while the arrow points the antenna boresight. The black dashed lines depict the measurement tracks, along which the UE cart moved. The tracks are numbered from 1 to 24; in this work we utilized the datasets of tracks 6 and 12, which are parallel and at $1~m$ distance from each other. Source: \cite{Nokia_Measurement}.} }
    \label{fig:Nokiatrack}
    \label{fig:Nokia}
\end{figure*}

A massive multiple-input multiple-output (MIMO) channel measurement campaign was
conducted on the Nokia campus in Stuttgart, Germany. The campaign area consisted of multiple roads with high buildings  ($15$~m high approximately), acting as reflectors and blockers for the radio
wave propagation. The transmit antenna array was placed on the roof-top of one of these buildings. The geometry of $64$-element transmit-array was such that there were $4$ rows with $16$ single-polarization patch antennas, with a horizontal spacing of $\lambda/2$, and vertical spacing of $\lambda$.

The transmit antenna array transmitted $64$ time-frequency orthogonal pilot signals at $2.18$~GHz carrier frequency, using orthogonal frequency division multiplexing (OFDM) waveforms according to the $10$~MHz LTE numerology (i.e., $600$ subcarriers with $15$~kHz spacing). The pilot signals have been arranged so that the sounding on 50 separate subbands (each consisting of 12 consecutive subcarriers) required $0.5$~ms. Within that pilot burst period, the propagation channel was assumed to be time-invariant.
The pilot bursts were sent continuously with a periodicity of $0.5$~ms. 

The receiver user equipment (UE) was mounted on a mobile cart and consisted of a single monopole antenna mounted at $1.5$~m height, a Rohde and Schwarz TSMW receiver and a Rohde and Schwarz IQR hard
disc recorder, which continuously captured the received base-band signal. Both the transmit array and the receiver were frequency synchronized via GPS. During the measurements, the receiver cart moved along several routes at walking speed ($3.6$~kmph), which corresponded to a spatial channel sampling
distance of less than $0.5$~mm. Post-processing was used to extract, for each pilot burst and subband, the $64$-dimensional CSI vector.

In this work, we used datasets on tracks $6$ and $12$, depicted in Fig. \ref{fig:Nokia}, that are parallel at a vertical distance of $1$~m. In detail, we assumed that Bob is the base station and Alice walks along track-$6$, while Eve performs an "on the shoulder attack'' and walks in parallel to Alice on track-$12$, i.e., legitimate and adversarial nodes are at all times $1$~m away. In order to remove frequency domain (within the coherence bandwidth) and antenna domain correlations, we have downsampled the dataset. In detail, we kept the measurements from every $10$th subcarrier and every $4$th antenna, i.e., we used sub-sampling factors of $5$ and $6$, respectively. 


Furthermore, as the Nokia dataset consists of only uplink data, we used alternate consecutive measurements to approximate downlink data and have further downsampled the data in the time domain, keeping every $5$th channel sample. In detail, starting from sample index $1$, we labeled odd index samples as uplink and even indexed samples as downlink. As a result, for each Alice and Eve, we used CSI vectors of length $M=800$, concatenating real and imaginary parts. Before presenting the details of the pre-processing, we discuss the statistical analysis of the two datasets.

\begin{figure}[t]
    \centering
    \begin{subfigure}[t]{0.25\textwidth}
    \includegraphics[width=\textwidth]{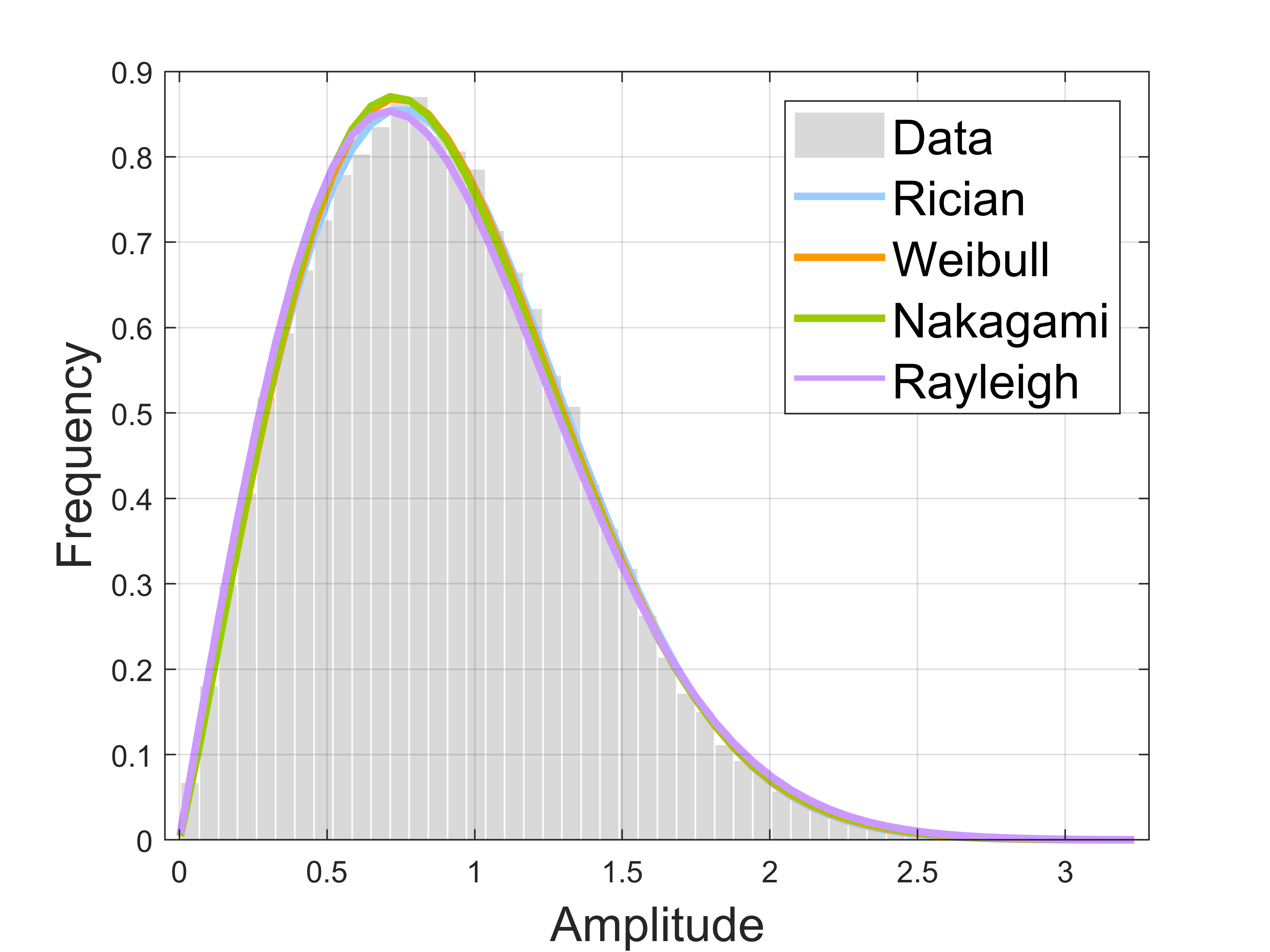}
    \caption{Quadriga channel amplitude}
    \end{subfigure}%
    \begin{subfigure}[t]{0.25\textwidth}
    \includegraphics[width=\textwidth]{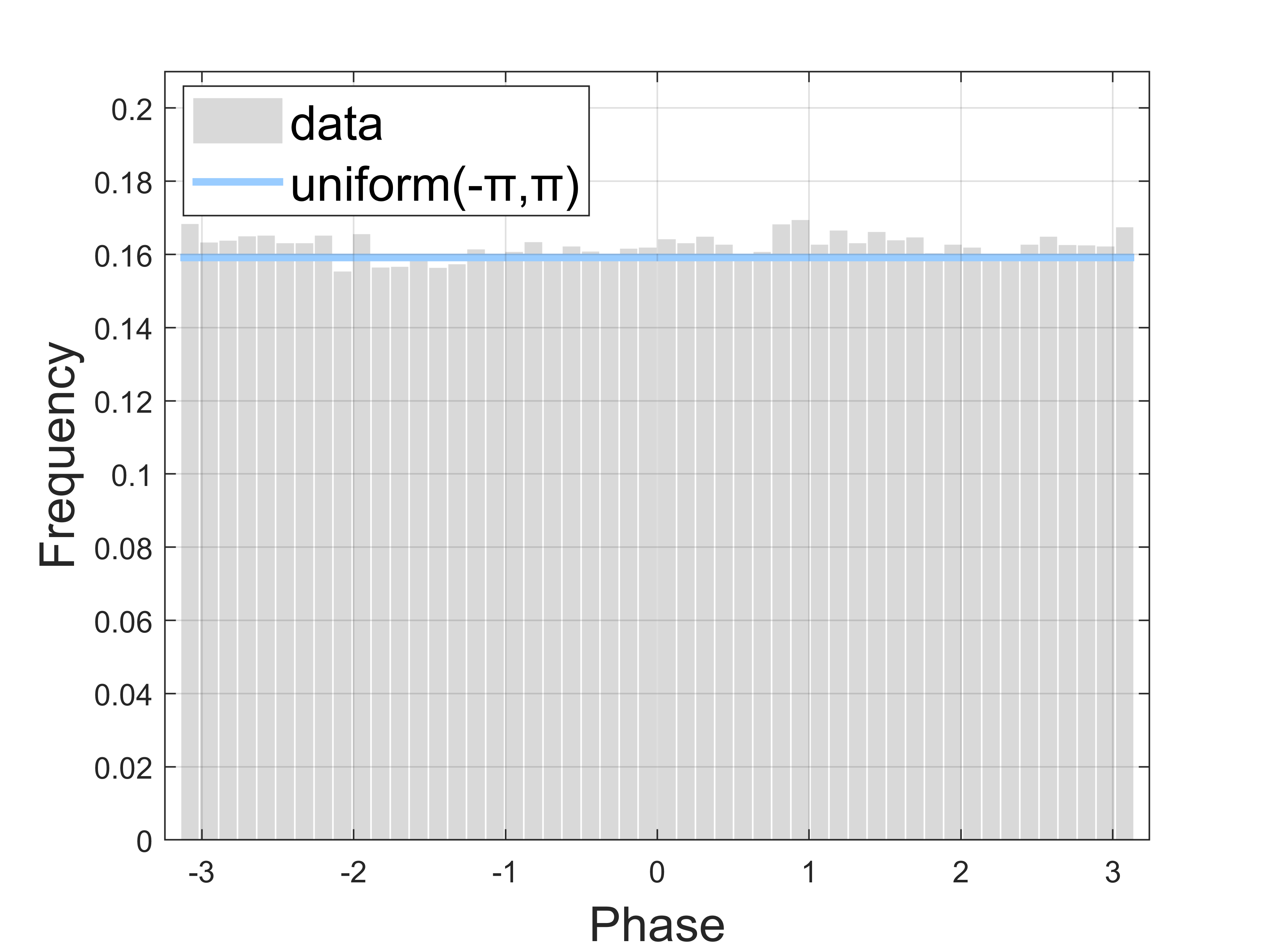}
    \caption{Quadriga channel phase}
    \end{subfigure}
    \begin{subfigure}[t]{0.25\textwidth}
    \includegraphics[width=\textwidth]{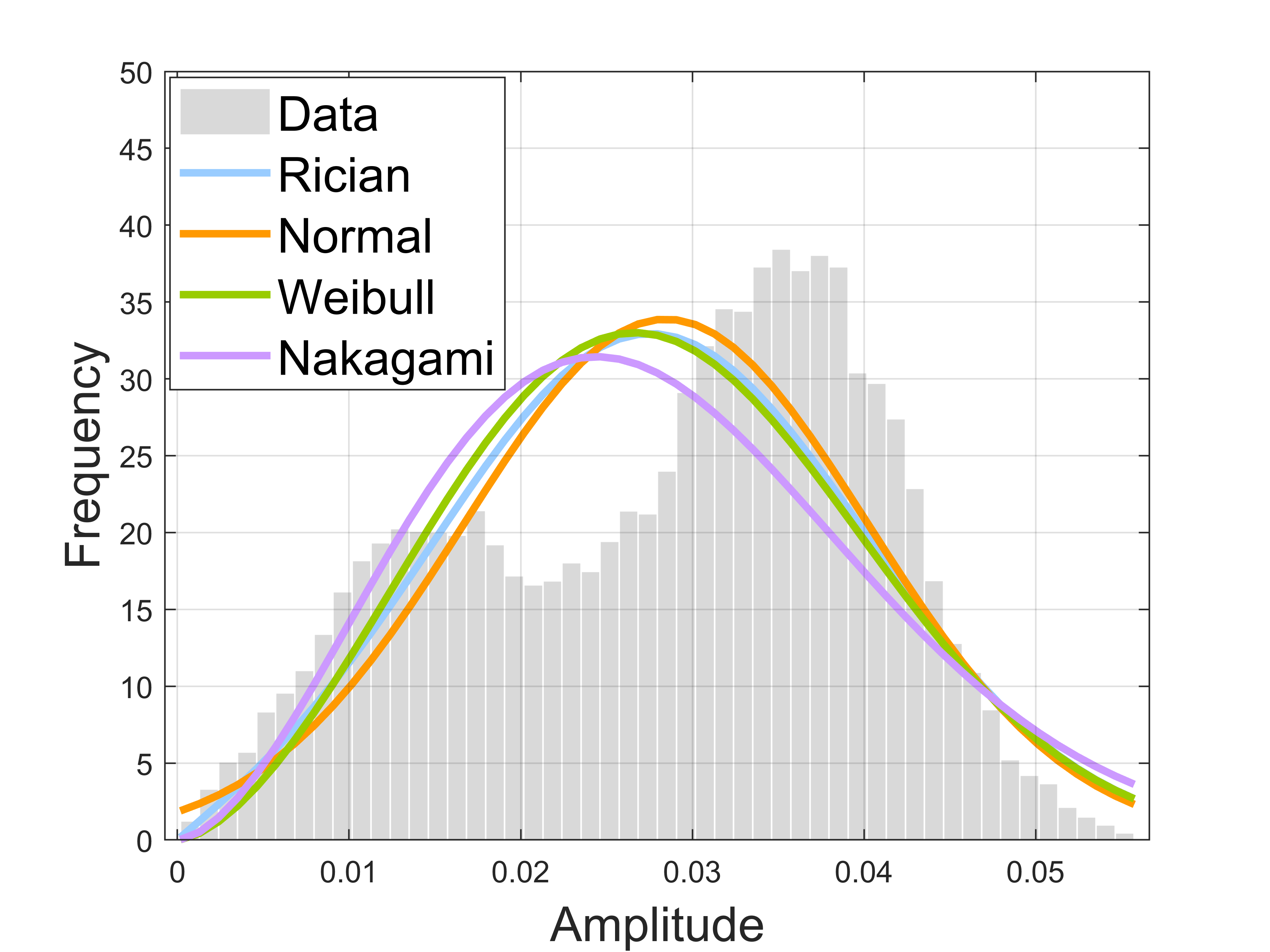}
    \caption{Nokia track 6 amplitude}
    \end{subfigure}%
    \begin{subfigure}[t]{0.25\textwidth}
    \includegraphics[width=\textwidth]{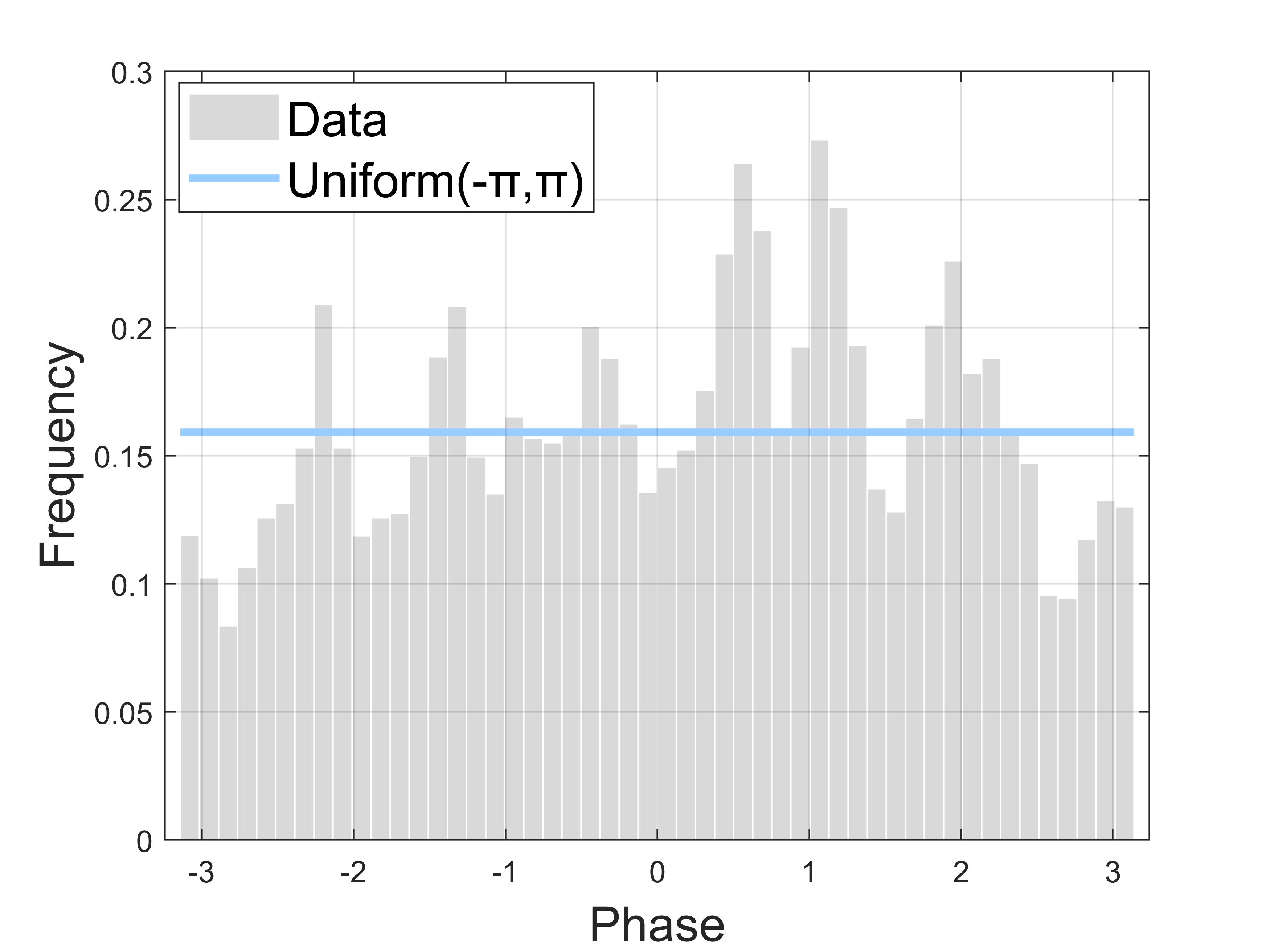}
    \caption{Nokia track 6 phase}
    \end{subfigure}
    \begin{subfigure}[t]{0.25\textwidth}
    \includegraphics[width=\textwidth]{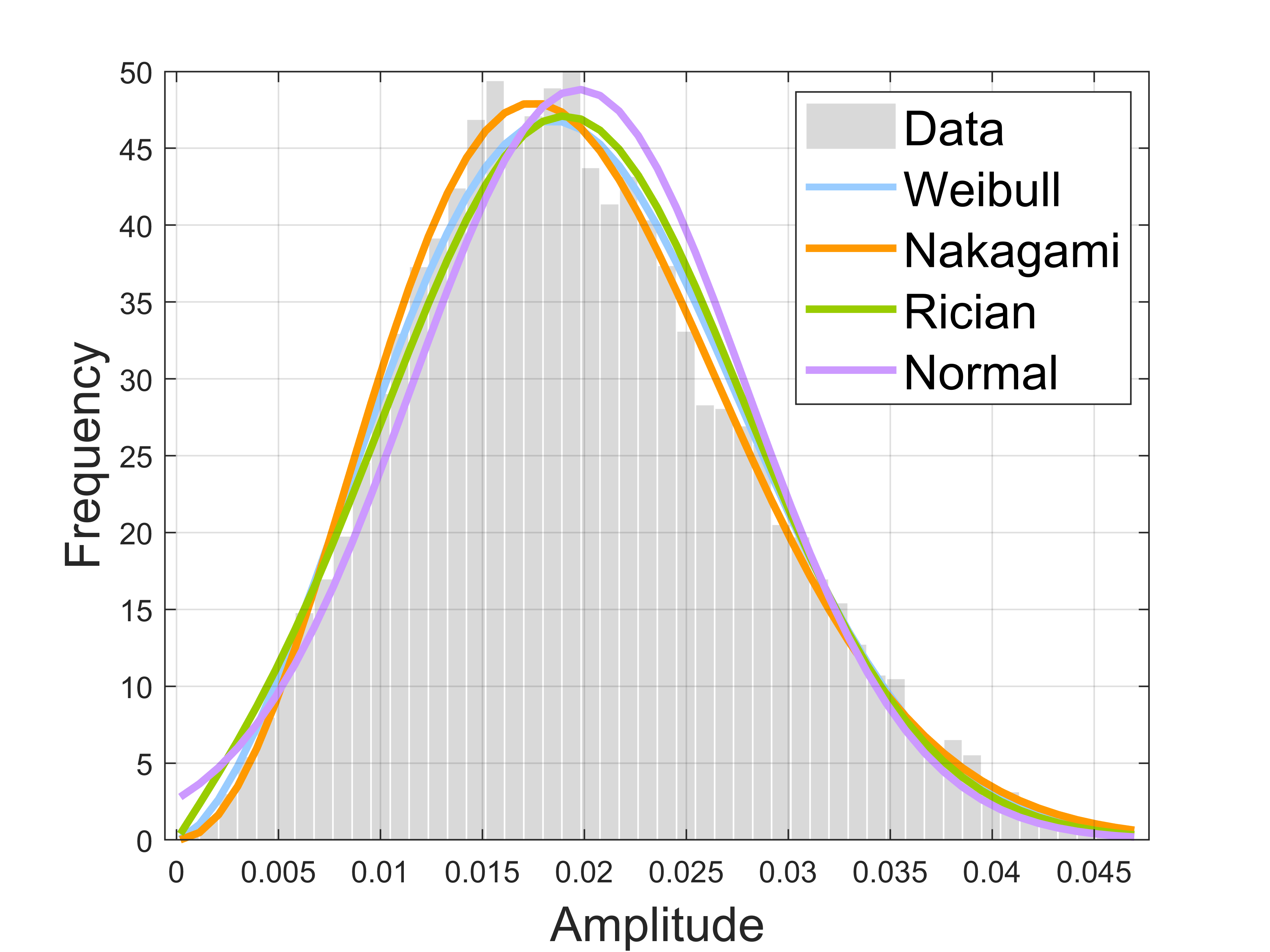}
    \caption{Nokia track 12 amplitude}
    \end{subfigure}%
    \begin{subfigure}[t]{0.25\textwidth}
    \includegraphics[width=\textwidth]{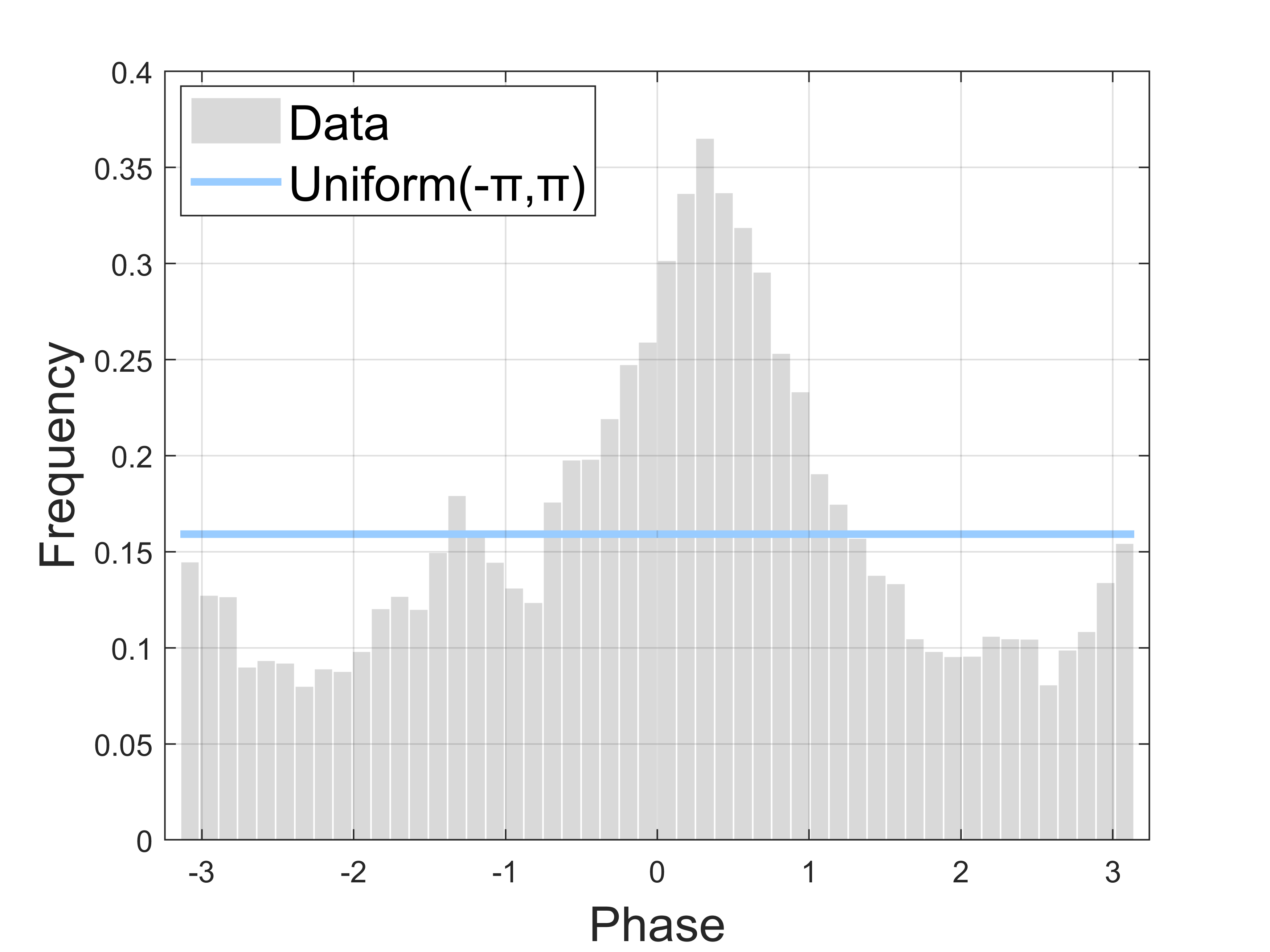}
   \caption{Nokia track 12 phase}
    \end{subfigure}
    \caption{Fitting of the underlying channel distribution of the amplitude and the phase.}
    \label{fig:est}
\end{figure} 

 \begin{table}[t]
 \renewcommand{\arraystretch}{1.5}
 \caption{Estimated pdf parameters for the Quadriga and the NOKIA datasets.} 
\setlength{\tabcolsep}{1.25pt}
\centering 
\label{Table:est}
{\footnotesize
\begin{tabular}{|l|cc|c|l|cc|c|l|cc|c|}
\hline
\multicolumn{4}{|c|}{$\text{Quadriga}$} & \multicolumn{4}{|c|}{$\text{NOKIA track 6}$} & \multicolumn{4}{|c|}{$\text{NOKIA track 12}$} \\
\hline
 & $\widehat{\alpha}$ & $\widehat{\beta}$ & $p$-val & & $\widehat{\alpha}$ & $\widehat{\beta}$ & $p$-val & & $\widehat{\alpha}$ & $\widehat{\beta}$ & $p$-val\\
\hline
Rician & $0.56$ & $0.59$ & $0.11$ & Rician & $0.02$ & $0.01$ & $0$ & Weibull & $0.02$ & $2.59$ & $0$\\
\hline
Weibull & $1.01$ & $2.07$ & $0$ & Normal & $0.03$ & $0.01$ & $0$ & Nakag. & $1.52$ & $0$ & $0.4$\\
\hline
Nakag. & $1.05$ & $1.01$ & $0$ & Weibull & $0.03 $ & $2.64$ & $0$ & Rician & $0.02$ & $0.01$ & $0$\\
\hline
Rayleigh & $0.71$ & - & $0$ & Nakag. & $1.34$ & $0$ & $0$ & Normal & $0.02$ & $0.01$ & $0$ \\
\hline
\end{tabular}
}
\end{table}  

\subsubsection{Analysis of datasets}

We fitted the empirical distributions of the amplitude and the phase for the Quadriga and the Nokia datasets to $16$ parametric probability density functions (pdf) and selected the ones with the lowest values of the Akaike information criterion (AIC),
\begin{equation}
    \text{AIC} = -2\ln({\widehat{L}}) +2k,
\end{equation}
where $\widehat{L}$ is the maximum value of the likelihood function and $k$ is the number of the estimated parameters. Finally, we applied the Kolmogorov-Smirnov (KS) as a goodness-of-fit test to extrapolate $p$-values.

In Table \ref{Table:est}, we summarize the estimated parameters of the chosen parametric distributions and also depict the $p$-values of the KS test (the null hypothesis being that the data follow the specified distribution).  
In Fig. \ref{fig:est} the fitted distributions of the amplitude and the phase are shown for the Quadriga and the NOKIA datasets (tracks 6 and 12).

According to the $p$-values of the KS test, the amplitude of the Quadriga-based channel follows a Rician distribution, while the phase follows a Uniform$(-\pi,\pi)$ distribution (see Fig. \ref{fig:est}). However, the amplitude of Nokia's track 6 CSIs does not fit any of the chosen distributions since it is bimodal, as shown in Fig. \ref{fig:est}; this is confirmed by the $p$-values of the KS test on the Nokia's track 6 channel which indicates that none of the selected distributions is a good fit. On the other hand, for Nokia track 12, the $p$-value of the Nakagami(1.52,0.0004) distribution ($p$-value $ =0.4$) implies that the channel is likely to follow the above distribution while the phase is uniformly distributed. In conclusion, Table \ref{Table:est} and Fig. \ref{fig:est} indicate that real datasets might not be well fitted to any of the usual distributions, as is the case for track $6$, confirming the point raised in Section II. In the future, we will consider non-parametric analysis and the use of generative models for fitting real datasets.         

Next, we present the proposed power domain pre-processing.
We will demonstrate how, from the observed CSI matrices at Bob, we can learn the functional mapping that captures the predictable, potentially spatially correlated components and the unpredictable, spatially decorrelated components by applying: (i) PCA; (ii) KPCA; and (iii) AEs. PCA is a linear approach, while KPCA and AE can capture non-linear dependencies.

\subsection{Pre-processing using PCA}
Let ${\bm{H}}_u= \left[  {\bm{h}}_{1u}, \cdots,   {\bm{h}}_{Nu} \right]$ denote the observed CSI matrix at Bob (aggregating the CSI vectors from all Alices) and $\bm U$ the $M\times{M}$ matrix whose rows are the eigenvectors of the matrix Cov$\left( {\mathbf{H}}_u\right)$, sorted in decreasing order. In many scenarios, e.g., Rician and generally line of sight settings, it is plausible to assume that the first few PCs correspond to the dominant large-scale fading terms, while the rest of the PCs correspond to small scale fading terms and noise. Using the eigenvectors $\widehat D \times M$ matrix $\bm{U}_{1:\widehat{D}}$ corresponding to the first $\widehat {D}$ PCs, we want to isolate the predictable part of the observed channel that will be used for RF fingerprinting, as follows,
\begin{equation}
     \widehat{\bm{H}}_{u} = \bm U^{H}_{1:\widehat{D}}   {\bm{W}}_{u},
\end{equation}
where the $\widehat D \times M$ matrix $  {\bm{W}}_{u}$ is
\begin{equation}
      {\bm{W}}_{u} =  \bm{U}_{1:\widehat{D}} {\bm{H}}_u,
\end{equation}
and $  \widehat{\bm{H}}_{u} =\left[ \widehat{\bm{h}}_{1u}, \cdots,  \widehat{\bm{h}}_{Nu} \right]$ for $u\in\{a, b\}$ is a $M\times{N}$ matrix. Furthermore, we want to identify a region of PCs with indices $\{\tilde{D}_1, \ldots, \tilde{D}_2\}$, corresponding to components $   \tilde{\bm{H}}_{u}= \left[   \tilde{\bm{h}}_{1u}, \cdots,     \tilde{\bm{h}}_{Nu} \right]$ for $u\in\{a, b\}$ over which low dependence and correlation between Alices and Eves is achieved while keeping the MP below a threshold. 


Note that the PCs beyond $\tilde D_2+1$ are dominated by noise and should be neglected (denoising). To efficiently disentangle the CSI matrix into predictable and unpredictable parts, the triplet $\{\widehat{D}, \tilde{D}_1, \tilde{D}_2\}$ is chosen such that the TVD is maximized for the first $\widehat{D}$ PCs while $\overline\Delta$ and the MP are kept as low as possible for the range $\{\tilde{D}_1, \ldots, \tilde{D}_2\}$. We discuss the trade-off between minimizing $\overline\Delta$ (as a measure of independence) and the MP in detail in Section \ref{sec:Numerical}. 

\subsection{Pre-processing using Kernel PCA (KPCA)}
PCA performs feature extraction using an orthogonal transformation based on the covariance matrix to convert the original data to a new feature space. However, PCA is a linear transformation; here, we extend to KPCA to extract potential nonlinear structures in the data. 
KPCA assumes a feature mapping $\phi:\mathbb{C}_{M,N}\xrightarrow[]{} \mathbb{F}$, where $\mathbb{F}$ is the new feature space through a (possible) nonlinear function $\phi$ \cite{kpca}. Instead of defining an explicit form of the function $\phi$, KPCA applies a kernel function for the mapping. Here, we choose to work with the complex Gausssian kernel, since the channel observations $\mathbf{H}_{u}$ are given in the complex space. In this case, the Gram matrix $\mathbf{K} = \left({K}_{i,j}\right) = (\kappa(h_i,h_j))\in\mathbb{C}_{N\times{N}}$ is based on the positive semi-definite complex Gaussian kernel, defined $\forall{h_i,h_j}\in\mathbb{C}$ as, 
\begin{equation}
    \kappa(h_i,h_j)=  \text{exp}\left( -\frac{\|h_i - h_j^{*}\|^2}{2\sigma^2}\right),
\end{equation} 
where $^*$ denotes the complex conjugate and $\sigma$ is the median bandwidth. Then, the centralized Gram matrix is given by,
\begin{equation}
    \widetilde{\mathbf{K}} = \mathbf{K} -\frac{1}{N}\mathbf{1}_{N\times{N}}\mathbf{K}-\frac{1}{N}\mathbf{K}\mathbf{1}_{N\times{N}}+\frac{1}{n^2}\mathbf{1}_{N\times{N}}\mathbf{K}\mathbf{1}_{N\times{N}}
\end{equation}
where $\mathbf{1}_{N\times{N}}$ is an $N\times{N}$ matrix of ones. 

Consequently, in the feature space, the eigenvalue problem has to be solved for the KPCA decomposition, 
\begin{equation}\label{dec}
   \widetilde{\mathbf{K}}\mathbf{V} = {N}\lambda\mathbf{V},
\end{equation}
where matrix $\mathbf{V}$ denotes the eigenvectors while $\lambda\geqslant{0}$ denotes the vector of eigenvalues. (\ref{dec}) can be also expressed as
\begin{equation}\label{dec2}
    \widetilde{\mathbf{K}}\boldsymbol{{\alpha}} = {N}\lambda\boldsymbol{\alpha},
\end{equation}
where $\boldsymbol{\alpha}_i=\frac{\mathbf{V_i}}{\sqrt{\lambda_i}}, \text{ } i=1,\hdots,N$. Then, the  first $\widehat{D}$ nonlinear PCs could be extracted through computing projections of the original data on the eigenvectors $\mathbf{V}_i$ in feature space $\mathbb{F}$, as above,
\begin{equation}
    \mathbf{Y}_{1:\widehat{D}}= \boldsymbol{\alpha}^{H}_{1:\widehat{D}}\widetilde{\mathbf{K}}.
\end{equation}

In order to estimate the predictable and the unpredictable parts as before, we apply KPCA reconstruction; unlike in PCA, this can not be done explicitly since $\phi$ is unknown. Hence, we approximate the reconstruction matrix based on the kernel ridge regression \cite{kpca_rec}. It follows that the predictable part of the observed channel is computed as follows
\begin{equation}
    \widehat{\mathbf{H}}_u = \boldsymbol{\beta}{\mathbf{K}}_{Y_{1:\widehat{D}}}, \quad \boldsymbol{\beta} = \mathbf{H}_u(\mathbf{K}_{Y_{1:\widehat{D}}}+\gamma\mathbf{I})^{-1},
\end{equation}
where $\mathbf{K}_{Y_{1:\widehat{D}}}=\left(k(y_i,y_j)\right)$ is the complex Gaussian Gram matrix of the first $\widehat{D}$ PCs and $\gamma$ is the hyperparameter of the ridge regression. Subsequently, we derive the unpredictable part of the observed CSI directly as the residual of removing the predictable part, i.e.,
\begin{equation}
    \tilde{\mathbf{H}}_u = \mathbf{H}_u - \widehat{\mathbf{H}}_u,
\end{equation}
In the case of KPCA, unlike in PCA, denoising is not performed.

\subsection{Pre-processing using autoencoders}
 AEs are unsupervised learning architectures that utilise and learn two functions, an encoder that maps the $M$ dimensional input  matrix $ {\bm{h}}_{nu}$ into $\widehat D$ dimensional encoded values $  {\bm{w}}_{nu}$ $\forall \: n=1, \hdots, N$ for $u\in\{a, b\}$ and a decoder that maps the encoded values back to an $M$ dimensional output $ \widehat{\bm{h}}_{nu}$, $\forall \: n=1, \hdots, N$ and for $u\in\{a, b\}$, such that the loss-function 
\begin{align} 
E_1 = \frac{1}{N}\sum_{n=1}^N\| {\bm{h}}_{nu} -  \widehat{\bm{h}}_{nu}\|_2^2, \text{ for} \: u\in\{a, b\},
\end{align} corresponding to the mean square error (MSE) between the AE input and output, is minimal. 
An AE can be used to extrapolate a $\widehat D$-dimensional representation $  {\bm{w}}_{nu}$, $\forall \: n=1, \hdots, N$ that can capture the dominant components. We treat the output of the decoder $ \widehat{\bm{h}}_{nu}, \forall n=1, \hdots,N, \text{ for} \: u\in\{a, b\}$ as the dominant predictable components under the conjecture that most of the received signal strength is due to large scale fading effects. Here again, we assume that the residuals \begin{equation}
\left \{    \tilde{\bm{h}}_{nu}(\widehat {D})\right \}_{n=1}^{N}= \{ {\bm{h}}_{nu}- \widehat{\bm{h}}_{nu}\}_{n=1}^{N}, \text{ for} \: u\in\{a, b\}
\end{equation} correspond to the unpredictable components of the CSI matrix.

In light of this, the value of $\widehat D$ is a hyperparameter that can be tuned to a more fine-grained loss function focusing on SKG; in particular, we build an alternative loss function to balance the spatial correlation with the reciprocity of the residuals in the uplink and the downlink.
Since we want to lower correlation, the loss function can also explicitly specify a correlation term instead of the MSE. Consequently, the following loss function is proposed:
\begin{equation}
\begin{aligned}
E_2 &= \frac{1}{N} \sum_{\underset{n_2 \in \mathcal{U}(n_1)}{n_1=1} }^N   \tilde{\bm{h}}_{n_1u}^H    \tilde{\bm{h}}_{n_2u}, \qquad  \text{ for} \: u\in\{a, b\},
\end{aligned}
\end{equation}
 as the inner product of the residual at each location and that from the neighbouring locations. Here, $\mathcal{U}(n_1)$ is the nearest neighbours of the $n_1$-th Alice-Bob pair.
 
\section{Numerical Results}\label{sec:Numerical}The proposed approaches are evaluated on the synthetic and real datasets. The results for RF fingerprinting are similar for all proposed approaches and are therefore only presented for the PCA for compactness.
\subsection{PCA}
\subsubsection{RF Fingerprinting}
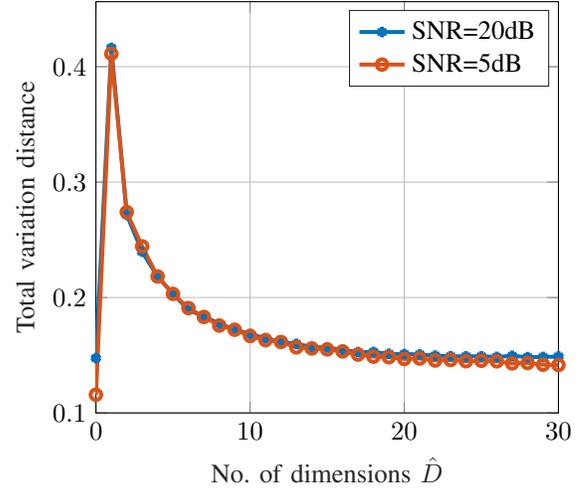
\begin{figure}[t]
    \centering
    \input{TIFS_figures_tikz/tifs_quadriga_tvd}
    \caption{Total Variation Distance vs $\widehat{D}$}
    \label{fig:TBerlindist1TVD}
\end{figure}
In Fig. \ref{fig:TBerlindist1TVD}, for the Quadriga dataset, the average TVD between the first $\widehat{D}$ PCs at any Alice and any of her neighbours is depicted. We observe that $\widehat{D}=1$ results in the largest value of TVD, while the point  $\widehat{D}=0$ corresponds to the original measurements. With an increase in the SNR, there is a slight increase in the TVD; with a decrease in noise, the variance of the first PCs increases, and hence the TVD decreases. 
\begin{figure}[t]
    \centering
    \begin{subfigure}[t]{0.24\textwidth}
    \includegraphics[width=\textwidth]{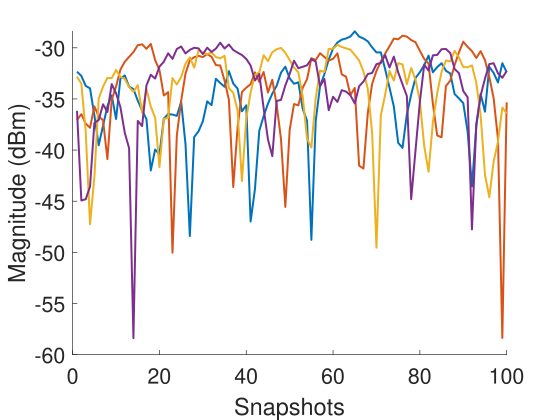}
    \caption{Original signals}
    \label{fig:SepOrig}
    \end{subfigure}%
    \begin{subfigure}[t]{0.24\textwidth}
    \includegraphics[width=\textwidth]{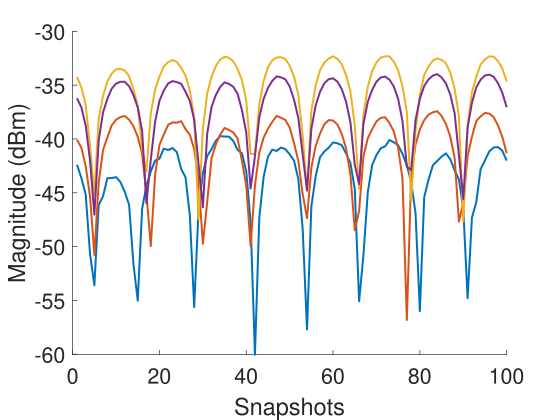}
    \caption{Predictable components}
    \label{fig:SepPred}
    \end{subfigure}%
    \caption{Separability of $6$ neighbours for the original signal and the $\widehat{D}=1$ PC for SNR$=20~\text{dB}$}
    \label{fig:TBerlinSepSNR20}
\end{figure} 

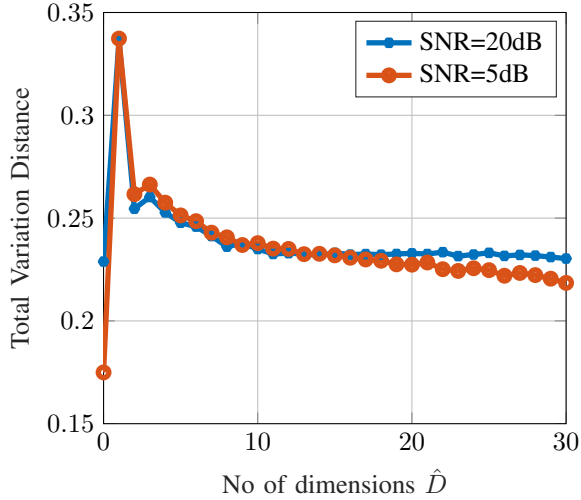
\begin{figure}[t]
    \centering
    \input{TIFS_figures_tikz/tifs_nokia_tvd}
    \caption{Total Variation Distance vs $\widehat{D}$}
    \label{fig:NokiaTVD}
\end{figure}

\begin{figure}[t]
    \centering
    \begin{subfigure}[t]{0.24\textwidth}
    \includegraphics[width=\textwidth]{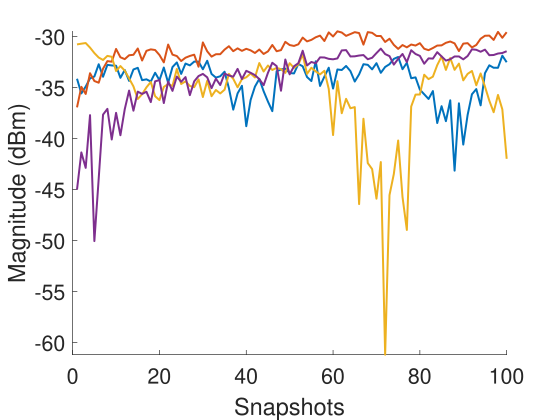}
    \caption{Original signals}
    \label{fig:SepOrigNokia}
    \end{subfigure}%
    \begin{subfigure}[t]{0.24\textwidth}
    \includegraphics[width=\textwidth]{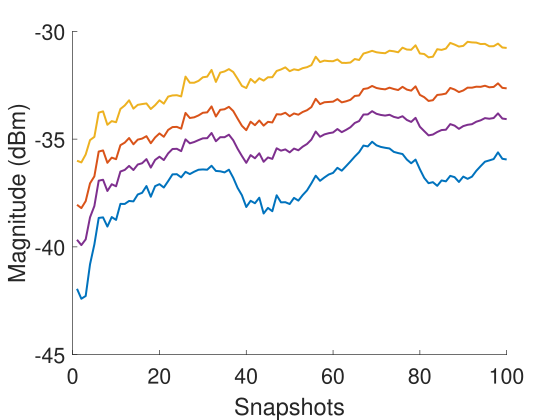}
    \caption{Predictable components}
    \label{fig:SepPredNokia}
    \end{subfigure}%
    \caption{Original signals and first PC of 4 neighbouring Alices in the Nokia dataset for SNR$=20~\text{dB}$}
    \label{fig:NokiaOrigPred}
\end{figure}

To showcase the impact of increasing TVD,  in Fig. \ref{fig:TBerlinSepSNR20}, we show the variation of the amplitude of the original CSI vs. time and that of the first PC vs. time for four neighbouring Alices. We observe that when compared to the original signal in Fig. \ref{fig:TBerlinSepSNR20}(\subref{fig:SepOrig}), the time series corresponding to the first PC in Fig. \ref{fig:TBerlinSepSNR20}(\subref{fig:SepPred}) are clearly distinguishable. Similar results are shown for the Nokia dataset in Fig. \ref{fig:NokiaTVD} and \ref{fig:NokiaOrigPred}, for which, notably, the increase in the separability is more accentuated.

\subsubsection{Secret key generation}
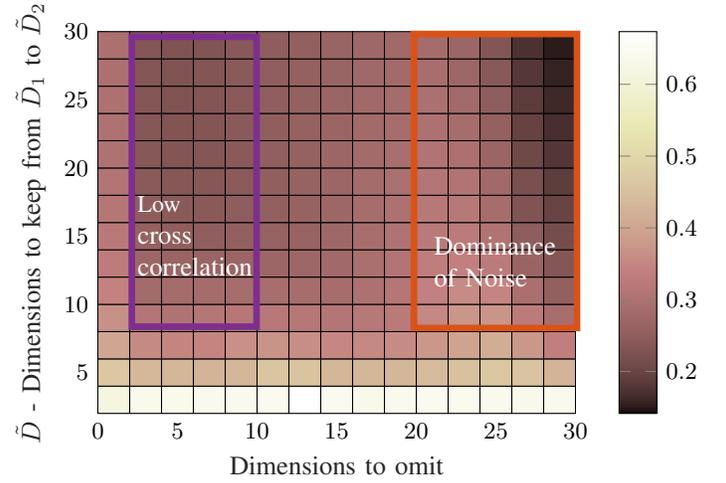
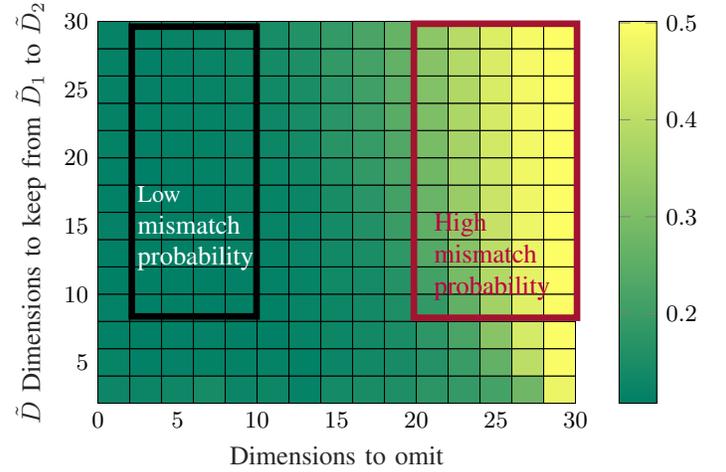
\begin{figure}[t]
    \centering
    \begin{subfigure}[t]{0.45\textwidth}
    \input{TIFS_figures_tikz/TBerlin_dist1_SNR20_corr}
    \caption{Average CC (Original= (0.30) }
    \label{fig:20corr}
    \end{subfigure}
    
    \begin{subfigure}[t]{0.45\textwidth}
    \input{TIFS_figures_tikz/TBerlin_dist1_SNR20_mismatch}
    \caption{Average MP}
    \label{fig:20mp}
    \end{subfigure}
    \caption{Trade-off between CC and MP for SNR $=20~\text{dB}$ for the Quadriga dataset. Darker colours indicate lower values.}
    \label{fig:tradeoff20}
\end{figure}

First, we study the effect of pre-processing using PCA for SNR $= 20~\text{dB}$ in Fig. \ref{fig:tradeoff20}, starting with the Quadriga dataset. Figs. \ref{fig:tradeoff20}(\subref{fig:20corr}), and (\subref{fig:20mp}) illustrate the variation of two metrics: i) the average CC between the locations and their nearest neighbours; and ii) the average MP between the Alices and Bob, respectively, with respect to the variation in the pair $\{\tilde{D}_1, \tilde{D}_2\}$ in steps of $2$. With no pre-processing, the average CC is approximately $0.30$. However, with a sufficient number of dimensions $\tilde{D}_1-\tilde{D}_2$ retained, an increase in the number of dimensions omitted $\tilde{D}_1-1$ results in a decrease in the CC. Specifically, for $\tilde{D}_1=2$ and $\tilde{D}_2 =20$, we observe a drop in the CC to $0.18$, with no significant increase in MP. 

\begin{figure}
    \centering
    \begin{subfigure}[t]{0.5\textwidth}
    \input{TIFS_figures_tikz/TBerlin_dist1_SNR5_corr}
    \caption{Average CC (Original= (0.21)}
    \label{fig:5corr}
    \end{subfigure}
        \begin{subfigure}[t]{0.5\textwidth}
    \input{TIFS_figures_tikz/TBerlin_dist1_SNR5_mismatch.tex}
    \caption{Average MP}
    \label{fig:5mp}
    \end{subfigure}%
    \caption{Trade-off between CC and MP for SNR $=  5~\text{dB}$ for the Quadriga dataset}
    \label{fig:TBerlindist1SNR5}
\end{figure}
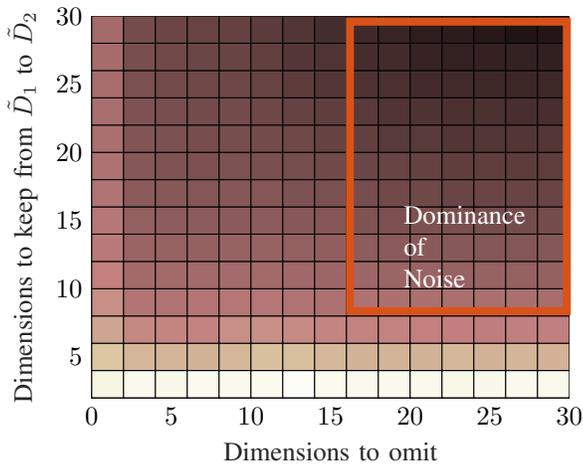
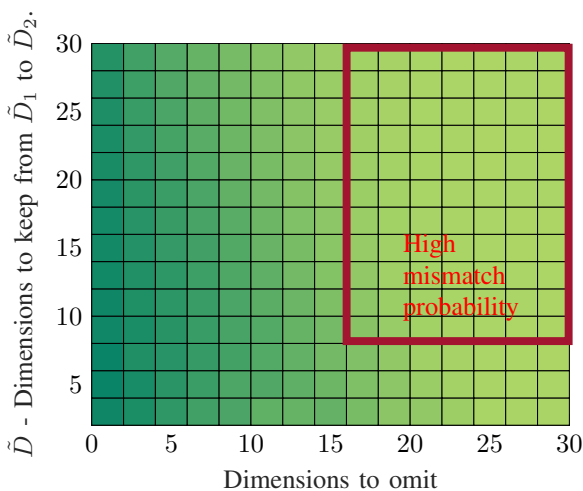

We posit that this range of PCs captures sufficiently well small scale fading terms. This regime is indicated as "Dominance of uncorrelated components" in Fig. \ref{fig:tradeoff20}(\subref{fig:20corr}) while the corresponding region is referred to as "Low Mismatch Probability" in Fig. \ref{fig:tradeoff20}(\subref{fig:20mp}).
Note that the drop in CC is more pronounced beyond $\tilde{D}_1=14$, beyond which noise becomes dominant, resulting in an increase of the MP. This regime is referred to as "Dominance of Noise" in Fig. \ref{fig:tradeoff20}(\subref{fig:20corr}). The corresponding region is marked as "High Mismatch Probability" in Fig. \ref{fig:tradeoff20}(\subref{fig:20mp}). 
In Fig. \ref{fig:TBerlindist1SNR5}, the trade-off between CC and MP is shown for SNR$=5~\text{dB}$. As expected, with a decrease in SNR, the effect of noise is more pronounced. Therefore, the regime of noise dominance and high MP is seen even at $\tilde{D}_1=10$. An important conclusion of this analysis is that for low SNRs it is possible to omit any pre-processing to avoid compromising the MP. 


A trend similar to CC is observed for $\overline{\Delta}$ in Fig. \ref{fig:quadrigahsic}, especially for higher values of $\tilde D_1$ indicating likely independence. On the other hand, $\overline{\Delta}$ for $\tilde D_1 <8$ does not follow the CC drop, indicating the limitations of CC compared to $\overline{\Delta}$ to capture dependence. For example, omitting the first $6$ PCs may guarantee a low CC but not a significant decrease in $\overline{}\Delta$ and statistical independence. The impact of the observation vector length on $\overline{\Delta}$ will be investigated in detail in future work.

\begin{figure}[t]
\centering
   \input{TIFS_figures_tikz/quadriga_hsic_new}
    \caption{Evolution of $\bar{\Delta}$  with $\tilde{D}_1$ for $\tilde D_2=30$ for the Quadriga dataset}
        \label{fig:quadrigahsic}
\end{figure}
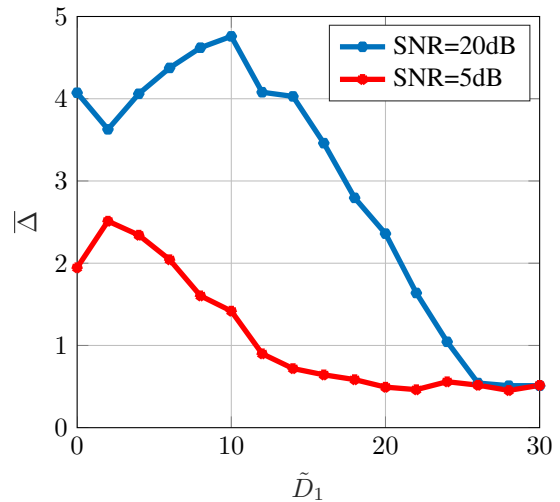

\begin{figure}[t]
    \centering
    \begin{subfigure}{0.5\textwidth}
    \input{TIFS_figures_tikz/Nokia_snr20_corr}
    \caption{Average CC (Original= (0.38) }
    \label{fig:Nokia20corr}
    \end{subfigure}
    
    \begin{subfigure}{0.5\textwidth}
    \input{TIFS_figures_tikz/Nokia_snr20_mp}
    \caption{Average MP}
    \label{fig:Nokia20mp}
    \end{subfigure}%
    \caption{Trade-off between CC and MP for SNR $=  20~\text{dB}$ for the Nokia dataset}
    \label{fig:NokiaSNR20}
\end{figure}

\begin{figure}
    \centering
    \begin{subfigure}[b]{0.5\textwidth}
    \input{TIFS_figures_tikz/Nokia_snr5_corr}
    \caption{Average CC (Original= (0.26) }
    \label{fig:Nokia5corr}
    \end{subfigure}
    
    \begin{subfigure}[b]{0.5\textwidth}
    \input{TIFS_figures_tikz/Nokia_snr5_mp}
    \caption{Average MP}
    \label{fig:Nokia5mp}
    \end{subfigure}%
    \caption{Trade-off between CC and MP for SNR $=  5~\text{dB}$ for the Nokia dataset}
    \label{fig:NokiaSNR5}
\end{figure}

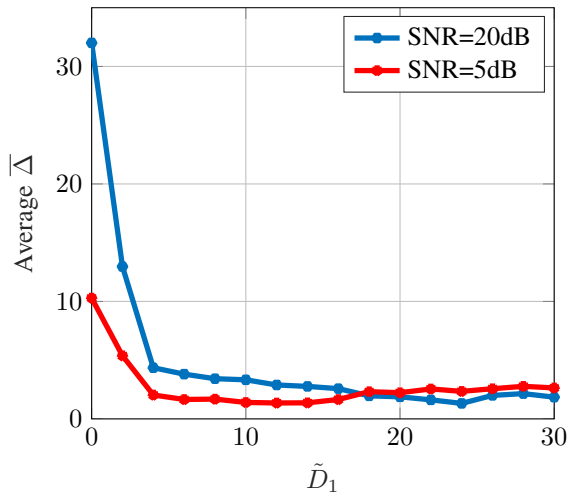
\begin{figure}[t]
\centering
   \input{TIFS_figures_tikz/nokia_hsic_new}
    \caption{Evolution of $\overline{\Delta}$ with $\tilde{D}_1$ for $\tilde{D}_2=30$ for the Nokia dataset}
        \label{fig:nokiahsic}
\end{figure}

Next, we present results for the Nokia dataset starting with SNR $= 20~\text{dB}$ in Fig. \ref{fig:NokiaSNR20}. With no pre-processing, the average CC is approximately $0.38$. However, with a sufficient number of dimensions retained, an increase in the number of dimensions omitted decreases the CC. Specifically, for $\tilde{D}_1=6$ and $\tilde{D}_2 =30$, we observe a drop in the CC to $0.15$, with no significant increase in MP. As in the Quadriga dataset, we posit that this is the regime in which the predominant large scale predictable components have been removed, and the small scale fading components have been retained. Importantly, a trend similar to CC is observed in the average $\overline{\Delta}$ in Fig. \ref{fig:nokiahsic}, especially for $\tilde{D}_1\geq 4$ after which value the dependence level collapses. Finally, similarly to Quadriga, for a low SNR=5 dB, any pre-processing would induce high MP as shown in Fig. 12 and is therefore advised to be omitted.

\subsection{KPCA}
 \begin{table}[t]
 \renewcommand{\arraystretch}{2.2}
 \caption{KPCA results for the Quadriga dataset, considering  SNR=$\{5,20\}$ dB. $\widehat{D} = 0$ denotes no pre-rpocessing}
\setlength{\tabcolsep}{1.2pt}
\centering 
\label{Table:KPCA1}
{\footnotesize
\begin{tabular}{|l|l|c|c|c|c|c|c|c|c|c|c|c|}
\hline
   & \multicolumn{1}{c|}{$\widehat{D}$} & 
  \multicolumn{1}{c|}{$0$} & \multicolumn{1}{c|}{$1$} & \multicolumn{1}{c|}{$2$} & \multicolumn{1}{c|}{$3$} & \multicolumn{1}{c|}{$4$} & \multicolumn{1}{c|}{$5$} & \multicolumn{1}{c|}{$6$} & \multicolumn{1}{c|}{$7$} & \multicolumn{1}{c|}{$8$} &
 \multicolumn{1}{c|}{$9$} &
 \multicolumn{1}{c|}{$10$}\\
\hline
\multirow{2}{*}{\STAB{\rotatebox[origin=c]{90}{SNR=5dB}}}  
& $\overline{\Delta}$ & $1.93$ & $1.7$ & $0.75$ & $0.66$ & $0.69$ & $0.61$ & $0.53$ & $0.54$ & $0.49$ & $0.49$ & $0.43$\\

& MP & $0.36$ & $0.38$ & $0.4$ & $0.41$ & $0.41$ & $0.42$ & $0.43$ & $0.43$ & $0.43$ & $0.43$ & $0.43$\\
\hline
\multirow{2}{*}{\STAB{\rotatebox[origin=c]{90}{SNR=20dB}}}  
& $\overline{\Delta}$ & $4.07$ & $3.93$ & $3.11$ & $3.04$ & $3.21$ & $3.2$ & $3.16$ & $3.04$ & $3.05$ & $3.04$ & $2.93$\\
& MP & $0.14$ & $0.14$ & $0.15$ & $0.15$ & $0.15$ & $0.15$ & $0.15$ & $0.16$ & $0.16$ & $0.16$ & $0.16$\\
\hline
\end{tabular}
}
\end{table}

 \begin{table}[t]
 \renewcommand{\arraystretch}{2.2}
 \caption{KPCA results for the NOKIA dataset, considering  SNR=$\{5,20\}$ dB.} 
\setlength{\tabcolsep}{0.8pt}
\centering 
\label{Table:KPCA2}
{\footnotesize
\begin{tabular}{|l|l|c|c|c|c|c|c|c|c|c|c|c|}
\hline
   & \multicolumn{1}{c|}{$\widehat{D}$} & 
  \multicolumn{1}{c|}{$0$} & \multicolumn{1}{c|}{$1$} & \multicolumn{1}{c|}{$2$} & \multicolumn{1}{c|}{$3$} & \multicolumn{1}{c|}{$4$} & \multicolumn{1}{c|}{$5$} & \multicolumn{1}{c|}{$6$} & \multicolumn{1}{c|}{$7$} & \multicolumn{1}{c|}{$8$} &
  \multicolumn{1}{c|}{$9$} &
  \multicolumn{1}{c|}{$10$}
\\
\hline
\multirow{2}{*}{\STAB{\rotatebox[origin=c]{90}{SNR=5dB}}}  
& Residual-$\overline{\Delta}$ & $10.22$ & $4.37$ & $2.74$ & $1.15$ & $0.83$ & $0.88$ & $0.71$ & $0.73$ & $0.72$ & $0.74$ & $0.73$\\
& MP & $0.39$ & $0.39$ & $0.46$ & $0.45$ & $0.44$ & $0.45$ & $0.46$ & $0.44$ & $0.46$ & $0.45$ & $0.46$\\
\hline
\multirow{2}{*}{\STAB{\rotatebox[origin=c]{90}{SNR=20dB}}}  
& Residual-$\overline{\Delta}$ & $32.11$ & $25.13$ & $17.78$ & $10.8$ & $9.7$ & $10.6$ & $9.6$ & $9.34$ & $9.55$ & $9.67$ & $9.74$\\

& MP & $0.1$ & $0.14$ & $0.16$ & $0.15$ & $0.14$ & $0.14$ & $0.16$ & $0.14$ & $0.2$ & $0.17$ & $0.22$ \\
\hline
\end{tabular}
}
\end{table} 

Next, we evaluate the performance of KPCA. Recall that, in this case, we only apply the parameter $\widehat{D}$ to derive the residuals that will be used as SKG seeds. Tables \ref{Table:KPCA1} and \ref{Table:KPCA2} indicate that an increase in $\widehat{D}$ leads to lower values of $\overline{\Delta}$ and higher values of MP. 
More precisely, focusing on Table \ref{Table:KPCA1}, $\overline{\Delta}$ undergoes a significant decrease for $\widehat{D}\geqslant{2}$, leading to a slight increase in MP. 
Moreover, comparing the outcomes of the PCA and the KPCA, KPCA seems to lead to a "faster" decrease of $\overline{\Delta}$. 
The results provided in Table \ref{Table:KPCA2} concern the NOKIA dataset and similarly show a significant decrease in $\overline{\Delta}$; overall, compared to PCA, KPCA  seems slightly more efficient in decreasing dependencies. 

\subsection{AE}
\begin{table}[t]
    \centering
    \caption{The layers and activation function for AE1. For AE2 the only change is that the dimensions of the input and the output layers are 400. }
    \begin{tabular}{|c|c|c|}
        \hline
        Layer & Dimensions & Activation \\
        \hline
        Input & 200  & Linear \\
                \hline
        1 & 100  & tanh \\
                \hline
2 & 50 & softplus \\
        \hline
3 & 20 & tanh \\
        \hline
Intermediate & $\widehat D$  & linear \\
        \hline
4 & 20  & relu \\
        \hline
5 & 50 & softplus \\
        \hline
6 & 100 & tanh \\
        \hline
        Output & 200  & Linear \\
                \hline
    \end{tabular}
    
    \label{tab:AE_Details}
\end{table}

The layers and the activation function of the AE are given in Table \ref{tab:AE_Details} and follow \cite{studer2018channel}. For brevity, the AE with MSE loss function is referred to as AE1 and that with dot-product loss function is referred to as AE2. The input to the AE2 is formed by grouping the $200 \times 1$ CSI vector ($100$ real and $100$ imaginary) of each spatial location with $200 \times 1$ long CSI vector from each of the $8$ nearest neighbours surrounding the location. In other words, the dimension at the input and the output is $400 \times 1$. This ensures that the loss function can minimize the correlation between the users while minimizing the reconstruction error between the input and the output. Two types of training are possible. Either Bob and the set of Alices train separate AEs with their local datasets in localized training, or, Bob trains a global AE whose parameters are distributed to the Alices in centralized training.

\begin{table}[t]
 \renewcommand{\arraystretch}{1.1}
 \caption{AE: Key results for Quadriga}
\setlength{\tabcolsep}{2.5pt}
\centering 
\label{Table:AEQuadriga}
{\footnotesize
\begin{tabular}{|l|c|c|c|c|c|c|c|c|}
\hline
\multicolumn{1}{|c|}{$\widehat{D}$} &  \multicolumn{4}{|c|}{$1$} &  \multicolumn{4}{|c|}{$8$}\\
\hline
\multicolumn{1}{|c|}{SNR (dB)} &  \multicolumn{2}{|c|}{$5$} &  \multicolumn{2}{|c|}{$20$} &  \multicolumn{2}{|c|}{$5$} &  \multicolumn{2}{|c|}{$20$}\\
\hline
\multicolumn{1}{|c|}{AE type} & 
 \multicolumn{1}{|c|}{$AE1$} 
& \multicolumn{1}{|c|}{$AE2$} &  
 \multicolumn{1}{|c|}{$AE1$} 
& \multicolumn{1}{|c|}{$AE2$} & 
 \multicolumn{1}{|c|}{$AE1$} 
& \multicolumn{1}{|c|}{$AE2$} &  
 \multicolumn{1}{|c|}{$AE1$} 
& \multicolumn{1}{|c|}{$AE2$}\\
\hline
Original-CC & $0.21$ & $0.21$ & $0.30$ & $0.30$ & $0.21$ & $0.21$ & $0.30$ & $0.30$   \\
\hline
Residual-CC & $0.20$ & $0.19$ & $0.27$ & $0.24$ & $0.18$ & $0.15$ & $0.25$ & $0.21$ \\
\hline
Original-$\overline{\Delta}$ & $2.18$ & $2.18$ & $4.5$ & $4.5$  & $2.18$ & $2.18$ & $4.5$ & $4.5$   \\
\hline
Residual-$\overline{\Delta}$ & $1.57$ & $1.11$ & $3.38$ & $2.8$ & $0.98$ & $0.65$ & $2.65$ & $2.6$   \\
\hline
MP for Localized & $0.37$ & $0.39$ & $0.19$ & $0.27$   & $0.39$ & $0.45$ & $0.17$ & $0.36$\\
\hline
MP for Centralized & $0.35$ & $0.36$ & $0.14$ & $0.15$   & $0.34$ & $0.40$ & $0.15$ & $0.15$\\
\hline
\end{tabular}
}
\end{table}

From Tables \ref{Table:AEQuadriga} and \ref{Table:AENokia}, note that, as in the case of PCA, the lower the SNR, the lower the CC and the higher the MP. Moreover, with an increase in the encoding dimensions $\widehat D$, the AE has more freedom to represent the predictable components. Therefore, with an increase in $\widehat D$, we observe a drop in the CC. For the Quadriga dataset, AE2 achieves a CC of $0.22$ for $\widehat D=8$ and SNR$=20~\text{dB}$ for the residual components, without a significant increase in MP for centralized training. This CC is almost equal to what the PCA achieves for the Quadriga dataset $\tilde{D}_1=2$ and $\tilde {D}_2=20$. For the Nokia dataset, AE2 achieves a CC of $0.07$ which is smaller than what PCA achieves in all cases. 

Also, the $\overline{\Delta}$ of the residuals shows a significant decrease from that of the original components, especially for SNR$=20~\text{dB}$. Observe that in AE2, since the loss function is the dot product between residuals instead of the MSE, we can observe a significant drop in CC of the residuals for AE2 compared to AE1. However, this is accompanied by an increase in the MP, especially in the case of localised training. Also, as expected, centralised training results in a much lower mismatch probability when compared to localised training.

 \begin{table}[t]
 \renewcommand{\arraystretch}{1.1}
 \caption{AE: Key results for Nokia dataset}
\setlength{\tabcolsep}{2.5pt}
\centering 
\label{Table:AENokia}
{\footnotesize
\begin{tabular}{|l|c|c|c|c|c|c|c|c|}
\hline
\multicolumn{1}{|c|}{$\widehat{D}$} &  \multicolumn{4}{|c|}{$1$} &  \multicolumn{4}{|c|}{$8$}\\
\hline
\multicolumn{1}{|c|}{SNR (dB)} &  \multicolumn{2}{|c|}{$5$} &  \multicolumn{2}{|c|}{$20$} &  \multicolumn{2}{|c|}{$5$} &  \multicolumn{2}{|c|}{$20$}\\
\hline
\multicolumn{1}{|c|}{AE type} & 
 \multicolumn{1}{|c|}{$AE1$} 
& \multicolumn{1}{|c|}{$AE2$} &  
 \multicolumn{1}{|c|}{$AE1$} 
& \multicolumn{1}{|c|}{$AE2$} & 
 \multicolumn{1}{|c|}{$AE1$} 
& \multicolumn{1}{|c|}{$AE2$} &  
 \multicolumn{1}{|c|}{$AE1$} 
& \multicolumn{1}{|c|}{$AE2$}\\
\hline
Original-CC & $0.30$ & $0.30$ & $0.39$ & $0.39$ & $0.30$ & $0.30$ & $0.39$ & $0.39$   \\
\hline
Residual-CC & $0.13$ & $0.11$ & $0.25$ & $0.17$ & $0.04$ & $0.05$ & $0.10$ & $0.07$ \\
\hline
Original-$\overline{\Delta}$ & $15.7$ & $15.7$ & $33.1$ & $33.1$ & $15.7$ & $15.7$ & $33.1$ & $33.1$   \\
\hline
Residual-$\overline{\Delta}$ & $2.51$ & $2.17$ & $11.2$ & $8.81$ & $0.14$ & $1.20$ & $1.92$ & $4.16$   \\
\hline
MP for Centralised & $0.43$ & $0.34$ & $0.15$ & $0.13$   & $0.49$ & $0.36$ & $0.29$ & $0.13$\\
\hline
MP for Localised & $0.47$ & $0.44$ & $0.37$ & $0.40$   & $0.49$ & $0.42$ & $0.36$ & $0.34$\\
\hline
\end{tabular}
}
\end{table}

A very similar trend is observable in the case of the Nokia dataset also. In this case, the residual CC for $\widehat D=8$ and SNR$=20~\text{dB}$, is much lower than what PCA attains for the same data set for $\tilde{D}_1=6$ and $\tilde {D}_2=30$.


\section{Discussion and Conclusions}
In this paper, we provided generic guidelines for the incorporation of PLS in 6G security protocols. To this end, we identified online learning, dimensionality reduction and the design of hybrid PLS-crypto systems as promising key approaches. To showcase some of these elements, we built and evaluated PCA and AE based pre-processing approaches for disentangling predictable from unpredictable components of observed CSI vectors, in order to perform jointly RF fingerprinting and SKG. We discussed in detail the trade-off between correlations and dependencies at different locations (generalised to include time, frequency and antenna domains) and reciprocity, or, the lack of mismatch between the uplink and downlink for the unpredictable components used for SKG. We have proposed the use of TVD as a separability measure of empirical fingerprints and a novel metric for statistical dependence, in order to systematise pre-processing criteria.

Overall, PCA has proven to be a sufficiently straightforward approach for the disentanglement of large from small scale fading terms in observed CSI and can be incorporated as a pre-processing tool to allow for the simultaneous use of CSI for authentication and key distillation. While KPCA and AE have been shown to increase performance as they could capture nonlinear structures, this came at a loss in explainability of the results. 
In future work, we will explore in more detail time, frequency and antenna domain dependencies, the use of time domain separation techniques (preliminary results using Kalman filters have provided promising performance) and finally the incorporation of $\overline{\Delta}$ in the AE loss-function.   


\bibliography{library.bib}
\bibliographystyle{IEEEtran}

\end{document}

%% file: TIFS_figures_tikz/tifs_quadriga_tvd.tex
%
%
\definecolor{mycolor1}{rgb}{0.00000,0.44700,0.74100}%
\definecolor{mycolor2}{rgb}{0.85098,0.32549,0.09804}%
\begin{tikzpicture}

\begin{axis}[%
width=2.421in,
height=2.154in,
at={(0.758in,0.692in)},
scale only axis,
xmin=0,
xmax=30,
xlabel style={font=\color{white!15!black}},
xlabel={No. of dimensions $\hat{D}$},
ymin=0.1,
ymax=0.456294511290081,
ylabel style={font=\color{white!15!black}},
ylabel={Total variation distance},
axis background/.style={fill=white},
xmajorgrids,
ymajorgrids,
legend style={legend cell align=left, align=left, draw=white!15!black}
]
\addplot [color=mycolor1, line width=1.5pt, mark=asterisk, mark options={solid, mycolor1}]
  table[row sep=crcr]{%
0	0.147518852523433\\
1	0.416294511290079\\
2	0.271572536892361\\
3	0.239636498387132\\
4	0.218075180576989\\
5	0.203330376184198\\
6	0.18914753059628\\
7	0.183791256081747\\
8	0.177251954464591\\
9	0.172363783596964\\
10	0.168337054049854\\
11	0.164658618398491\\
12	0.161442676852925\\
13	0.159814235441313\\
14	0.157158996833203\\
15	0.156670883032191\\
16	0.154607660322359\\
17	0.152737506296081\\
18	0.152839315280993\\
19	0.15121397167567\\
20	0.151005330236195\\
21	0.150704926751114\\
22	0.149647821287722\\
23	0.149105621463477\\
24	0.149041572225435\\
25	0.149070624624915\\
26	0.148523819953489\\
27	0.149470074186599\\
28	0.148305717646178\\
29	0.148432309081361\\
30	0.148891370482897\\
};
\addlegendentry{SNR=20dB}

\addplot [color=mycolor2, line width=1.5pt, mark=o, mark options={solid, mycolor2}]
  table[row sep=crcr]{%
0	0.115787210625822\\
1	0.411187400870197\\
2	0.274006742835862\\
3	0.244311004345636\\
4	0.218243466810272\\
5	0.203113278570815\\
6	0.190877194921338\\
7	0.183181574315199\\
8	0.175685134280691\\
9	0.172086153174298\\
10	0.166672778554741\\
11	0.16305763862097\\
12	0.16148638103888\\
13	0.156809781968022\\
14	0.155859174061213\\
15	0.155059772590878\\
16	0.153300134896906\\
17	0.15055246444723\\
18	0.148748871393817\\
19	0.148253138663836\\
20	0.146752125932355\\
21	0.147175939321844\\
22	0.145443512088477\\
23	0.145866488233025\\
24	0.144814155065802\\
25	0.145205232178068\\
26	0.144723146540638\\
27	0.142938056600439\\
28	0.143271531260716\\
29	0.141810873735427\\
30	0.141531150200404\\
};
\addlegendentry{SNR=5dB}

\end{axis}

\end{tikzpicture}%

%% file: TIFS_figures_tikz/tifs_nokia_tvd.tex
%
%
\definecolor{mycolor1}{rgb}{0.00000,0.44700,0.74100}%
\definecolor{mycolor2}{rgb}{0.85098,0.32549,0.09804}%
\begin{tikzpicture}

\begin{axis}[%
width=2.421in,
height=2.154in,
at={(0.758in,0.692in)},
scale only axis,
xmin=0,
xmax=30,
xlabel style={font=\color{white!15!black}},
xlabel={ No of dimensions $\hat{D}$},
ymin=0.15,
ymax=0.35,
ylabel style={font=\color{white!15!black}},
ylabel={Total Variation Distance},
axis background/.style={fill=white},
xmajorgrids,
ymajorgrids,
legend style={legend cell align=left, align=left, draw=white!15!black}
]
\addplot [color=mycolor1, line width=2.0pt, mark=asterisk, mark options={solid, mycolor1}]
  table[row sep=crcr]{%
0	0.228866522879464\\
1	0.33649955203683\\
2	0.254553633091518\\
3	0.26024619656808\\
4	0.252738869001117\\
5	0.247740222377232\\
6	0.245849941266741\\
7	0.241272139369421\\
8	0.236085875976563\\
9	0.237695409179686\\
10	0.234965461216518\\
11	0.232562541992188\\
12	0.232934175223214\\
13	0.23255350013951\\
14	0.231876077706474\\
15	0.233109443219867\\
16	0.232257107979912\\
17	0.232506396344867\\
18	0.232231022739956\\
19	0.232542044642859\\
20	0.232873643973214\\
21	0.232525369838168\\
22	0.233441921595983\\
23	0.231515534737724\\
24	0.2322865078125\\
25	0.233111185825894\\
26	0.231614341099331\\
27	0.232104881696429\\
28	0.231762384486608\\
29	0.231052432198659\\
30	0.230313756277901\\
};
\addlegendentry{SNR=20dB}

\addplot [color=mycolor2, line width=2.0pt, mark=o, mark options={solid, mycolor2}]
  table[row sep=crcr]{%
0	0.174951345424109\\
1	0.337317793387278\\
2	0.261628530691965\\
3	0.26631505984933\\
4	0.257515328543526\\
5	0.251309777901785\\
6	0.248522161272323\\
7	0.242925131835939\\
8	0.240646509626117\\
9	0.236915663085938\\
10	0.237822964006696\\
11	0.235164848493305\\
12	0.235010468470982\\
13	0.23246756752232\\
14	0.232635776227678\\
15	0.231883209821429\\
16	0.23072251422991\\
17	0.22988702078683\\
18	0.229256564732143\\
19	0.227510562639509\\
20	0.227347090262278\\
21	0.228414868722098\\
22	0.225148081612723\\
23	0.224304107142856\\
24	0.22560049497768\\
25	0.22468987234933\\
26	0.221966053152901\\
27	0.223322282505581\\
28	0.222284413364957\\
29	0.220536790178571\\
30	0.218485758370537\\
};
\addlegendentry{SNR=5dB}

\end{axis}

\end{tikzpicture}%

%% file: TIFS_figures_tikz/TBerlin_dist1_SNR20_corr.tex
%
%
\definecolor{mycolor1}{rgb}{0.85098,0.32549,0.09804}%
\definecolor{mycolor2}{rgb}{0.49412,0.18431,0.55686}%
\begin{tikzpicture}

\begin{axis}[%
width=2.5in,
height=2.000in,
at={(0.666in,0.569in)},
scale only axis,
point meta min=0.141548192733656,
point meta max=0.673381198915271,
xmin=0,
xmax=30,
xlabel style={font=\color{white!15!black}},
xlabel={Dimensions to omit},
ymin=2,
ymax=30,
ylabel style={font=\color{white!15!black}},
ylabel={$\tilde{D}$ - Dimensions to keep from $\tilde{D}_1$ to $\tilde{D}_2$},
tick label style={font=\small},
axis background/.style={fill=white},
axis x line*=bottom,
axis y line*=left,
colormap={mymap}{[1pt] rgb(0pt)=(0.0589256,0,0); rgb(1pt)=(0.0977692,0.051131,0.051131); rgb(2pt)=(0.125082,0.0723102,0.0723102); rgb(3pt)=(0.147418,0.0885615,0.0885615); rgb(4pt)=(0.166789,0.102262,0.102262); rgb(5pt)=(0.184134,0.114332,0.114332); rgb(6pt)=(0.19998,0.125245,0.125245); rgb(7pt)=(0.214659,0.13528,0.13528); rgb(8pt)=(0.228397,0.14462,0.14462); rgb(9pt)=(0.241354,0.153393,0.153393); rgb(10pt)=(0.25365,0.16169,0.16169); rgb(11pt)=(0.265377,0.169582,0.169582); rgb(12pt)=(0.276607,0.177123,0.177123); rgb(13pt)=(0.287399,0.184355,0.184355); rgb(14pt)=(0.2978,0.191315,0.191315); rgb(15pt)=(0.307849,0.19803,0.19803); rgb(16pt)=(0.317581,0.204524,0.204524); rgb(17pt)=(0.327024,0.210819,0.210819); rgb(18pt)=(0.336201,0.21693,0.21693); rgb(19pt)=(0.345134,0.222875,0.222875); rgb(20pt)=(0.353842,0.228665,0.228665); rgb(21pt)=(0.362341,0.234312,0.234312); rgb(22pt)=(0.370645,0.239826,0.239826); rgb(23pt)=(0.378766,0.245216,0.245216); rgb(24pt)=(0.386718,0.25049,0.25049); rgb(25pt)=(0.394509,0.255655,0.255655); rgb(26pt)=(0.402149,0.260718,0.260718); rgb(27pt)=(0.409647,0.265684,0.265684); rgb(28pt)=(0.41701,0.27056,0.27056); rgb(29pt)=(0.424245,0.275349,0.275349); rgb(30pt)=(0.431359,0.280056,0.280056); rgb(31pt)=(0.438357,0.284685,0.284685); rgb(32pt)=(0.445245,0.289241,0.289241); rgb(33pt)=(0.452029,0.293725,0.293725); rgb(34pt)=(0.458712,0.298142,0.298142); rgb(35pt)=(0.465299,0.302495,0.302495); rgb(36pt)=(0.471794,0.306786,0.306786); rgb(37pt)=(0.478201,0.311018,0.311018); rgb(38pt)=(0.484524,0.315193,0.315193); rgb(39pt)=(0.490764,0.319313,0.319313); rgb(40pt)=(0.496927,0.323381,0.323381); rgb(41pt)=(0.503014,0.327398,0.327398); rgb(42pt)=(0.509028,0.331367,0.331367); rgb(43pt)=(0.514972,0.335288,0.335288); rgb(44pt)=(0.520848,0.339165,0.339165); rgb(45pt)=(0.526659,0.342997,0.342997); rgb(46pt)=(0.532406,0.346787,0.346787); rgb(47pt)=(0.538092,0.350536,0.350536); rgb(48pt)=(0.543718,0.354246,0.354246); rgb(49pt)=(0.549287,0.357917,0.357917); rgb(50pt)=(0.554799,0.361551,0.361551); rgb(51pt)=(0.560258,0.365148,0.365148); rgb(52pt)=(0.565664,0.368711,0.368711); rgb(53pt)=(0.571018,0.372239,0.372239); rgb(54pt)=(0.576323,0.375735,0.375735); rgb(55pt)=(0.58158,0.379198,0.379198); rgb(56pt)=(0.586789,0.382629,0.382629); rgb(57pt)=(0.591953,0.386031,0.386031); rgb(58pt)=(0.597072,0.389402,0.389402); rgb(59pt)=(0.602148,0.392745,0.392745); rgb(60pt)=(0.607181,0.396059,0.396059); rgb(61pt)=(0.612172,0.399346,0.399346); rgb(62pt)=(0.617124,0.402606,0.402606); rgb(63pt)=(0.622035,0.40584,0.40584); rgb(64pt)=(0.626909,0.409048,0.409048); rgb(65pt)=(0.631745,0.412231,0.412231); rgb(66pt)=(0.636544,0.41539,0.41539); rgb(67pt)=(0.641307,0.418525,0.418525); rgb(68pt)=(0.646035,0.421637,0.421637); rgb(69pt)=(0.650729,0.424726,0.424726); rgb(70pt)=(0.655389,0.427793,0.427793); rgb(71pt)=(0.660016,0.430837,0.430837); rgb(72pt)=(0.664611,0.433861,0.433861); rgb(73pt)=(0.669174,0.436863,0.436863); rgb(74pt)=(0.673707,0.439845,0.439845); rgb(75pt)=(0.678209,0.442807,0.442807); rgb(76pt)=(0.682681,0.44575,0.44575); rgb(77pt)=(0.687125,0.448673,0.448673); rgb(78pt)=(0.69154,0.451577,0.451577); rgb(79pt)=(0.695927,0.454462,0.454462); rgb(80pt)=(0.700286,0.45733,0.45733); rgb(81pt)=(0.704618,0.460179,0.460179); rgb(82pt)=(0.708924,0.463011,0.463011); rgb(83pt)=(0.713204,0.465826,0.465826); rgb(84pt)=(0.717459,0.468623,0.468623); rgb(85pt)=(0.721688,0.471405,0.471405); rgb(86pt)=(0.725893,0.474169,0.474169); rgb(87pt)=(0.730073,0.476918,0.476918); rgb(88pt)=(0.73423,0.479651,0.479651); rgb(89pt)=(0.738363,0.482369,0.482369); rgb(90pt)=(0.742473,0.485071,0.485071); rgb(91pt)=(0.746561,0.487759,0.487759); rgb(92pt)=(0.750626,0.490431,0.490431); rgb(93pt)=(0.75467,0.493089,0.493089); rgb(94pt)=(0.758691,0.495733,0.495733); rgb(95pt)=(0.762692,0.498363,0.498363); rgb(96pt)=(0.764404,0.504433,0.500979); rgb(97pt)=(0.766112,0.51043,0.503582); rgb(98pt)=(0.767817,0.516358,0.506171); rgb(99pt)=(0.769517,0.522219,0.508747); rgb(100pt)=(0.771214,0.528014,0.51131); rgb(101pt)=(0.772907,0.533747,0.51386); rgb(102pt)=(0.774597,0.539418,0.516398); rgb(103pt)=(0.776282,0.545031,0.518923); rgb(104pt)=(0.777964,0.550586,0.521436); rgb(105pt)=(0.779643,0.556086,0.523937); rgb(106pt)=(0.781318,0.561532,0.526426); rgb(107pt)=(0.782989,0.566926,0.528903); rgb(108pt)=(0.784657,0.572269,0.531369); rgb(109pt)=(0.786321,0.577562,0.533823); rgb(110pt)=(0.787982,0.582808,0.536266); rgb(111pt)=(0.789639,0.588006,0.538699); rgb(112pt)=(0.791292,0.59316,0.54112); rgb(113pt)=(0.792943,0.598268,0.54353); rgb(114pt)=(0.79459,0.603334,0.54593); rgb(115pt)=(0.796233,0.608357,0.548319); rgb(116pt)=(0.797873,0.613339,0.550698); rgb(117pt)=(0.79951,0.618281,0.553066); rgb(118pt)=(0.801143,0.623184,0.555425); rgb(119pt)=(0.802773,0.628048,0.557773); rgb(120pt)=(0.8044,0.632875,0.560112); rgb(121pt)=(0.806023,0.637666,0.562441); rgb(122pt)=(0.807643,0.642421,0.56476); rgb(123pt)=(0.80926,0.647141,0.56707); rgb(124pt)=(0.810874,0.651826,0.569371); rgb(125pt)=(0.812484,0.656479,0.571662); rgb(126pt)=(0.814092,0.661098,0.573944); rgb(127pt)=(0.815696,0.665686,0.576217); rgb(128pt)=(0.817297,0.670242,0.578481); rgb(129pt)=(0.818895,0.674767,0.580737); rgb(130pt)=(0.820489,0.679262,0.582983); rgb(131pt)=(0.822081,0.683728,0.585221); rgb(132pt)=(0.823669,0.688164,0.58745); rgb(133pt)=(0.825255,0.692573,0.589671); rgb(134pt)=(0.826837,0.696953,0.591884); rgb(135pt)=(0.828417,0.701306,0.594089); rgb(136pt)=(0.829993,0.705632,0.596285); rgb(137pt)=(0.831567,0.709932,0.598473); rgb(138pt)=(0.833137,0.714206,0.600653); rgb(139pt)=(0.834705,0.718454,0.602826); rgb(140pt)=(0.836269,0.722678,0.60499); rgb(141pt)=(0.837831,0.726877,0.607147); rgb(142pt)=(0.83939,0.731051,0.609296); rgb(143pt)=(0.840946,0.735203,0.611438); rgb(144pt)=(0.842499,0.73933,0.613572); rgb(145pt)=(0.844049,0.743435,0.615699); rgb(146pt)=(0.845596,0.747518,0.617818); rgb(147pt)=(0.847141,0.751578,0.61993); rgb(148pt)=(0.848682,0.755616,0.622035); rgb(149pt)=(0.850221,0.759633,0.624133); rgb(150pt)=(0.851757,0.763629,0.626224); rgb(151pt)=(0.85329,0.767604,0.628308); rgb(152pt)=(0.854821,0.771558,0.630385); rgb(153pt)=(0.856349,0.775493,0.632456); rgb(154pt)=(0.857874,0.779407,0.634519); rgb(155pt)=(0.859396,0.783302,0.636576); rgb(156pt)=(0.860916,0.787178,0.638626); rgb(157pt)=(0.862433,0.791034,0.64067); rgb(158pt)=(0.863947,0.794872,0.642707); rgb(159pt)=(0.865459,0.798692,0.644737); rgb(160pt)=(0.866968,0.802493,0.646762); rgb(161pt)=(0.868475,0.806276,0.64878); rgb(162pt)=(0.869979,0.810042,0.650791); rgb(163pt)=(0.87148,0.81379,0.652797); rgb(164pt)=(0.872979,0.817522,0.654796); rgb(165pt)=(0.874475,0.821236,0.65679); rgb(166pt)=(0.875968,0.824933,0.658777); rgb(167pt)=(0.877459,0.828614,0.660758); rgb(168pt)=(0.878948,0.832279,0.662733); rgb(169pt)=(0.880434,0.835927,0.664703); rgb(170pt)=(0.881917,0.83956,0.666667); rgb(171pt)=(0.883398,0.843177,0.668625); rgb(172pt)=(0.884877,0.846779,0.670577); rgb(173pt)=(0.886353,0.850365,0.672523); rgb(174pt)=(0.887826,0.853936,0.674464); rgb(175pt)=(0.889297,0.857493,0.6764); rgb(176pt)=(0.890766,0.861035,0.678329); rgb(177pt)=(0.892232,0.864562,0.680254); rgb(178pt)=(0.893696,0.868075,0.682173); rgb(179pt)=(0.895158,0.871574,0.684086); rgb(180pt)=(0.896617,0.875058,0.685994); rgb(181pt)=(0.898073,0.878529,0.687897); rgb(182pt)=(0.899528,0.881987,0.689795); rgb(183pt)=(0.90098,0.88543,0.691687); rgb(184pt)=(0.90243,0.888861,0.693575); rgb(185pt)=(0.903877,0.892278,0.695457); rgb(186pt)=(0.905322,0.895682,0.697334); rgb(187pt)=(0.906765,0.899074,0.699206); rgb(188pt)=(0.908205,0.902452,0.701073); rgb(189pt)=(0.909643,0.905818,0.702935); rgb(190pt)=(0.911079,0.909172,0.704792); rgb(191pt)=(0.912513,0.912513,0.706644); rgb(192pt)=(0.913944,0.913944,0.712158); rgb(193pt)=(0.915373,0.915373,0.717629); rgb(194pt)=(0.9168,0.9168,0.723059); rgb(195pt)=(0.918225,0.918225,0.728449); rgb(196pt)=(0.919648,0.919648,0.733798); rgb(197pt)=(0.921068,0.921068,0.739109); rgb(198pt)=(0.922486,0.922486,0.744383); rgb(199pt)=(0.923902,0.923902,0.749619); rgb(200pt)=(0.925316,0.925316,0.754818); rgb(201pt)=(0.926727,0.926727,0.759983); rgb(202pt)=(0.928137,0.928137,0.765112); rgb(203pt)=(0.929544,0.929544,0.770207); rgb(204pt)=(0.930949,0.930949,0.775269); rgb(205pt)=(0.932352,0.932352,0.780298); rgb(206pt)=(0.933753,0.933753,0.785294); rgb(207pt)=(0.935152,0.935152,0.790259); rgb(208pt)=(0.936549,0.936549,0.795193); rgb(209pt)=(0.937944,0.937944,0.800097); rgb(210pt)=(0.939336,0.939336,0.804971); rgb(211pt)=(0.940727,0.940727,0.809815); rgb(212pt)=(0.942116,0.942116,0.814631); rgb(213pt)=(0.943502,0.943502,0.819418); rgb(214pt)=(0.944886,0.944886,0.824178); rgb(215pt)=(0.946269,0.946269,0.82891); rgb(216pt)=(0.947649,0.947649,0.833615); rgb(217pt)=(0.949028,0.949028,0.838294); rgb(218pt)=(0.950404,0.950404,0.842947); rgb(219pt)=(0.951779,0.951779,0.847574); rgb(220pt)=(0.953151,0.953151,0.852177); rgb(221pt)=(0.954521,0.954521,0.856754); rgb(222pt)=(0.95589,0.95589,0.861307); rgb(223pt)=(0.957256,0.957256,0.865837); rgb(224pt)=(0.958621,0.958621,0.870342); rgb(225pt)=(0.959984,0.959984,0.874825); rgb(226pt)=(0.961344,0.961344,0.879285); rgb(227pt)=(0.962703,0.962703,0.883722); rgb(228pt)=(0.96406,0.96406,0.888137); rgb(229pt)=(0.965415,0.965415,0.89253); rgb(230pt)=(0.966768,0.966768,0.896901); rgb(231pt)=(0.968119,0.968119,0.901252); rgb(232pt)=(0.969469,0.969469,0.905581); rgb(233pt)=(0.970816,0.970816,0.90989); rgb(234pt)=(0.972162,0.972162,0.914179); rgb(235pt)=(0.973505,0.973505,0.918447); rgb(236pt)=(0.974847,0.974847,0.922696); rgb(237pt)=(0.976187,0.976187,0.926926); rgb(238pt)=(0.977525,0.977525,0.931136); rgb(239pt)=(0.978862,0.978862,0.935327); rgb(240pt)=(0.980196,0.980196,0.9395); rgb(241pt)=(0.981529,0.981529,0.943654); rgb(242pt)=(0.98286,0.98286,0.947789); rgb(243pt)=(0.984189,0.984189,0.951907); rgb(244pt)=(0.985516,0.985516,0.956007); rgb(245pt)=(0.986842,0.986842,0.96009); rgb(246pt)=(0.988165,0.988165,0.964155); rgb(247pt)=(0.989487,0.989487,0.968204); rgb(248pt)=(0.990807,0.990807,0.972235); rgb(249pt)=(0.992126,0.992126,0.97625); rgb(250pt)=(0.993443,0.993443,0.980248); rgb(251pt)=(0.994757,0.994757,0.98423); rgb(252pt)=(0.996071,0.996071,0.988196); rgb(253pt)=(0.997382,0.997382,0.992146); rgb(254pt)=(0.998692,0.998692,0.996081); rgb(255pt)=(1,1,1)},
colorbar
]

\addplot[%
surf,
shader=flat corner, draw=black, colormap={mymap}{[1pt] rgb(0pt)=(0.0589256,0,0); rgb(1pt)=(0.0977692,0.051131,0.051131); rgb(2pt)=(0.125082,0.0723102,0.0723102); rgb(3pt)=(0.147418,0.0885615,0.0885615); rgb(4pt)=(0.166789,0.102262,0.102262); rgb(5pt)=(0.184134,0.114332,0.114332); rgb(6pt)=(0.19998,0.125245,0.125245); rgb(7pt)=(0.214659,0.13528,0.13528); rgb(8pt)=(0.228397,0.14462,0.14462); rgb(9pt)=(0.241354,0.153393,0.153393); rgb(10pt)=(0.25365,0.16169,0.16169); rgb(11pt)=(0.265377,0.169582,0.169582); rgb(12pt)=(0.276607,0.177123,0.177123); rgb(13pt)=(0.287399,0.184355,0.184355); rgb(14pt)=(0.2978,0.191315,0.191315); rgb(15pt)=(0.307849,0.19803,0.19803); rgb(16pt)=(0.317581,0.204524,0.204524); rgb(17pt)=(0.327024,0.210819,0.210819); rgb(18pt)=(0.336201,0.21693,0.21693); rgb(19pt)=(0.345134,0.222875,0.222875); rgb(20pt)=(0.353842,0.228665,0.228665); rgb(21pt)=(0.362341,0.234312,0.234312); rgb(22pt)=(0.370645,0.239826,0.239826); rgb(23pt)=(0.378766,0.245216,0.245216); rgb(24pt)=(0.386718,0.25049,0.25049); rgb(25pt)=(0.394509,0.255655,0.255655); rgb(26pt)=(0.402149,0.260718,0.260718); rgb(27pt)=(0.409647,0.265684,0.265684); rgb(28pt)=(0.41701,0.27056,0.27056); rgb(29pt)=(0.424245,0.275349,0.275349); rgb(30pt)=(0.431359,0.280056,0.280056); rgb(31pt)=(0.438357,0.284685,0.284685); rgb(32pt)=(0.445245,0.289241,0.289241); rgb(33pt)=(0.452029,0.293725,0.293725); rgb(34pt)=(0.458712,0.298142,0.298142); rgb(35pt)=(0.465299,0.302495,0.302495); rgb(36pt)=(0.471794,0.306786,0.306786); rgb(37pt)=(0.478201,0.311018,0.311018); rgb(38pt)=(0.484524,0.315193,0.315193); rgb(39pt)=(0.490764,0.319313,0.319313); rgb(40pt)=(0.496927,0.323381,0.323381); rgb(41pt)=(0.503014,0.327398,0.327398); rgb(42pt)=(0.509028,0.331367,0.331367); rgb(43pt)=(0.514972,0.335288,0.335288); rgb(44pt)=(0.520848,0.339165,0.339165); rgb(45pt)=(0.526659,0.342997,0.342997); rgb(46pt)=(0.532406,0.346787,0.346787); rgb(47pt)=(0.538092,0.350536,0.350536); rgb(48pt)=(0.543718,0.354246,0.354246); rgb(49pt)=(0.549287,0.357917,0.357917); rgb(50pt)=(0.554799,0.361551,0.361551); rgb(51pt)=(0.560258,0.365148,0.365148); rgb(52pt)=(0.565664,0.368711,0.368711); rgb(53pt)=(0.571018,0.372239,0.372239); rgb(54pt)=(0.576323,0.375735,0.375735); rgb(55pt)=(0.58158,0.379198,0.379198); rgb(56pt)=(0.586789,0.382629,0.382629); rgb(57pt)=(0.591953,0.386031,0.386031); rgb(58pt)=(0.597072,0.389402,0.389402); rgb(59pt)=(0.602148,0.392745,0.392745); rgb(60pt)=(0.607181,0.396059,0.396059); rgb(61pt)=(0.612172,0.399346,0.399346); rgb(62pt)=(0.617124,0.402606,0.402606); rgb(63pt)=(0.622035,0.40584,0.40584); rgb(64pt)=(0.626909,0.409048,0.409048); rgb(65pt)=(0.631745,0.412231,0.412231); rgb(66pt)=(0.636544,0.41539,0.41539); rgb(67pt)=(0.641307,0.418525,0.418525); rgb(68pt)=(0.646035,0.421637,0.421637); rgb(69pt)=(0.650729,0.424726,0.424726); rgb(70pt)=(0.655389,0.427793,0.427793); rgb(71pt)=(0.660016,0.430837,0.430837); rgb(72pt)=(0.664611,0.433861,0.433861); rgb(73pt)=(0.669174,0.436863,0.436863); rgb(74pt)=(0.673707,0.439845,0.439845); rgb(75pt)=(0.678209,0.442807,0.442807); rgb(76pt)=(0.682681,0.44575,0.44575); rgb(77pt)=(0.687125,0.448673,0.448673); rgb(78pt)=(0.69154,0.451577,0.451577); rgb(79pt)=(0.695927,0.454462,0.454462); rgb(80pt)=(0.700286,0.45733,0.45733); rgb(81pt)=(0.704618,0.460179,0.460179); rgb(82pt)=(0.708924,0.463011,0.463011); rgb(83pt)=(0.713204,0.465826,0.465826); rgb(84pt)=(0.717459,0.468623,0.468623); rgb(85pt)=(0.721688,0.471405,0.471405); rgb(86pt)=(0.725893,0.474169,0.474169); rgb(87pt)=(0.730073,0.476918,0.476918); rgb(88pt)=(0.73423,0.479651,0.479651); rgb(89pt)=(0.738363,0.482369,0.482369); rgb(90pt)=(0.742473,0.485071,0.485071); rgb(91pt)=(0.746561,0.487759,0.487759); rgb(92pt)=(0.750626,0.490431,0.490431); rgb(93pt)=(0.75467,0.493089,0.493089); rgb(94pt)=(0.758691,0.495733,0.495733); rgb(95pt)=(0.762692,0.498363,0.498363); rgb(96pt)=(0.764404,0.504433,0.500979); rgb(97pt)=(0.766112,0.51043,0.503582); rgb(98pt)=(0.767817,0.516358,0.506171); rgb(99pt)=(0.769517,0.522219,0.508747); rgb(100pt)=(0.771214,0.528014,0.51131); rgb(101pt)=(0.772907,0.533747,0.51386); rgb(102pt)=(0.774597,0.539418,0.516398); rgb(103pt)=(0.776282,0.545031,0.518923); rgb(104pt)=(0.777964,0.550586,0.521436); rgb(105pt)=(0.779643,0.556086,0.523937); rgb(106pt)=(0.781318,0.561532,0.526426); rgb(107pt)=(0.782989,0.566926,0.528903); rgb(108pt)=(0.784657,0.572269,0.531369); rgb(109pt)=(0.786321,0.577562,0.533823); rgb(110pt)=(0.787982,0.582808,0.536266); rgb(111pt)=(0.789639,0.588006,0.538699); rgb(112pt)=(0.791292,0.59316,0.54112); rgb(113pt)=(0.792943,0.598268,0.54353); rgb(114pt)=(0.79459,0.603334,0.54593); rgb(115pt)=(0.796233,0.608357,0.548319); rgb(116pt)=(0.797873,0.613339,0.550698); rgb(117pt)=(0.79951,0.618281,0.553066); rgb(118pt)=(0.801143,0.623184,0.555425); rgb(119pt)=(0.802773,0.628048,0.557773); rgb(120pt)=(0.8044,0.632875,0.560112); rgb(121pt)=(0.806023,0.637666,0.562441); rgb(122pt)=(0.807643,0.642421,0.56476); rgb(123pt)=(0.80926,0.647141,0.56707); rgb(124pt)=(0.810874,0.651826,0.569371); rgb(125pt)=(0.812484,0.656479,0.571662); rgb(126pt)=(0.814092,0.661098,0.573944); rgb(127pt)=(0.815696,0.665686,0.576217); rgb(128pt)=(0.817297,0.670242,0.578481); rgb(129pt)=(0.818895,0.674767,0.580737); rgb(130pt)=(0.820489,0.679262,0.582983); rgb(131pt)=(0.822081,0.683728,0.585221); rgb(132pt)=(0.823669,0.688164,0.58745); rgb(133pt)=(0.825255,0.692573,0.589671); rgb(134pt)=(0.826837,0.696953,0.591884); rgb(135pt)=(0.828417,0.701306,0.594089); rgb(136pt)=(0.829993,0.705632,0.596285); rgb(137pt)=(0.831567,0.709932,0.598473); rgb(138pt)=(0.833137,0.714206,0.600653); rgb(139pt)=(0.834705,0.718454,0.602826); rgb(140pt)=(0.836269,0.722678,0.60499); rgb(141pt)=(0.837831,0.726877,0.607147); rgb(142pt)=(0.83939,0.731051,0.609296); rgb(143pt)=(0.840946,0.735203,0.611438); rgb(144pt)=(0.842499,0.73933,0.613572); rgb(145pt)=(0.844049,0.743435,0.615699); rgb(146pt)=(0.845596,0.747518,0.617818); rgb(147pt)=(0.847141,0.751578,0.61993); rgb(148pt)=(0.848682,0.755616,0.622035); rgb(149pt)=(0.850221,0.759633,0.624133); rgb(150pt)=(0.851757,0.763629,0.626224); rgb(151pt)=(0.85329,0.767604,0.628308); rgb(152pt)=(0.854821,0.771558,0.630385); rgb(153pt)=(0.856349,0.775493,0.632456); rgb(154pt)=(0.857874,0.779407,0.634519); rgb(155pt)=(0.859396,0.783302,0.636576); rgb(156pt)=(0.860916,0.787178,0.638626); rgb(157pt)=(0.862433,0.791034,0.64067); rgb(158pt)=(0.863947,0.794872,0.642707); rgb(159pt)=(0.865459,0.798692,0.644737); rgb(160pt)=(0.866968,0.802493,0.646762); rgb(161pt)=(0.868475,0.806276,0.64878); rgb(162pt)=(0.869979,0.810042,0.650791); rgb(163pt)=(0.87148,0.81379,0.652797); rgb(164pt)=(0.872979,0.817522,0.654796); rgb(165pt)=(0.874475,0.821236,0.65679); rgb(166pt)=(0.875968,0.824933,0.658777); rgb(167pt)=(0.877459,0.828614,0.660758); rgb(168pt)=(0.878948,0.832279,0.662733); rgb(169pt)=(0.880434,0.835927,0.664703); rgb(170pt)=(0.881917,0.83956,0.666667); rgb(171pt)=(0.883398,0.843177,0.668625); rgb(172pt)=(0.884877,0.846779,0.670577); rgb(173pt)=(0.886353,0.850365,0.672523); rgb(174pt)=(0.887826,0.853936,0.674464); rgb(175pt)=(0.889297,0.857493,0.6764); rgb(176pt)=(0.890766,0.861035,0.678329); rgb(177pt)=(0.892232,0.864562,0.680254); rgb(178pt)=(0.893696,0.868075,0.682173); rgb(179pt)=(0.895158,0.871574,0.684086); rgb(180pt)=(0.896617,0.875058,0.685994); rgb(181pt)=(0.898073,0.878529,0.687897); rgb(182pt)=(0.899528,0.881987,0.689795); rgb(183pt)=(0.90098,0.88543,0.691687); rgb(184pt)=(0.90243,0.888861,0.693575); rgb(185pt)=(0.903877,0.892278,0.695457); rgb(186pt)=(0.905322,0.895682,0.697334); rgb(187pt)=(0.906765,0.899074,0.699206); rgb(188pt)=(0.908205,0.902452,0.701073); rgb(189pt)=(0.909643,0.905818,0.702935); rgb(190pt)=(0.911079,0.909172,0.704792); rgb(191pt)=(0.912513,0.912513,0.706644); rgb(192pt)=(0.913944,0.913944,0.712158); rgb(193pt)=(0.915373,0.915373,0.717629); rgb(194pt)=(0.9168,0.9168,0.723059); rgb(195pt)=(0.918225,0.918225,0.728449); rgb(196pt)=(0.919648,0.919648,0.733798); rgb(197pt)=(0.921068,0.921068,0.739109); rgb(198pt)=(0.922486,0.922486,0.744383); rgb(199pt)=(0.923902,0.923902,0.749619); rgb(200pt)=(0.925316,0.925316,0.754818); rgb(201pt)=(0.926727,0.926727,0.759983); rgb(202pt)=(0.928137,0.928137,0.765112); rgb(203pt)=(0.929544,0.929544,0.770207); rgb(204pt)=(0.930949,0.930949,0.775269); rgb(205pt)=(0.932352,0.932352,0.780298); rgb(206pt)=(0.933753,0.933753,0.785294); rgb(207pt)=(0.935152,0.935152,0.790259); rgb(208pt)=(0.936549,0.936549,0.795193); rgb(209pt)=(0.937944,0.937944,0.800097); rgb(210pt)=(0.939336,0.939336,0.804971); rgb(211pt)=(0.940727,0.940727,0.809815); rgb(212pt)=(0.942116,0.942116,0.814631); rgb(213pt)=(0.943502,0.943502,0.819418); rgb(214pt)=(0.944886,0.944886,0.824178); rgb(215pt)=(0.946269,0.946269,0.82891); rgb(216pt)=(0.947649,0.947649,0.833615); rgb(217pt)=(0.949028,0.949028,0.838294); rgb(218pt)=(0.950404,0.950404,0.842947); rgb(219pt)=(0.951779,0.951779,0.847574); rgb(220pt)=(0.953151,0.953151,0.852177); rgb(221pt)=(0.954521,0.954521,0.856754); rgb(222pt)=(0.95589,0.95589,0.861307); rgb(223pt)=(0.957256,0.957256,0.865837); rgb(224pt)=(0.958621,0.958621,0.870342); rgb(225pt)=(0.959984,0.959984,0.874825); rgb(226pt)=(0.961344,0.961344,0.879285); rgb(227pt)=(0.962703,0.962703,0.883722); rgb(228pt)=(0.96406,0.96406,0.888137); rgb(229pt)=(0.965415,0.965415,0.89253); rgb(230pt)=(0.966768,0.966768,0.896901); rgb(231pt)=(0.968119,0.968119,0.901252); rgb(232pt)=(0.969469,0.969469,0.905581); rgb(233pt)=(0.970816,0.970816,0.90989); rgb(234pt)=(0.972162,0.972162,0.914179); rgb(235pt)=(0.973505,0.973505,0.918447); rgb(236pt)=(0.974847,0.974847,0.922696); rgb(237pt)=(0.976187,0.976187,0.926926); rgb(238pt)=(0.977525,0.977525,0.931136); rgb(239pt)=(0.978862,0.978862,0.935327); rgb(240pt)=(0.980196,0.980196,0.9395); rgb(241pt)=(0.981529,0.981529,0.943654); rgb(242pt)=(0.98286,0.98286,0.947789); rgb(243pt)=(0.984189,0.984189,0.951907); rgb(244pt)=(0.985516,0.985516,0.956007); rgb(245pt)=(0.986842,0.986842,0.96009); rgb(246pt)=(0.988165,0.988165,0.964155); rgb(247pt)=(0.989487,0.989487,0.968204); rgb(248pt)=(0.990807,0.990807,0.972235); rgb(249pt)=(0.992126,0.992126,0.97625); rgb(250pt)=(0.993443,0.993443,0.980248); rgb(251pt)=(0.994757,0.994757,0.98423); rgb(252pt)=(0.996071,0.996071,0.988196); rgb(253pt)=(0.997382,0.997382,0.992146); rgb(254pt)=(0.998692,0.998692,0.996081); rgb(255pt)=(1,1,1)}, mesh/rows=16]
table[row sep=crcr, point meta=\thisrow{c}] {%
x	y	c\\
0	0	0.293215479053174\\
2	0	0.293274230702635\\
4	0	0.293256436547817\\
6	0	0.2931636308072\\
8	0	0.293227388817239\\
10	0	0.293344314325386\\
12	0	0.293144252980838\\
14	0	0.293504795222235\\
16	0	0.293347182182795\\
18	0	0.293325869090793\\
20	0	0.293363379249511\\
22	0	0.293402377991301\\
24	0	0.293129630864662\\
26	0	0.293388405999297\\
28	0	0.293291157108819\\
30	0	0.293211137662563\\
0	2	0.61246864366996\\
2	2	0.635952556663902\\
4	2	0.634783061279098\\
6	2	0.639335021640915\\
8	2	0.6364402860949\\
10	2	0.640918775981026\\
12	2	0.673381198915271\\
14	2	0.643415678747727\\
16	2	0.633308991911989\\
18	2	0.638086829623914\\
20	2	0.640844675000241\\
22	2	0.64003345189563\\
24	2	0.638204085118805\\
26	2	0.633445106292532\\
28	2	0.636104630190787\\
30	2	0.639904955865721\\
0	4	0.463087504448196\\
2	4	0.430209842450027\\
4	4	0.428556137310516\\
6	4	0.425318107891795\\
8	4	0.428179118801132\\
10	4	0.459203092600911\\
12	4	0.456350400311085\\
14	4	0.428423181680667\\
16	4	0.42855327760111\\
18	4	0.428275565373125\\
20	4	0.439297098328509\\
22	4	0.451388359192914\\
24	4	0.465761597108112\\
26	4	0.454728752441515\\
28	4	0.424607576959231\\
30	4	0.422721376380304\\
0	6	0.400527663766868\\
2	6	0.354624343477549\\
4	6	0.347213835391524\\
6	6	0.347234976193312\\
8	6	0.364371724639197\\
10	6	0.368640534973638\\
12	6	0.361446463759154\\
14	6	0.350593959414788\\
16	6	0.351610406130783\\
18	6	0.356440985160438\\
20	6	0.375723689087875\\
22	6	0.395494834303316\\
24	6	0.412918759605025\\
26	6	0.376092081207973\\
28	6	0.337391760697581\\
30	6	0.337305353541741\\
0	8	0.365541074010269\\
2	8	0.309838045187926\\
4	8	0.304509003016235\\
6	8	0.312508138337078\\
8	8	0.316711130887444\\
10	8	0.318133347853314\\
12	8	0.314369156947151\\
14	8	0.309192295790082\\
16	8	0.31772283243138\\
18	8	0.325349500843082\\
20	8	0.352089353773812\\
22	8	0.373730531062878\\
24	8	0.371510585868348\\
26	8	0.326050869289041\\
28	8	0.289431107502574\\
30	8	0.288966109318321\\
0	10	0.342521188438074\\
2	10	0.28290969882546\\
4	10	0.283500577297381\\
6	10	0.283855142673428\\
8	10	0.286191752792715\\
10	10	0.288718954760736\\
12	10	0.286435660265266\\
14	10	0.287996556568984\\
16	10	0.298909654753143\\
18	10	0.312337172735008\\
20	10	0.341048349521003\\
22	10	0.356128729664436\\
24	10	0.347686604653373\\
26	10	0.293253698779869\\
28	10	0.254642937114296\\
30	10	0.258333869504999\\
0	12	0.328056327727195\\
2	12	0.270009245997678\\
4	12	0.264788801412134\\
6	12	0.263036809199642\\
8	12	0.266991886704641\\
10	12	0.270249863110312\\
12	12	0.271509907139643\\
14	12	0.276863015744065\\
16	12	0.291931666943794\\
18	12	0.306448618319728\\
20	12	0.333806917911889\\
22	12	0.33887554407331\\
24	12	0.327266616959529\\
26	12	0.264805938054989\\
28	12	0.233130336439535\\
30	12	0.230515789081144\\
0	14	0.320173327070341\\
2	14	0.257493473189961\\
4	14	0.250937770981349\\
6	14	0.249091873751888\\
8	14	0.254656674348134\\
10	14	0.259745628063078\\
12	14	0.262902874572504\\
14	14	0.273305444699919\\
16	14	0.289011689697799\\
18	14	0.302821784881745\\
20	14	0.323414123106207\\
22	14	0.327263664182886\\
24	14	0.304969335806862\\
26	14	0.246844487407939\\
28	14	0.214235064023262\\
30	14	0.214459713827973\\
0	16	0.312466918307122\\
2	16	0.247742495526791\\
4	16	0.241293737388285\\
6	16	0.240135278910149\\
8	16	0.24697841302641\\
10	16	0.254117561921672\\
12	16	0.259691852956967\\
14	16	0.270710313806121\\
16	16	0.287009777227274\\
18	16	0.298483021293383\\
20	16	0.316860121698161\\
22	16	0.31681690548383\\
24	16	0.288162092861091\\
26	16	0.228233180384872\\
28	16	0.197617577716725\\
30	16	0.19947178536484\\
0	18	0.306480465035831\\
2	18	0.24066375122815\\
4	18	0.234698883150078\\
6	18	0.234591615588803\\
8	18	0.242344487818994\\
10	18	0.251944876865054\\
12	18	0.258872927463062\\
14	18	0.269319944898216\\
16	18	0.284558029043151\\
18	18	0.294326049473028\\
20	18	0.309408382285406\\
22	18	0.306101693716214\\
24	18	0.277722318395215\\
26	18	0.217284966394822\\
28	18	0.187120911739158\\
30	18	0.187137854340119\\
0	20	0.302309003728638\\
2	20	0.236618296750911\\
4	20	0.231154089688628\\
6	20	0.2315663046992\\
8	20	0.240941545129399\\
10	20	0.25141596760973\\
12	20	0.25749948562526\\
14	20	0.268685790978202\\
16	20	0.28138343814932\\
18	20	0.290300357779901\\
20	20	0.305065859992376\\
22	20	0.296296724978891\\
24	20	0.26626393886498\\
26	20	0.205213976500475\\
28	20	0.177645330679593\\
30	20	0.177234318095232\\
0	22	0.29960868080316\\
2	22	0.233965009363945\\
4	22	0.228868619002688\\
6	22	0.230605996800064\\
8	22	0.240714098691294\\
10	22	0.250717556597054\\
12	22	0.257395651438359\\
14	22	0.266311570737533\\
16	22	0.280262913980777\\
18	22	0.287636557267202\\
20	22	0.29844705011998\\
22	22	0.286228489366194\\
24	22	0.255780076330413\\
26	22	0.195256869227353\\
28	22	0.167691846439285\\
30	22	0.167940481417038\\
0	24	0.297666791434549\\
2	24	0.232474133233432\\
4	24	0.228598050708006\\
6	24	0.230191890096325\\
8	24	0.240431134606658\\
10	24	0.250059010977208\\
12	24	0.256256802292727\\
14	24	0.265546338599273\\
16	24	0.276950955872714\\
18	24	0.284288720334411\\
20	24	0.294091673503412\\
22	24	0.279091636526164\\
24	24	0.24725038486282\\
26	24	0.185916570144472\\
28	24	0.16124022271975\\
30	24	0.163227032266269\\
0	26	0.296659635052304\\
2	26	0.231915421344657\\
4	26	0.227953378926363\\
6	26	0.230302417289615\\
8	26	0.239967017588332\\
10	26	0.249303312526253\\
12	26	0.255230635160487\\
14	26	0.263609964657957\\
16	26	0.276073064976476\\
18	26	0.280677922984681\\
20	26	0.288999107094004\\
22	26	0.272483607663745\\
24	26	0.238361977517402\\
26	26	0.178391001568392\\
28	26	0.155393231357751\\
30	26	0.154131004069502\\
0	28	0.296322095963899\\
2	28	0.231951402817461\\
4	28	0.228013445966456\\
6	28	0.229653391604729\\
8	28	0.238993556770092\\
10	28	0.248573779767125\\
12	28	0.25478769658056\\
14	28	0.263351749778856\\
16	28	0.273705483273884\\
18	28	0.278070476100909\\
20	28	0.285213992036459\\
22	28	0.268136288993756\\
24	28	0.231441739753824\\
26	28	0.17140706411745\\
28	28	0.150273395091729\\
30	28	0.147281997786882\\
0	30	0.296298420418385\\
2	30	0.231668110419566\\
4	30	0.227469265650732\\
6	30	0.229630441978237\\
8	30	0.238984538944121\\
10	30	0.248343971362841\\
12	30	0.253000084743319\\
14	30	0.261677039550996\\
16	30	0.272081125177815\\
18	30	0.274442647325307\\
20	30	0.280290245843218\\
22	30	0.262480284643262\\
24	30	0.226743357951173\\
26	30	0.166781394368155\\
28	30	0.143339472768327\\
30	30	0.141548192733656\\
};
\end{axis}

\begin{axis}[%
width=3.633in,
height=2.425in,
at={(0.27in,0.3in)},
scale only axis,
point meta min=0,
point meta max=1,
xmin=0,
xmax=1,
ymin=0,
ymax=1,
axis line style={draw=none},
ticks=none,
axis x line*=bottom,
axis y line*=left
]
\node[below right, align=left, font=\color{white}]
at (rel axis cs:0.153,0.601) {\small Low\\cross\\correlation};
\node[below right, align=left, font=\color{white}]
at (rel axis cs:0.58,0.543) {\\Dominance\\of Noise};
\draw[line width=2.5pt, draw=mycolor1] (axis cs:0.564285714285714,0.295477909058985) rectangle (axis cs:0.799332596283025,0.930634920634921);
\draw[line width=2.5pt, draw=mycolor2] (axis cs:0.157953929690997,0.298412698412698) rectangle (axis cs:0.337619047619048,0.924339986556226);
\end{axis}
\end{tikzpicture}%

%% file: TIFS_figures_tikz/TBerlin_dist1_SNR20_mismatch.tex
%
%
\definecolor{mycolor1}{rgb}{0.63529,0.07843,0.18431}%
\begin{tikzpicture}

\begin{axis}[%
width=2.5in,
height=2.000in,
at={(0.666in,0.569in)},
scale only axis,
point meta min=0.107445068359375,
point meta max=0.501531982421875,
xmin=0,
xmax=30,
xlabel style={font=\color{white!15!black}},
xlabel={Dimensions to omit},
ymin=2,
ymax=30,
ylabel style={font=\color{white!15!black}},
ylabel={ $\tilde{D}$ Dimensions to keep from $\tilde{D}_1$ to $\tilde{D}_2$},
tick label style={font=\small},
axis background/.style={fill=white},
axis x line*=bottom,
axis y line*=left,
colormap={mymap}{[1pt] rgb(0pt)=(0,0.5,0.4); rgb(255pt)=(1,1,0.4)},
colorbar
]

\addplot[%
surf,
shader=flat corner, draw=black, colormap={mymap}{[1pt] rgb(0pt)=(0,0.5,0.4); rgb(255pt)=(1,1,0.4)}, mesh/rows=16]
table[row sep=crcr, point meta=\thisrow{c}] {%
x	y	c\\
0	0	0.14045654296875\\
2	0	0.140477294921875\\
4	0	0.139161376953125\\
6	0	0.140634765625\\
8	0	0.13903564453125\\
10	0	0.140135498046875\\
12	0	0.139300537109375\\
14	0	0.138714599609375\\
16	0	0.140345458984375\\
18	0	0.1408837890625\\
20	0	0.141136474609375\\
22	0	0.1395263671875\\
24	0	0.14016357421875\\
26	0	0.139039306640625\\
28	0	0.13771484375\\
30	0	0.139381103515625\\
0	2	0.126453857421875\\
2	2	0.1243115234375\\
4	2	0.130325927734375\\
6	2	0.143719482421875\\
8	2	0.124888916015625\\
10	2	0.127130126953125\\
12	2	0.117392578125\\
14	2	0.138909912109375\\
16	2	0.146322021484375\\
18	2	0.151898193359375\\
20	2	0.1740283203125\\
22	2	0.187545166015625\\
24	2	0.213428955078125\\
26	2	0.2785595703125\\
28	2	0.469910888671875\\
30	2	0.49919677734375\\
0	4	0.118955078125\\
2	4	0.11285400390625\\
4	4	0.1135302734375\\
6	4	0.1141796875\\
8	4	0.11590576171875\\
10	4	0.11595458984375\\
12	4	0.111083984375\\
14	4	0.125897216796875\\
16	4	0.1355517578125\\
18	4	0.1444189453125\\
20	4	0.1623291015625\\
22	4	0.1949462890625\\
24	4	0.259088134765625\\
26	4	0.399114990234375\\
28	4	0.49212646484375\\
30	4	0.501531982421875\\
0	6	0.11099853515625\\
2	6	0.1087890625\\
4	6	0.111251220703125\\
6	6	0.1134521484375\\
8	6	0.110323486328125\\
10	6	0.109251708984375\\
12	6	0.116866455078125\\
14	6	0.12772705078125\\
16	6	0.134976806640625\\
18	6	0.148001708984375\\
20	6	0.175645751953125\\
22	6	0.220328369140625\\
24	6	0.318204345703125\\
26	6	0.436629638671875\\
28	6	0.49480224609375\\
30	6	0.49845458984375\\
0	8	0.113087158203125\\
2	8	0.107945556640625\\
4	8	0.110308837890625\\
6	8	0.107445068359375\\
8	8	0.1099658203125\\
10	8	0.111661376953125\\
12	8	0.119190673828125\\
14	8	0.127835693359375\\
16	8	0.13720458984375\\
18	8	0.152955322265625\\
20	8	0.184744873046875\\
22	8	0.25694091796875\\
24	8	0.35969970703125\\
26	8	0.454266357421875\\
28	8	0.49690185546875\\
30	8	0.5010498046875\\
0	10	0.109765625\\
2	10	0.1104541015625\\
4	10	0.1086181640625\\
6	10	0.108677978515625\\
8	10	0.1118310546875\\
10	10	0.11329345703125\\
12	10	0.11862060546875\\
14	10	0.134007568359375\\
16	10	0.141319580078125\\
18	10	0.164337158203125\\
20	10	0.2124951171875\\
22	10	0.290711669921875\\
24	10	0.383150634765625\\
26	10	0.462177734375\\
28	10	0.4978125\\
30	10	0.499830322265625\\
0	12	0.1086572265625\\
2	12	0.108760986328125\\
4	12	0.10988525390625\\
6	12	0.11327880859375\\
8	12	0.116298828125\\
10	12	0.115572509765625\\
12	12	0.12267333984375\\
14	12	0.134539794921875\\
16	12	0.149154052734375\\
18	12	0.179256591796875\\
20	12	0.2365966796875\\
22	12	0.315108642578125\\
24	12	0.39882568359375\\
26	12	0.470206298828125\\
28	12	0.49804443359375\\
30	12	0.499765625\\
0	14	0.11351806640625\\
2	14	0.109051513671875\\
4	14	0.10848876953125\\
6	14	0.10947509765625\\
8	14	0.113089599609375\\
10	14	0.11781005859375\\
12	14	0.1228662109375\\
14	14	0.139884033203125\\
16	14	0.1598291015625\\
18	14	0.196187744140625\\
20	14	0.255115966796875\\
22	14	0.331505126953125\\
24	14	0.411690673828125\\
26	14	0.4732666015625\\
28	14	0.49833251953125\\
30	14	0.50025390625\\
0	16	0.112418212890625\\
2	16	0.10936279296875\\
4	16	0.107818603515625\\
6	16	0.111761474609375\\
8	16	0.11808837890625\\
10	16	0.11828369140625\\
12	16	0.125367431640625\\
14	16	0.1466845703125\\
16	16	0.17071044921875\\
18	16	0.2115673828125\\
20	16	0.269000244140625\\
22	16	0.343817138671875\\
24	16	0.417877197265625\\
26	16	0.476605224609375\\
28	16	0.4983984375\\
30	16	0.4989990234375\\
0	18	0.111287841796875\\
2	18	0.10817626953125\\
4	18	0.111746826171875\\
6	18	0.114224853515625\\
8	18	0.12148193359375\\
10	18	0.123079833984375\\
12	18	0.132921142578125\\
14	18	0.155518798828125\\
16	18	0.183792724609375\\
18	18	0.223214111328125\\
20	18	0.28060546875\\
22	18	0.355350341796875\\
24	18	0.427242431640625\\
26	18	0.480054931640625\\
28	18	0.498560791015625\\
30	18	0.49868896484375\\
0	20	0.11373291015625\\
2	20	0.111466064453125\\
4	20	0.112718505859375\\
6	20	0.1140380859375\\
8	20	0.1214990234375\\
10	20	0.127900390625\\
12	20	0.139739990234375\\
14	20	0.169351806640625\\
16	20	0.191192626953125\\
18	20	0.233341064453125\\
20	20	0.292919921875\\
22	20	0.365430908203125\\
24	20	0.43220947265625\\
26	20	0.4814306640625\\
28	20	0.497841796875\\
30	20	0.499351806640625\\
0	22	0.1154638671875\\
2	22	0.11390625\\
4	22	0.114840087890625\\
6	22	0.117374267578125\\
8	22	0.122701416015625\\
10	22	0.130462646484375\\
12	22	0.14517333984375\\
14	22	0.17066650390625\\
16	22	0.20010009765625\\
18	22	0.2444189453125\\
20	22	0.302080078125\\
22	22	0.373458251953125\\
24	22	0.436888427734375\\
26	22	0.4828515625\\
28	22	0.4979052734375\\
30	22	0.500439453125\\
0	24	0.114521484375\\
2	24	0.112891845703125\\
4	24	0.114326171875\\
6	24	0.1196142578125\\
8	24	0.126842041015625\\
10	24	0.13545654296875\\
12	24	0.14938720703125\\
14	24	0.173089599609375\\
16	24	0.206087646484375\\
18	24	0.251832275390625\\
20	24	0.30976806640625\\
22	24	0.379669189453125\\
24	24	0.441424560546875\\
26	24	0.484168701171875\\
28	24	0.499217529296875\\
30	24	0.499075927734375\\
0	26	0.11507568359375\\
2	26	0.11206298828125\\
4	26	0.1154296875\\
6	26	0.1209033203125\\
8	26	0.129739990234375\\
10	26	0.139246826171875\\
12	26	0.153758544921875\\
14	26	0.18302490234375\\
16	26	0.21303955078125\\
18	26	0.258929443359375\\
20	26	0.316541748046875\\
22	26	0.384654541015625\\
24	26	0.444156494140625\\
26	26	0.485772705078125\\
28	26	0.49865478515625\\
30	26	0.49919921875\\
0	28	0.11689453125\\
2	28	0.11537109375\\
4	28	0.119515380859375\\
6	28	0.122469482421875\\
8	28	0.133011474609375\\
10	28	0.144888916015625\\
12	28	0.160245361328125\\
14	28	0.1865234375\\
16	28	0.2209619140625\\
18	28	0.265828857421875\\
20	28	0.324947509765625\\
22	28	0.389696044921875\\
24	28	0.448016357421875\\
26	28	0.486361083984375\\
28	28	0.4983544921875\\
30	28	0.499453125\\
0	30	0.116778564453125\\
2	30	0.117010498046875\\
4	30	0.12046142578125\\
6	30	0.12560302734375\\
8	30	0.137071533203125\\
10	30	0.14638427734375\\
12	30	0.164927978515625\\
14	30	0.193602294921875\\
16	30	0.225023193359375\\
18	30	0.273134765625\\
20	30	0.328658447265625\\
22	30	0.395281982421875\\
24	30	0.4502197265625\\
26	30	0.488056640625\\
28	30	0.499249267578125\\
30	30	0.5001220703125\\
};
\end{axis}

\begin{axis}[%
width=3.633in,
height=2.425in,
at={(0.27in,0.3in)},
scale only axis,
point meta min=0,
point meta max=1,
xmin=0,
xmax=1,
ymin=0,
ymax=1,
axis line style={draw=none},
ticks=none,
axis x line*=bottom,
axis y line*=left
]
\node[below right, align=left, font=\color{white}]
at (rel axis cs:0.153,0.601) {\small Low\\mismatch\\probability};
\node[below right, align=left, font=\color{purple}]
at (rel axis cs:0.58,0.543) {High\\mismatch\\probability};
\draw[line width=2.5pt, draw=mycolor1] (axis cs:0.564285714285714,0.295477909058985) rectangle (axis cs:0.799332596283025,0.930634920634921);
\draw[line width=2.5pt, draw=black] (axis cs:0.157953929690997,0.298412698412698) rectangle (axis cs:0.337619047619048,0.924339986556226);
\end{axis}
\end{tikzpicture}%

%% file: TIFS_figures_tikz/TBerlin_dist1_SNR5_corr.tex
%
%
\definecolor{mycolor1}{rgb}{0.85098,0.32549,0.09804}%
\begin{tikzpicture}

\begin{axis}[%
width=2.5in,
height=2.000in,
at={(0.666in,0.569in)},
scale only axis,
point meta min=0.141548192733656,
point meta max=0.673381198915271,
xmin=0,
xmax=30,
xlabel style={font=\color{white!15!black}},
xlabel={Dimensions to omit},
ymin=2,
ymax=30,
ylabel style={font=\color{white!15!black}},
ylabel={Dimensions to keep from $\tilde{D}_1$ to $\tilde{D}_2$},
axis background/.style={fill=white},
axis x line*=bottom,
axis y line*=left,
colormap={mymap}{[1pt] rgb(0pt)=(0.0589256,0,0); rgb(1pt)=(0.0977692,0.051131,0.051131); rgb(2pt)=(0.125082,0.0723102,0.0723102); rgb(3pt)=(0.147418,0.0885615,0.0885615); rgb(4pt)=(0.166789,0.102262,0.102262); rgb(5pt)=(0.184134,0.114332,0.114332); rgb(6pt)=(0.19998,0.125245,0.125245); rgb(7pt)=(0.214659,0.13528,0.13528); rgb(8pt)=(0.228397,0.14462,0.14462); rgb(9pt)=(0.241354,0.153393,0.153393); rgb(10pt)=(0.25365,0.16169,0.16169); rgb(11pt)=(0.265377,0.169582,0.169582); rgb(12pt)=(0.276607,0.177123,0.177123); rgb(13pt)=(0.287399,0.184355,0.184355); rgb(14pt)=(0.2978,0.191315,0.191315); rgb(15pt)=(0.307849,0.19803,0.19803); rgb(16pt)=(0.317581,0.204524,0.204524); rgb(17pt)=(0.327024,0.210819,0.210819); rgb(18pt)=(0.336201,0.21693,0.21693); rgb(19pt)=(0.345134,0.222875,0.222875); rgb(20pt)=(0.353842,0.228665,0.228665); rgb(21pt)=(0.362341,0.234312,0.234312); rgb(22pt)=(0.370645,0.239826,0.239826); rgb(23pt)=(0.378766,0.245216,0.245216); rgb(24pt)=(0.386718,0.25049,0.25049); rgb(25pt)=(0.394509,0.255655,0.255655); rgb(26pt)=(0.402149,0.260718,0.260718); rgb(27pt)=(0.409647,0.265684,0.265684); rgb(28pt)=(0.41701,0.27056,0.27056); rgb(29pt)=(0.424245,0.275349,0.275349); rgb(30pt)=(0.431359,0.280056,0.280056); rgb(31pt)=(0.438357,0.284685,0.284685); rgb(32pt)=(0.445245,0.289241,0.289241); rgb(33pt)=(0.452029,0.293725,0.293725); rgb(34pt)=(0.458712,0.298142,0.298142); rgb(35pt)=(0.465299,0.302495,0.302495); rgb(36pt)=(0.471794,0.306786,0.306786); rgb(37pt)=(0.478201,0.311018,0.311018); rgb(38pt)=(0.484524,0.315193,0.315193); rgb(39pt)=(0.490764,0.319313,0.319313); rgb(40pt)=(0.496927,0.323381,0.323381); rgb(41pt)=(0.503014,0.327398,0.327398); rgb(42pt)=(0.509028,0.331367,0.331367); rgb(43pt)=(0.514972,0.335288,0.335288); rgb(44pt)=(0.520848,0.339165,0.339165); rgb(45pt)=(0.526659,0.342997,0.342997); rgb(46pt)=(0.532406,0.346787,0.346787); rgb(47pt)=(0.538092,0.350536,0.350536); rgb(48pt)=(0.543718,0.354246,0.354246); rgb(49pt)=(0.549287,0.357917,0.357917); rgb(50pt)=(0.554799,0.361551,0.361551); rgb(51pt)=(0.560258,0.365148,0.365148); rgb(52pt)=(0.565664,0.368711,0.368711); rgb(53pt)=(0.571018,0.372239,0.372239); rgb(54pt)=(0.576323,0.375735,0.375735); rgb(55pt)=(0.58158,0.379198,0.379198); rgb(56pt)=(0.586789,0.382629,0.382629); rgb(57pt)=(0.591953,0.386031,0.386031); rgb(58pt)=(0.597072,0.389402,0.389402); rgb(59pt)=(0.602148,0.392745,0.392745); rgb(60pt)=(0.607181,0.396059,0.396059); rgb(61pt)=(0.612172,0.399346,0.399346); rgb(62pt)=(0.617124,0.402606,0.402606); rgb(63pt)=(0.622035,0.40584,0.40584); rgb(64pt)=(0.626909,0.409048,0.409048); rgb(65pt)=(0.631745,0.412231,0.412231); rgb(66pt)=(0.636544,0.41539,0.41539); rgb(67pt)=(0.641307,0.418525,0.418525); rgb(68pt)=(0.646035,0.421637,0.421637); rgb(69pt)=(0.650729,0.424726,0.424726); rgb(70pt)=(0.655389,0.427793,0.427793); rgb(71pt)=(0.660016,0.430837,0.430837); rgb(72pt)=(0.664611,0.433861,0.433861); rgb(73pt)=(0.669174,0.436863,0.436863); rgb(74pt)=(0.673707,0.439845,0.439845); rgb(75pt)=(0.678209,0.442807,0.442807); rgb(76pt)=(0.682681,0.44575,0.44575); rgb(77pt)=(0.687125,0.448673,0.448673); rgb(78pt)=(0.69154,0.451577,0.451577); rgb(79pt)=(0.695927,0.454462,0.454462); rgb(80pt)=(0.700286,0.45733,0.45733); rgb(81pt)=(0.704618,0.460179,0.460179); rgb(82pt)=(0.708924,0.463011,0.463011); rgb(83pt)=(0.713204,0.465826,0.465826); rgb(84pt)=(0.717459,0.468623,0.468623); rgb(85pt)=(0.721688,0.471405,0.471405); rgb(86pt)=(0.725893,0.474169,0.474169); rgb(87pt)=(0.730073,0.476918,0.476918); rgb(88pt)=(0.73423,0.479651,0.479651); rgb(89pt)=(0.738363,0.482369,0.482369); rgb(90pt)=(0.742473,0.485071,0.485071); rgb(91pt)=(0.746561,0.487759,0.487759); rgb(92pt)=(0.750626,0.490431,0.490431); rgb(93pt)=(0.75467,0.493089,0.493089); rgb(94pt)=(0.758691,0.495733,0.495733); rgb(95pt)=(0.762692,0.498363,0.498363); rgb(96pt)=(0.764404,0.504433,0.500979); rgb(97pt)=(0.766112,0.51043,0.503582); rgb(98pt)=(0.767817,0.516358,0.506171); rgb(99pt)=(0.769517,0.522219,0.508747); rgb(100pt)=(0.771214,0.528014,0.51131); rgb(101pt)=(0.772907,0.533747,0.51386); rgb(102pt)=(0.774597,0.539418,0.516398); rgb(103pt)=(0.776282,0.545031,0.518923); rgb(104pt)=(0.777964,0.550586,0.521436); rgb(105pt)=(0.779643,0.556086,0.523937); rgb(106pt)=(0.781318,0.561532,0.526426); rgb(107pt)=(0.782989,0.566926,0.528903); rgb(108pt)=(0.784657,0.572269,0.531369); rgb(109pt)=(0.786321,0.577562,0.533823); rgb(110pt)=(0.787982,0.582808,0.536266); rgb(111pt)=(0.789639,0.588006,0.538699); rgb(112pt)=(0.791292,0.59316,0.54112); rgb(113pt)=(0.792943,0.598268,0.54353); rgb(114pt)=(0.79459,0.603334,0.54593); rgb(115pt)=(0.796233,0.608357,0.548319); rgb(116pt)=(0.797873,0.613339,0.550698); rgb(117pt)=(0.79951,0.618281,0.553066); rgb(118pt)=(0.801143,0.623184,0.555425); rgb(119pt)=(0.802773,0.628048,0.557773); rgb(120pt)=(0.8044,0.632875,0.560112); rgb(121pt)=(0.806023,0.637666,0.562441); rgb(122pt)=(0.807643,0.642421,0.56476); rgb(123pt)=(0.80926,0.647141,0.56707); rgb(124pt)=(0.810874,0.651826,0.569371); rgb(125pt)=(0.812484,0.656479,0.571662); rgb(126pt)=(0.814092,0.661098,0.573944); rgb(127pt)=(0.815696,0.665686,0.576217); rgb(128pt)=(0.817297,0.670242,0.578481); rgb(129pt)=(0.818895,0.674767,0.580737); rgb(130pt)=(0.820489,0.679262,0.582983); rgb(131pt)=(0.822081,0.683728,0.585221); rgb(132pt)=(0.823669,0.688164,0.58745); rgb(133pt)=(0.825255,0.692573,0.589671); rgb(134pt)=(0.826837,0.696953,0.591884); rgb(135pt)=(0.828417,0.701306,0.594089); rgb(136pt)=(0.829993,0.705632,0.596285); rgb(137pt)=(0.831567,0.709932,0.598473); rgb(138pt)=(0.833137,0.714206,0.600653); rgb(139pt)=(0.834705,0.718454,0.602826); rgb(140pt)=(0.836269,0.722678,0.60499); rgb(141pt)=(0.837831,0.726877,0.607147); rgb(142pt)=(0.83939,0.731051,0.609296); rgb(143pt)=(0.840946,0.735203,0.611438); rgb(144pt)=(0.842499,0.73933,0.613572); rgb(145pt)=(0.844049,0.743435,0.615699); rgb(146pt)=(0.845596,0.747518,0.617818); rgb(147pt)=(0.847141,0.751578,0.61993); rgb(148pt)=(0.848682,0.755616,0.622035); rgb(149pt)=(0.850221,0.759633,0.624133); rgb(150pt)=(0.851757,0.763629,0.626224); rgb(151pt)=(0.85329,0.767604,0.628308); rgb(152pt)=(0.854821,0.771558,0.630385); rgb(153pt)=(0.856349,0.775493,0.632456); rgb(154pt)=(0.857874,0.779407,0.634519); rgb(155pt)=(0.859396,0.783302,0.636576); rgb(156pt)=(0.860916,0.787178,0.638626); rgb(157pt)=(0.862433,0.791034,0.64067); rgb(158pt)=(0.863947,0.794872,0.642707); rgb(159pt)=(0.865459,0.798692,0.644737); rgb(160pt)=(0.866968,0.802493,0.646762); rgb(161pt)=(0.868475,0.806276,0.64878); rgb(162pt)=(0.869979,0.810042,0.650791); rgb(163pt)=(0.87148,0.81379,0.652797); rgb(164pt)=(0.872979,0.817522,0.654796); rgb(165pt)=(0.874475,0.821236,0.65679); rgb(166pt)=(0.875968,0.824933,0.658777); rgb(167pt)=(0.877459,0.828614,0.660758); rgb(168pt)=(0.878948,0.832279,0.662733); rgb(169pt)=(0.880434,0.835927,0.664703); rgb(170pt)=(0.881917,0.83956,0.666667); rgb(171pt)=(0.883398,0.843177,0.668625); rgb(172pt)=(0.884877,0.846779,0.670577); rgb(173pt)=(0.886353,0.850365,0.672523); rgb(174pt)=(0.887826,0.853936,0.674464); rgb(175pt)=(0.889297,0.857493,0.6764); rgb(176pt)=(0.890766,0.861035,0.678329); rgb(177pt)=(0.892232,0.864562,0.680254); rgb(178pt)=(0.893696,0.868075,0.682173); rgb(179pt)=(0.895158,0.871574,0.684086); rgb(180pt)=(0.896617,0.875058,0.685994); rgb(181pt)=(0.898073,0.878529,0.687897); rgb(182pt)=(0.899528,0.881987,0.689795); rgb(183pt)=(0.90098,0.88543,0.691687); rgb(184pt)=(0.90243,0.888861,0.693575); rgb(185pt)=(0.903877,0.892278,0.695457); rgb(186pt)=(0.905322,0.895682,0.697334); rgb(187pt)=(0.906765,0.899074,0.699206); rgb(188pt)=(0.908205,0.902452,0.701073); rgb(189pt)=(0.909643,0.905818,0.702935); rgb(190pt)=(0.911079,0.909172,0.704792); rgb(191pt)=(0.912513,0.912513,0.706644); rgb(192pt)=(0.913944,0.913944,0.712158); rgb(193pt)=(0.915373,0.915373,0.717629); rgb(194pt)=(0.9168,0.9168,0.723059); rgb(195pt)=(0.918225,0.918225,0.728449); rgb(196pt)=(0.919648,0.919648,0.733798); rgb(197pt)=(0.921068,0.921068,0.739109); rgb(198pt)=(0.922486,0.922486,0.744383); rgb(199pt)=(0.923902,0.923902,0.749619); rgb(200pt)=(0.925316,0.925316,0.754818); rgb(201pt)=(0.926727,0.926727,0.759983); rgb(202pt)=(0.928137,0.928137,0.765112); rgb(203pt)=(0.929544,0.929544,0.770207); rgb(204pt)=(0.930949,0.930949,0.775269); rgb(205pt)=(0.932352,0.932352,0.780298); rgb(206pt)=(0.933753,0.933753,0.785294); rgb(207pt)=(0.935152,0.935152,0.790259); rgb(208pt)=(0.936549,0.936549,0.795193); rgb(209pt)=(0.937944,0.937944,0.800097); rgb(210pt)=(0.939336,0.939336,0.804971); rgb(211pt)=(0.940727,0.940727,0.809815); rgb(212pt)=(0.942116,0.942116,0.814631); rgb(213pt)=(0.943502,0.943502,0.819418); rgb(214pt)=(0.944886,0.944886,0.824178); rgb(215pt)=(0.946269,0.946269,0.82891); rgb(216pt)=(0.947649,0.947649,0.833615); rgb(217pt)=(0.949028,0.949028,0.838294); rgb(218pt)=(0.950404,0.950404,0.842947); rgb(219pt)=(0.951779,0.951779,0.847574); rgb(220pt)=(0.953151,0.953151,0.852177); rgb(221pt)=(0.954521,0.954521,0.856754); rgb(222pt)=(0.95589,0.95589,0.861307); rgb(223pt)=(0.957256,0.957256,0.865837); rgb(224pt)=(0.958621,0.958621,0.870342); rgb(225pt)=(0.959984,0.959984,0.874825); rgb(226pt)=(0.961344,0.961344,0.879285); rgb(227pt)=(0.962703,0.962703,0.883722); rgb(228pt)=(0.96406,0.96406,0.888137); rgb(229pt)=(0.965415,0.965415,0.89253); rgb(230pt)=(0.966768,0.966768,0.896901); rgb(231pt)=(0.968119,0.968119,0.901252); rgb(232pt)=(0.969469,0.969469,0.905581); rgb(233pt)=(0.970816,0.970816,0.90989); rgb(234pt)=(0.972162,0.972162,0.914179); rgb(235pt)=(0.973505,0.973505,0.918447); rgb(236pt)=(0.974847,0.974847,0.922696); rgb(237pt)=(0.976187,0.976187,0.926926); rgb(238pt)=(0.977525,0.977525,0.931136); rgb(239pt)=(0.978862,0.978862,0.935327); rgb(240pt)=(0.980196,0.980196,0.9395); rgb(241pt)=(0.981529,0.981529,0.943654); rgb(242pt)=(0.98286,0.98286,0.947789); rgb(243pt)=(0.984189,0.984189,0.951907); rgb(244pt)=(0.985516,0.985516,0.956007); rgb(245pt)=(0.986842,0.986842,0.96009); rgb(246pt)=(0.988165,0.988165,0.964155); rgb(247pt)=(0.989487,0.989487,0.968204); rgb(248pt)=(0.990807,0.990807,0.972235); rgb(249pt)=(0.992126,0.992126,0.97625); rgb(250pt)=(0.993443,0.993443,0.980248); rgb(251pt)=(0.994757,0.994757,0.98423); rgb(252pt)=(0.996071,0.996071,0.988196); rgb(253pt)=(0.997382,0.997382,0.992146); rgb(254pt)=(0.998692,0.998692,0.996081); rgb(255pt)=(1,1,1)},
colorbar
]

\addplot[%
surf,
shader=flat corner, draw=black, colormap={mymap}{[1pt] rgb(0pt)=(0.0589256,0,0); rgb(1pt)=(0.0977692,0.051131,0.051131); rgb(2pt)=(0.125082,0.0723102,0.0723102); rgb(3pt)=(0.147418,0.0885615,0.0885615); rgb(4pt)=(0.166789,0.102262,0.102262); rgb(5pt)=(0.184134,0.114332,0.114332); rgb(6pt)=(0.19998,0.125245,0.125245); rgb(7pt)=(0.214659,0.13528,0.13528); rgb(8pt)=(0.228397,0.14462,0.14462); rgb(9pt)=(0.241354,0.153393,0.153393); rgb(10pt)=(0.25365,0.16169,0.16169); rgb(11pt)=(0.265377,0.169582,0.169582); rgb(12pt)=(0.276607,0.177123,0.177123); rgb(13pt)=(0.287399,0.184355,0.184355); rgb(14pt)=(0.2978,0.191315,0.191315); rgb(15pt)=(0.307849,0.19803,0.19803); rgb(16pt)=(0.317581,0.204524,0.204524); rgb(17pt)=(0.327024,0.210819,0.210819); rgb(18pt)=(0.336201,0.21693,0.21693); rgb(19pt)=(0.345134,0.222875,0.222875); rgb(20pt)=(0.353842,0.228665,0.228665); rgb(21pt)=(0.362341,0.234312,0.234312); rgb(22pt)=(0.370645,0.239826,0.239826); rgb(23pt)=(0.378766,0.245216,0.245216); rgb(24pt)=(0.386718,0.25049,0.25049); rgb(25pt)=(0.394509,0.255655,0.255655); rgb(26pt)=(0.402149,0.260718,0.260718); rgb(27pt)=(0.409647,0.265684,0.265684); rgb(28pt)=(0.41701,0.27056,0.27056); rgb(29pt)=(0.424245,0.275349,0.275349); rgb(30pt)=(0.431359,0.280056,0.280056); rgb(31pt)=(0.438357,0.284685,0.284685); rgb(32pt)=(0.445245,0.289241,0.289241); rgb(33pt)=(0.452029,0.293725,0.293725); rgb(34pt)=(0.458712,0.298142,0.298142); rgb(35pt)=(0.465299,0.302495,0.302495); rgb(36pt)=(0.471794,0.306786,0.306786); rgb(37pt)=(0.478201,0.311018,0.311018); rgb(38pt)=(0.484524,0.315193,0.315193); rgb(39pt)=(0.490764,0.319313,0.319313); rgb(40pt)=(0.496927,0.323381,0.323381); rgb(41pt)=(0.503014,0.327398,0.327398); rgb(42pt)=(0.509028,0.331367,0.331367); rgb(43pt)=(0.514972,0.335288,0.335288); rgb(44pt)=(0.520848,0.339165,0.339165); rgb(45pt)=(0.526659,0.342997,0.342997); rgb(46pt)=(0.532406,0.346787,0.346787); rgb(47pt)=(0.538092,0.350536,0.350536); rgb(48pt)=(0.543718,0.354246,0.354246); rgb(49pt)=(0.549287,0.357917,0.357917); rgb(50pt)=(0.554799,0.361551,0.361551); rgb(51pt)=(0.560258,0.365148,0.365148); rgb(52pt)=(0.565664,0.368711,0.368711); rgb(53pt)=(0.571018,0.372239,0.372239); rgb(54pt)=(0.576323,0.375735,0.375735); rgb(55pt)=(0.58158,0.379198,0.379198); rgb(56pt)=(0.586789,0.382629,0.382629); rgb(57pt)=(0.591953,0.386031,0.386031); rgb(58pt)=(0.597072,0.389402,0.389402); rgb(59pt)=(0.602148,0.392745,0.392745); rgb(60pt)=(0.607181,0.396059,0.396059); rgb(61pt)=(0.612172,0.399346,0.399346); rgb(62pt)=(0.617124,0.402606,0.402606); rgb(63pt)=(0.622035,0.40584,0.40584); rgb(64pt)=(0.626909,0.409048,0.409048); rgb(65pt)=(0.631745,0.412231,0.412231); rgb(66pt)=(0.636544,0.41539,0.41539); rgb(67pt)=(0.641307,0.418525,0.418525); rgb(68pt)=(0.646035,0.421637,0.421637); rgb(69pt)=(0.650729,0.424726,0.424726); rgb(70pt)=(0.655389,0.427793,0.427793); rgb(71pt)=(0.660016,0.430837,0.430837); rgb(72pt)=(0.664611,0.433861,0.433861); rgb(73pt)=(0.669174,0.436863,0.436863); rgb(74pt)=(0.673707,0.439845,0.439845); rgb(75pt)=(0.678209,0.442807,0.442807); rgb(76pt)=(0.682681,0.44575,0.44575); rgb(77pt)=(0.687125,0.448673,0.448673); rgb(78pt)=(0.69154,0.451577,0.451577); rgb(79pt)=(0.695927,0.454462,0.454462); rgb(80pt)=(0.700286,0.45733,0.45733); rgb(81pt)=(0.704618,0.460179,0.460179); rgb(82pt)=(0.708924,0.463011,0.463011); rgb(83pt)=(0.713204,0.465826,0.465826); rgb(84pt)=(0.717459,0.468623,0.468623); rgb(85pt)=(0.721688,0.471405,0.471405); rgb(86pt)=(0.725893,0.474169,0.474169); rgb(87pt)=(0.730073,0.476918,0.476918); rgb(88pt)=(0.73423,0.479651,0.479651); rgb(89pt)=(0.738363,0.482369,0.482369); rgb(90pt)=(0.742473,0.485071,0.485071); rgb(91pt)=(0.746561,0.487759,0.487759); rgb(92pt)=(0.750626,0.490431,0.490431); rgb(93pt)=(0.75467,0.493089,0.493089); rgb(94pt)=(0.758691,0.495733,0.495733); rgb(95pt)=(0.762692,0.498363,0.498363); rgb(96pt)=(0.764404,0.504433,0.500979); rgb(97pt)=(0.766112,0.51043,0.503582); rgb(98pt)=(0.767817,0.516358,0.506171); rgb(99pt)=(0.769517,0.522219,0.508747); rgb(100pt)=(0.771214,0.528014,0.51131); rgb(101pt)=(0.772907,0.533747,0.51386); rgb(102pt)=(0.774597,0.539418,0.516398); rgb(103pt)=(0.776282,0.545031,0.518923); rgb(104pt)=(0.777964,0.550586,0.521436); rgb(105pt)=(0.779643,0.556086,0.523937); rgb(106pt)=(0.781318,0.561532,0.526426); rgb(107pt)=(0.782989,0.566926,0.528903); rgb(108pt)=(0.784657,0.572269,0.531369); rgb(109pt)=(0.786321,0.577562,0.533823); rgb(110pt)=(0.787982,0.582808,0.536266); rgb(111pt)=(0.789639,0.588006,0.538699); rgb(112pt)=(0.791292,0.59316,0.54112); rgb(113pt)=(0.792943,0.598268,0.54353); rgb(114pt)=(0.79459,0.603334,0.54593); rgb(115pt)=(0.796233,0.608357,0.548319); rgb(116pt)=(0.797873,0.613339,0.550698); rgb(117pt)=(0.79951,0.618281,0.553066); rgb(118pt)=(0.801143,0.623184,0.555425); rgb(119pt)=(0.802773,0.628048,0.557773); rgb(120pt)=(0.8044,0.632875,0.560112); rgb(121pt)=(0.806023,0.637666,0.562441); rgb(122pt)=(0.807643,0.642421,0.56476); rgb(123pt)=(0.80926,0.647141,0.56707); rgb(124pt)=(0.810874,0.651826,0.569371); rgb(125pt)=(0.812484,0.656479,0.571662); rgb(126pt)=(0.814092,0.661098,0.573944); rgb(127pt)=(0.815696,0.665686,0.576217); rgb(128pt)=(0.817297,0.670242,0.578481); rgb(129pt)=(0.818895,0.674767,0.580737); rgb(130pt)=(0.820489,0.679262,0.582983); rgb(131pt)=(0.822081,0.683728,0.585221); rgb(132pt)=(0.823669,0.688164,0.58745); rgb(133pt)=(0.825255,0.692573,0.589671); rgb(134pt)=(0.826837,0.696953,0.591884); rgb(135pt)=(0.828417,0.701306,0.594089); rgb(136pt)=(0.829993,0.705632,0.596285); rgb(137pt)=(0.831567,0.709932,0.598473); rgb(138pt)=(0.833137,0.714206,0.600653); rgb(139pt)=(0.834705,0.718454,0.602826); rgb(140pt)=(0.836269,0.722678,0.60499); rgb(141pt)=(0.837831,0.726877,0.607147); rgb(142pt)=(0.83939,0.731051,0.609296); rgb(143pt)=(0.840946,0.735203,0.611438); rgb(144pt)=(0.842499,0.73933,0.613572); rgb(145pt)=(0.844049,0.743435,0.615699); rgb(146pt)=(0.845596,0.747518,0.617818); rgb(147pt)=(0.847141,0.751578,0.61993); rgb(148pt)=(0.848682,0.755616,0.622035); rgb(149pt)=(0.850221,0.759633,0.624133); rgb(150pt)=(0.851757,0.763629,0.626224); rgb(151pt)=(0.85329,0.767604,0.628308); rgb(152pt)=(0.854821,0.771558,0.630385); rgb(153pt)=(0.856349,0.775493,0.632456); rgb(154pt)=(0.857874,0.779407,0.634519); rgb(155pt)=(0.859396,0.783302,0.636576); rgb(156pt)=(0.860916,0.787178,0.638626); rgb(157pt)=(0.862433,0.791034,0.64067); rgb(158pt)=(0.863947,0.794872,0.642707); rgb(159pt)=(0.865459,0.798692,0.644737); rgb(160pt)=(0.866968,0.802493,0.646762); rgb(161pt)=(0.868475,0.806276,0.64878); rgb(162pt)=(0.869979,0.810042,0.650791); rgb(163pt)=(0.87148,0.81379,0.652797); rgb(164pt)=(0.872979,0.817522,0.654796); rgb(165pt)=(0.874475,0.821236,0.65679); rgb(166pt)=(0.875968,0.824933,0.658777); rgb(167pt)=(0.877459,0.828614,0.660758); rgb(168pt)=(0.878948,0.832279,0.662733); rgb(169pt)=(0.880434,0.835927,0.664703); rgb(170pt)=(0.881917,0.83956,0.666667); rgb(171pt)=(0.883398,0.843177,0.668625); rgb(172pt)=(0.884877,0.846779,0.670577); rgb(173pt)=(0.886353,0.850365,0.672523); rgb(174pt)=(0.887826,0.853936,0.674464); rgb(175pt)=(0.889297,0.857493,0.6764); rgb(176pt)=(0.890766,0.861035,0.678329); rgb(177pt)=(0.892232,0.864562,0.680254); rgb(178pt)=(0.893696,0.868075,0.682173); rgb(179pt)=(0.895158,0.871574,0.684086); rgb(180pt)=(0.896617,0.875058,0.685994); rgb(181pt)=(0.898073,0.878529,0.687897); rgb(182pt)=(0.899528,0.881987,0.689795); rgb(183pt)=(0.90098,0.88543,0.691687); rgb(184pt)=(0.90243,0.888861,0.693575); rgb(185pt)=(0.903877,0.892278,0.695457); rgb(186pt)=(0.905322,0.895682,0.697334); rgb(187pt)=(0.906765,0.899074,0.699206); rgb(188pt)=(0.908205,0.902452,0.701073); rgb(189pt)=(0.909643,0.905818,0.702935); rgb(190pt)=(0.911079,0.909172,0.704792); rgb(191pt)=(0.912513,0.912513,0.706644); rgb(192pt)=(0.913944,0.913944,0.712158); rgb(193pt)=(0.915373,0.915373,0.717629); rgb(194pt)=(0.9168,0.9168,0.723059); rgb(195pt)=(0.918225,0.918225,0.728449); rgb(196pt)=(0.919648,0.919648,0.733798); rgb(197pt)=(0.921068,0.921068,0.739109); rgb(198pt)=(0.922486,0.922486,0.744383); rgb(199pt)=(0.923902,0.923902,0.749619); rgb(200pt)=(0.925316,0.925316,0.754818); rgb(201pt)=(0.926727,0.926727,0.759983); rgb(202pt)=(0.928137,0.928137,0.765112); rgb(203pt)=(0.929544,0.929544,0.770207); rgb(204pt)=(0.930949,0.930949,0.775269); rgb(205pt)=(0.932352,0.932352,0.780298); rgb(206pt)=(0.933753,0.933753,0.785294); rgb(207pt)=(0.935152,0.935152,0.790259); rgb(208pt)=(0.936549,0.936549,0.795193); rgb(209pt)=(0.937944,0.937944,0.800097); rgb(210pt)=(0.939336,0.939336,0.804971); rgb(211pt)=(0.940727,0.940727,0.809815); rgb(212pt)=(0.942116,0.942116,0.814631); rgb(213pt)=(0.943502,0.943502,0.819418); rgb(214pt)=(0.944886,0.944886,0.824178); rgb(215pt)=(0.946269,0.946269,0.82891); rgb(216pt)=(0.947649,0.947649,0.833615); rgb(217pt)=(0.949028,0.949028,0.838294); rgb(218pt)=(0.950404,0.950404,0.842947); rgb(219pt)=(0.951779,0.951779,0.847574); rgb(220pt)=(0.953151,0.953151,0.852177); rgb(221pt)=(0.954521,0.954521,0.856754); rgb(222pt)=(0.95589,0.95589,0.861307); rgb(223pt)=(0.957256,0.957256,0.865837); rgb(224pt)=(0.958621,0.958621,0.870342); rgb(225pt)=(0.959984,0.959984,0.874825); rgb(226pt)=(0.961344,0.961344,0.879285); rgb(227pt)=(0.962703,0.962703,0.883722); rgb(228pt)=(0.96406,0.96406,0.888137); rgb(229pt)=(0.965415,0.965415,0.89253); rgb(230pt)=(0.966768,0.966768,0.896901); rgb(231pt)=(0.968119,0.968119,0.901252); rgb(232pt)=(0.969469,0.969469,0.905581); rgb(233pt)=(0.970816,0.970816,0.90989); rgb(234pt)=(0.972162,0.972162,0.914179); rgb(235pt)=(0.973505,0.973505,0.918447); rgb(236pt)=(0.974847,0.974847,0.922696); rgb(237pt)=(0.976187,0.976187,0.926926); rgb(238pt)=(0.977525,0.977525,0.931136); rgb(239pt)=(0.978862,0.978862,0.935327); rgb(240pt)=(0.980196,0.980196,0.9395); rgb(241pt)=(0.981529,0.981529,0.943654); rgb(242pt)=(0.98286,0.98286,0.947789); rgb(243pt)=(0.984189,0.984189,0.951907); rgb(244pt)=(0.985516,0.985516,0.956007); rgb(245pt)=(0.986842,0.986842,0.96009); rgb(246pt)=(0.988165,0.988165,0.964155); rgb(247pt)=(0.989487,0.989487,0.968204); rgb(248pt)=(0.990807,0.990807,0.972235); rgb(249pt)=(0.992126,0.992126,0.97625); rgb(250pt)=(0.993443,0.993443,0.980248); rgb(251pt)=(0.994757,0.994757,0.98423); rgb(252pt)=(0.996071,0.996071,0.988196); rgb(253pt)=(0.997382,0.997382,0.992146); rgb(254pt)=(0.998692,0.998692,0.996081); rgb(255pt)=(1,1,1)}, mesh/rows=16]
table[row sep=crcr, point meta=\thisrow{c}] {%
x	y	c\\
0	0	0.220783302450986\\
2	0	0.22001045046603\\
4	0	0.219285140953451\\
6	0	0.220268624923521\\
8	0	0.219789019903085\\
10	0	0.220176216307628\\
12	0	0.220328649548406\\
14	0	0.219748253712619\\
16	0	0.220796835437566\\
18	0	0.220513740004008\\
20	0	0.220024649443409\\
22	0	0.220151204825398\\
24	0	0.220488625619912\\
26	0	0.219808942673901\\
28	0	0.2201300933855\\
30	0	0.219811650065819\\
0	2	0.612730935453353\\
2	2	0.638063515810591\\
4	2	0.633394512870962\\
6	2	0.639511292186912\\
8	2	0.638952640169102\\
10	2	0.640515267499298\\
12	2	0.658528353267128\\
14	2	0.640454577803477\\
16	2	0.637875523093121\\
18	2	0.638741853476359\\
20	2	0.634479464471035\\
22	2	0.62924640802514\\
24	2	0.634074525184953\\
26	2	0.636770832074421\\
28	2	0.635630589421674\\
30	2	0.638679777090155\\
0	4	0.463041238817743\\
2	4	0.429466279187554\\
4	4	0.428084094501583\\
6	4	0.423953697644322\\
8	4	0.428212494549024\\
10	4	0.451133429738919\\
12	4	0.448347019704439\\
14	4	0.43034325748846\\
16	4	0.423445022801867\\
18	4	0.42492213336132\\
20	4	0.423490643016707\\
22	4	0.431473416073487\\
24	4	0.424069277418595\\
26	4	0.427922374562472\\
28	4	0.420809003237373\\
30	4	0.427288666422908\\
0	6	0.398607696789708\\
2	6	0.354187093469068\\
4	6	0.34712927233377\\
6	6	0.344922618322817\\
8	6	0.362679738449225\\
10	6	0.365176028283663\\
12	6	0.354874394966093\\
14	6	0.345731048937781\\
16	6	0.344578662155039\\
18	6	0.345306428888368\\
20	6	0.340415647619334\\
22	6	0.340096728635016\\
24	6	0.333055321592215\\
26	6	0.338278471007252\\
28	6	0.339193816626463\\
30	6	0.339983494996361\\
0	8	0.361980784977074\\
2	8	0.308051276203007\\
4	8	0.3024047372735\\
6	8	0.309102533897687\\
8	8	0.313353746237637\\
10	8	0.311931191820253\\
12	8	0.303722954872363\\
14	8	0.298502709086754\\
16	8	0.301028912090391\\
18	8	0.296089420948892\\
20	8	0.294420421565513\\
22	8	0.29292538555984\\
24	8	0.292407487076888\\
26	8	0.293215269160413\\
28	8	0.289968176849612\\
30	8	0.288251880741888\\
0	10	0.340617514803223\\
2	10	0.28069134128631\\
4	10	0.279619032448206\\
6	10	0.279400523887022\\
8	10	0.281044222755296\\
10	10	0.279744895824887\\
12	10	0.275433209292658\\
14	10	0.270255984622254\\
16	10	0.271940367102254\\
18	10	0.265984759579889\\
20	10	0.261341064730394\\
22	10	0.259901559765584\\
24	10	0.25866692260248\\
26	10	0.257170609292934\\
28	10	0.256032481754335\\
30	10	0.256218024397118\\
0	12	0.323242947355245\\
2	12	0.265402546610502\\
4	12	0.259844966905385\\
6	12	0.256089766349432\\
8	12	0.259270962684\\
10	12	0.258307421784025\\
12	12	0.254410986847132\\
14	12	0.250495202271269\\
16	12	0.244587234767491\\
18	12	0.24078365650408\\
20	12	0.238863754423416\\
22	12	0.236518622170643\\
24	12	0.231579417136362\\
26	12	0.232575504803321\\
28	12	0.2340391782061\\
30	12	0.232024238758318\\
0	14	0.315867344132747\\
2	14	0.253601016736053\\
4	14	0.244958742944089\\
6	14	0.243127982831848\\
8	14	0.242752512478935\\
10	14	0.243370731852028\\
12	14	0.236598838977071\\
14	14	0.232943407114847\\
16	14	0.226715108068012\\
18	14	0.221055226487425\\
20	14	0.222879379072062\\
22	14	0.217342542092765\\
24	14	0.214348371530402\\
26	14	0.213052426642349\\
28	14	0.213043146468085\\
30	14	0.212439362714385\\
0	16	0.306203303796463\\
2	16	0.241535237436364\\
4	16	0.235618101811895\\
6	16	0.232399462630045\\
8	16	0.23403015161884\\
10	16	0.23276566107785\\
12	16	0.223105485882011\\
14	16	0.220287947921446\\
16	16	0.214078842970825\\
18	16	0.207434867314314\\
20	16	0.204868681851253\\
22	16	0.19984856814178\\
24	16	0.200153937411253\\
26	16	0.19851704246583\\
28	16	0.198818462984884\\
30	16	0.199188713244375\\
0	18	0.300134486876905\\
2	18	0.234497084117858\\
4	18	0.226290340253032\\
6	18	0.222680777862802\\
8	18	0.22314549276399\\
10	18	0.217959523216986\\
12	18	0.213399801330885\\
14	18	0.208083974663067\\
16	18	0.200851091865695\\
18	18	0.197638525846682\\
20	18	0.193807812461234\\
22	18	0.188832693468427\\
24	18	0.189433209859641\\
26	18	0.186340784616367\\
28	18	0.187029557058081\\
30	18	0.185469741593564\\
0	20	0.293373938066366\\
2	20	0.229180246257013\\
4	20	0.221718440018776\\
6	20	0.214989585045724\\
8	20	0.215301281931312\\
10	20	0.210737711869405\\
12	20	0.204830837122824\\
14	20	0.195725033604503\\
16	20	0.191530892502988\\
18	20	0.182267706242488\\
20	20	0.182484023180149\\
22	20	0.17624430480692\\
24	20	0.176976458956327\\
26	20	0.176408838034469\\
28	20	0.177261360318359\\
30	20	0.176959350839553\\
0	22	0.29026379136964\\
2	22	0.223895083311363\\
4	22	0.216098473396616\\
6	22	0.210378541580131\\
8	22	0.210279704130582\\
10	22	0.203876115205314\\
12	22	0.195375617548667\\
14	22	0.18807880503749\\
16	22	0.182719854526816\\
18	22	0.174965536354851\\
20	22	0.173414488194255\\
22	22	0.171467877209816\\
24	22	0.170828692458523\\
26	22	0.168005228254502\\
28	22	0.167313749525187\\
30	22	0.167536997458199\\
0	24	0.287495799328422\\
2	24	0.220212768145261\\
4	24	0.210365708172574\\
6	24	0.205762972198033\\
8	24	0.203000523844413\\
10	24	0.196817282326973\\
12	24	0.186140933576279\\
14	24	0.177730483604781\\
16	24	0.173059441152037\\
18	24	0.170269956172407\\
20	24	0.164507654441074\\
22	24	0.163373156482319\\
24	24	0.161621485398027\\
26	24	0.160613538827601\\
28	24	0.160857315815315\\
30	24	0.161328195228817\\
0	26	0.28525640116096\\
2	26	0.216937432559096\\
4	26	0.207151485262323\\
6	26	0.201206485574477\\
8	26	0.198088388084644\\
10	26	0.192152331043958\\
12	26	0.178097956496856\\
14	26	0.174297289723464\\
16	26	0.166521072164376\\
18	26	0.164173010251025\\
20	26	0.156433430368758\\
22	26	0.154189133908425\\
24	26	0.155175524770955\\
26	26	0.154848523607525\\
28	26	0.15378973637562\\
30	26	0.152527014297736\\
0	28	0.283303616314706\\
2	28	0.212596583713179\\
4	28	0.201778767587584\\
6	28	0.196089973051249\\
8	28	0.193962814747459\\
10	28	0.184442980175321\\
12	28	0.175375008871662\\
14	28	0.16724874450034\\
16	28	0.162366362070837\\
18	28	0.157120906875999\\
20	28	0.152130664499119\\
22	28	0.151458390448871\\
24	28	0.148881742180771\\
26	28	0.146288330987017\\
28	28	0.147522082498773\\
30	28	0.146746337471485\\
0	30	0.28025321417017\\
2	30	0.210978722224357\\
4	30	0.199064258020724\\
6	30	0.191789316526668\\
8	30	0.18638200242781\\
10	30	0.179009343318386\\
12	30	0.169566996391809\\
14	30	0.162690844686534\\
16	30	0.155123026544778\\
18	30	0.149641306575323\\
20	30	0.147366687826568\\
22	30	0.143583792018337\\
24	30	0.142273727489196\\
26	30	0.143216598705731\\
28	30	0.142907974597684\\
30	30	0.141985493019674\\
};
\end{axis}

\begin{axis}[%
width=3.633in,
height=2.425in,
at={(0.27in,0.3in)},
scale only axis,
point meta min=0,
point meta max=1,
xmin=0,
xmax=1,
ymin=0,
ymax=1,
axis line style={draw=none},
ticks=none,
axis x line*=bottom,
axis y line*=left
]
\node[below right, align=left, font=\color{white}]
at (rel axis cs:0.545,0.545) {Dominance\\of\\Noise};
\draw[line width=3.0pt, draw=mycolor1] (axis cs:0.480952380952381,0.298039215686277) rectangle (axis cs:0.793499262949692,0.923436041083101);
\end{axis}
\end{tikzpicture}%

%% file: TIFS_figures_tikz/TBerlin_dist1_SNR5_mismatch.tex
%
%
\definecolor{mycolor1}{rgb}{0.63529,0.07843,0.18431}%
\begin{tikzpicture}

\begin{axis}[%
width=2.5in,
height=2.000in,
at={(0.666in,0.569in)},
scale only axis,
point meta min=0.141548192733656,
point meta max=0.673381198915271,
xmin=0,
xmax=30,
xlabel style={font=\color{white!15!black}},
xlabel={Dimensions to omit},
ymin=2,
ymax=30,
ylabel style={font=\color{white!15!black}},
ylabel={$\tilde{D}$ - Dimensions to keep from $\tilde{D}_1$ to $\tilde{D}_2$.},
axis background/.style={fill=white},
axis x line*=bottom,
axis y line*=left,
colormap={mymap}{[1pt] rgb(0pt)=(0,0.5,0.4); rgb(255pt)=(1,1,0.4)},
colorbar
]

\addplot[%
surf,
shader=flat corner, draw=black, colormap={mymap}{[1pt] rgb(0pt)=(0,0.5,0.4); rgb(255pt)=(1,1,0.4)}, mesh/rows=16]
table[row sep=crcr, point meta=\thisrow{c}] {%
x	y	c\\
0	0	0.3598046875\\
2	0	0.358450927734375\\
4	0	0.358876953125\\
6	0	0.359600830078125\\
8	0	0.360050048828125\\
10	0	0.359044189453125\\
12	0	0.359031982421875\\
14	0	0.358729248046875\\
16	0	0.3594775390625\\
18	0	0.359287109375\\
20	0	0.360042724609375\\
22	0	0.3589013671875\\
24	0	0.36073974609375\\
26	0	0.35935546875\\
28	0	0.359237060546875\\
30	0	0.359967041015625\\
0	2	0.173753662109375\\
2	2	0.20093017578125\\
4	2	0.238895263671875\\
6	2	0.277601318359375\\
8	2	0.29992431640625\\
10	2	0.34269287109375\\
12	2	0.354481201171875\\
14	2	0.353150634765625\\
16	2	0.39472900390625\\
18	2	0.4318994140625\\
20	2	0.445087890625\\
22	2	0.468529052734375\\
24	2	0.494432373046875\\
26	2	0.495867919921875\\
28	2	0.496986083984375\\
30	2	0.498004150390625\\
0	4	0.161094970703125\\
2	4	0.19211669921875\\
4	4	0.22875732421875\\
6	4	0.265604248046875\\
8	4	0.303968505859375\\
10	4	0.30763916015625\\
12	4	0.341455078125\\
14	4	0.36708984375\\
16	4	0.40701416015625\\
18	4	0.43628173828125\\
20	4	0.463726806640625\\
22	4	0.485657958984375\\
24	4	0.49703857421875\\
26	4	0.5009521484375\\
28	4	0.498013916015625\\
30	4	0.500069580078125\\
0	6	0.16190673828125\\
2	6	0.208070068359375\\
4	6	0.223837890625\\
6	6	0.252801513671875\\
8	6	0.282022705078125\\
10	6	0.314974365234375\\
12	6	0.3473779296875\\
14	6	0.38080810546875\\
16	6	0.4106591796875\\
18	6	0.446051025390625\\
20	6	0.476583251953125\\
22	6	0.489876708984375\\
24	6	0.4971630859375\\
26	6	0.500830078125\\
28	6	0.499447021484375\\
30	6	0.50039306640625\\
0	8	0.174820556640625\\
2	8	0.19856689453125\\
4	8	0.225174560546875\\
6	8	0.25901611328125\\
8	8	0.2908642578125\\
10	8	0.3148681640625\\
12	8	0.35594482421875\\
14	8	0.38526123046875\\
16	8	0.42565185546875\\
18	8	0.458802490234375\\
20	8	0.47960205078125\\
22	8	0.4953369140625\\
24	8	0.498826904296875\\
26	8	0.501390380859375\\
28	8	0.5000732421875\\
30	8	0.50070556640625\\
0	10	0.172369384765625\\
2	10	0.207188720703125\\
4	10	0.233475341796875\\
6	10	0.260205078125\\
8	10	0.2966650390625\\
10	10	0.3303662109375\\
12	10	0.366593017578125\\
14	10	0.405142822265625\\
16	10	0.438155517578125\\
18	10	0.467880859375\\
20	10	0.485159912109375\\
22	10	0.49404541015625\\
24	10	0.498094482421875\\
26	10	0.49767578125\\
28	10	0.500230712890625\\
30	10	0.498807373046875\\
0	12	0.177757568359375\\
2	12	0.212232666015625\\
4	12	0.238199462890625\\
6	12	0.269736328125\\
8	12	0.305758056640625\\
10	12	0.341387939453125\\
12	12	0.38238525390625\\
14	12	0.417430419921875\\
16	12	0.446934814453125\\
18	12	0.476671142578125\\
20	12	0.4884619140625\\
22	12	0.496080322265625\\
24	12	0.499083251953125\\
26	12	0.4985546875\\
28	12	0.500313720703125\\
30	12	0.499520263671875\\
0	14	0.181202392578125\\
2	14	0.216688232421875\\
4	14	0.248970947265625\\
6	14	0.278685302734375\\
8	14	0.313397216796875\\
10	14	0.358248291015625\\
12	14	0.393897705078125\\
14	14	0.43040283203125\\
16	14	0.457020263671875\\
18	14	0.478203125\\
20	14	0.490579833984375\\
22	14	0.497001953125\\
24	14	0.499124755859375\\
26	14	0.499599609375\\
28	14	0.499532470703125\\
30	14	0.4985595703125\\
0	16	0.1860986328125\\
2	16	0.22483154296875\\
4	16	0.257440185546875\\
6	16	0.29160400390625\\
8	16	0.32574462890625\\
10	16	0.3674169921875\\
12	16	0.408607177734375\\
14	16	0.43597412109375\\
16	16	0.463187255859375\\
18	16	0.4802978515625\\
20	16	0.49154541015625\\
22	16	0.497008056640625\\
24	16	0.49921630859375\\
26	16	0.499688720703125\\
28	16	0.499840087890625\\
30	16	0.499403076171875\\
0	18	0.18711669921875\\
2	18	0.231661376953125\\
4	18	0.2652099609375\\
6	18	0.298507080078125\\
8	18	0.3451708984375\\
10	18	0.381884765625\\
12	18	0.418135986328125\\
14	18	0.448492431640625\\
16	18	0.46699462890625\\
18	18	0.4846484375\\
20	18	0.493294677734375\\
22	18	0.4967626953125\\
24	18	0.497733154296875\\
26	18	0.498994140625\\
28	18	0.49825927734375\\
30	18	0.500491943359375\\
0	20	0.194300537109375\\
2	20	0.23953369140625\\
4	20	0.274337158203125\\
6	20	0.3130810546875\\
8	20	0.35787109375\\
10	20	0.392850341796875\\
12	20	0.4226318359375\\
14	20	0.45156494140625\\
16	20	0.47108154296875\\
18	20	0.48527099609375\\
20	20	0.49291748046875\\
22	20	0.497471923828125\\
24	20	0.49925048828125\\
26	20	0.4988671875\\
28	20	0.499283447265625\\
30	20	0.49801025390625\\
0	22	0.198883056640625\\
2	22	0.247802734375\\
4	22	0.288175048828125\\
6	22	0.32750244140625\\
8	22	0.369552001953125\\
10	22	0.39917724609375\\
12	22	0.43246826171875\\
14	22	0.457967529296875\\
16	22	0.474486083984375\\
18	22	0.487708740234375\\
20	22	0.4935888671875\\
22	22	0.496861572265625\\
24	22	0.499893798828125\\
26	22	0.5004052734375\\
28	22	0.500172119140625\\
30	22	0.499073486328125\\
0	24	0.203173828125\\
2	24	0.25887451171875\\
4	24	0.2968505859375\\
6	24	0.33692138671875\\
8	24	0.380186767578125\\
10	24	0.407723388671875\\
12	24	0.43661376953125\\
14	24	0.46079345703125\\
16	24	0.4767529296875\\
18	24	0.4898095703125\\
20	24	0.4944384765625\\
22	24	0.496669921875\\
24	24	0.49910888671875\\
26	24	0.498282470703125\\
28	24	0.499818115234375\\
30	24	0.498714599609375\\
0	26	0.210875244140625\\
2	26	0.268516845703125\\
4	26	0.30834228515625\\
6	26	0.347691650390625\\
8	26	0.381988525390625\\
10	26	0.415361328125\\
12	26	0.444334716796875\\
14	26	0.46304931640625\\
16	26	0.47769287109375\\
18	26	0.489495849609375\\
20	26	0.495255126953125\\
22	26	0.4983984375\\
24	26	0.498338623046875\\
26	26	0.498951416015625\\
28	26	0.500167236328125\\
30	26	0.500299072265625\\
0	28	0.2190185546875\\
2	28	0.2773779296875\\
4	28	0.316990966796875\\
6	28	0.3556005859375\\
8	28	0.390389404296875\\
10	28	0.421771240234375\\
12	28	0.4451953125\\
14	28	0.466494140625\\
16	28	0.48015869140625\\
18	28	0.49018310546875\\
20	28	0.494720458984375\\
22	28	0.4992431640625\\
24	28	0.499395751953125\\
26	28	0.4986669921875\\
28	28	0.499061279296875\\
30	28	0.499581298828125\\
0	30	0.22582763671875\\
2	30	0.288170166015625\\
4	30	0.326260986328125\\
6	30	0.36248046875\\
8	30	0.398848876953125\\
10	30	0.423338623046875\\
12	30	0.450679931640625\\
14	30	0.4699072265625\\
16	30	0.4818212890625\\
18	30	0.490738525390625\\
20	30	0.495516357421875\\
22	30	0.497958984375\\
24	30	0.49957763671875\\
26	30	0.50001220703125\\
28	30	0.49857177734375\\
30	30	0.50021484375\\
};
\end{axis}

\begin{axis}[%
width=3.633in,
height=2.425in,
at={(0.27in,0.3in)},
scale only axis,
point meta min=0,
point meta max=1,
xmin=0,
xmax=1,
ymin=0,
ymax=1,
axis line style={draw=none},
ticks=none,
axis x line*=bottom,
axis y line*=left
]
\node[below right, align=left, font=\color{red}]
at (rel axis cs:0.544,0.544) {High\\mismatch\\probability};
\draw[line width=3.0pt, draw=mycolor1] (axis cs:0.476190476190476,0.292623490913101) rectangle (axis cs:0.796056547619046,0.926378134614557);
\end{axis}
\end{tikzpicture}%

%% file: TIFS_figures_tikz/quadriga_hsic_new.tex
%
%
\definecolor{mycolor1}{rgb}{0.00000,0.44706,0.74118}%
\begin{tikzpicture}

\begin{axis}[%
width=2.421in,
height=2.154in,
at={(0.758in,0.692in)},
scale only axis,
xmin=0,
xmax=30,
xlabel style={font=\color{white!15!black}},
xlabel={$\tilde{D}_1$},
ymin=0,
ymax=5,
ylabel style={font=\color{white!15!black}},
ylabel={$\overline {\Delta}$},
axis background/.style={fill=white},
xmajorgrids,
ymajorgrids,
legend style={legend cell align=left, align=left, draw=white!15!black}
]
\addplot [color=mycolor1, line width=2.0pt, mark=asterisk, mark options={solid, mycolor1}]
  table[row sep=crcr]{%
0	4.07276742318096\\
2	3.62787566766614\\
4	4.06017321494125\\
6	4.37493734212492\\
8	4.62069357477874\\
10	4.75861917339629\\
12	4.07910139182567\\
14	4.02840393921884\\
16	3.4598725204214\\
18	2.79489722211277\\
20	2.35973749872358\\
22	1.63826578762343\\
24	1.04471182703371\\
26	0.543095489682958\\
28	0.51073603791551\\
30	0.510838645959705\\
};
\addlegendentry{SNR=20dB}

\addplot [color=red, line width=2.0pt, mark=star, mark options={solid, red}]
  table[row sep=crcr]{%
0	1.9458839942026\\
2	2.5118388955174\\
4	2.34178431726011\\
6	2.04345587549624\\
8	1.60414029694309\\
10	1.41813777237778\\
12	0.899484049953813\\
14	0.719498612569307\\
16	0.643174377091693\\
18	0.584922751030906\\
20	0.493355727939019\\
22	0.462740341616172\\
24	0.559335915688742\\
26	0.517134457779317\\
28	0.453284009034519\\
30	0.515365086506606\\
};
\addlegendentry{SNR=5dB}

\end{axis}
\end{tikzpicture}%

%% file: TIFS_figures_tikz/Nokia_snr20_corr.tex
%
%
\definecolor{mycolor1}{rgb}{0.85098,0.32549,0.09804}%
\definecolor{mycolor2}{rgb}{0.49412,0.18431,0.55686}%
\begin{tikzpicture}

\begin{axis}[%
width=2.5in,
height=2.000in,
at={(0.666in,0.569in)},
scale only axis,
point meta min=0.141548192733656,
point meta max=0.673381198915271,
xmin=0,
xmax=30,
xlabel style={font=\color{white!15!black}},
xlabel={Dimensions to omit},
ymin=2,
ymax=30,
ylabel style={font=\color{white!15!black}},
ylabel={$\tilde{D}$ - Dimensions to keep from $\tilde{D}_1$ to $\tilde{D}_2$},
axis background/.style={fill=white},
axis x line*=bottom,
axis y line*=left,
colormap={mymap}{[1pt] rgb(0pt)=(0.0589256,0,0); rgb(1pt)=(0.0977692,0.051131,0.051131); rgb(2pt)=(0.125082,0.0723102,0.0723102); rgb(3pt)=(0.147418,0.0885615,0.0885615); rgb(4pt)=(0.166789,0.102262,0.102262); rgb(5pt)=(0.184134,0.114332,0.114332); rgb(6pt)=(0.19998,0.125245,0.125245); rgb(7pt)=(0.214659,0.13528,0.13528); rgb(8pt)=(0.228397,0.14462,0.14462); rgb(9pt)=(0.241354,0.153393,0.153393); rgb(10pt)=(0.25365,0.16169,0.16169); rgb(11pt)=(0.265377,0.169582,0.169582); rgb(12pt)=(0.276607,0.177123,0.177123); rgb(13pt)=(0.287399,0.184355,0.184355); rgb(14pt)=(0.2978,0.191315,0.191315); rgb(15pt)=(0.307849,0.19803,0.19803); rgb(16pt)=(0.317581,0.204524,0.204524); rgb(17pt)=(0.327024,0.210819,0.210819); rgb(18pt)=(0.336201,0.21693,0.21693); rgb(19pt)=(0.345134,0.222875,0.222875); rgb(20pt)=(0.353842,0.228665,0.228665); rgb(21pt)=(0.362341,0.234312,0.234312); rgb(22pt)=(0.370645,0.239826,0.239826); rgb(23pt)=(0.378766,0.245216,0.245216); rgb(24pt)=(0.386718,0.25049,0.25049); rgb(25pt)=(0.394509,0.255655,0.255655); rgb(26pt)=(0.402149,0.260718,0.260718); rgb(27pt)=(0.409647,0.265684,0.265684); rgb(28pt)=(0.41701,0.27056,0.27056); rgb(29pt)=(0.424245,0.275349,0.275349); rgb(30pt)=(0.431359,0.280056,0.280056); rgb(31pt)=(0.438357,0.284685,0.284685); rgb(32pt)=(0.445245,0.289241,0.289241); rgb(33pt)=(0.452029,0.293725,0.293725); rgb(34pt)=(0.458712,0.298142,0.298142); rgb(35pt)=(0.465299,0.302495,0.302495); rgb(36pt)=(0.471794,0.306786,0.306786); rgb(37pt)=(0.478201,0.311018,0.311018); rgb(38pt)=(0.484524,0.315193,0.315193); rgb(39pt)=(0.490764,0.319313,0.319313); rgb(40pt)=(0.496927,0.323381,0.323381); rgb(41pt)=(0.503014,0.327398,0.327398); rgb(42pt)=(0.509028,0.331367,0.331367); rgb(43pt)=(0.514972,0.335288,0.335288); rgb(44pt)=(0.520848,0.339165,0.339165); rgb(45pt)=(0.526659,0.342997,0.342997); rgb(46pt)=(0.532406,0.346787,0.346787); rgb(47pt)=(0.538092,0.350536,0.350536); rgb(48pt)=(0.543718,0.354246,0.354246); rgb(49pt)=(0.549287,0.357917,0.357917); rgb(50pt)=(0.554799,0.361551,0.361551); rgb(51pt)=(0.560258,0.365148,0.365148); rgb(52pt)=(0.565664,0.368711,0.368711); rgb(53pt)=(0.571018,0.372239,0.372239); rgb(54pt)=(0.576323,0.375735,0.375735); rgb(55pt)=(0.58158,0.379198,0.379198); rgb(56pt)=(0.586789,0.382629,0.382629); rgb(57pt)=(0.591953,0.386031,0.386031); rgb(58pt)=(0.597072,0.389402,0.389402); rgb(59pt)=(0.602148,0.392745,0.392745); rgb(60pt)=(0.607181,0.396059,0.396059); rgb(61pt)=(0.612172,0.399346,0.399346); rgb(62pt)=(0.617124,0.402606,0.402606); rgb(63pt)=(0.622035,0.40584,0.40584); rgb(64pt)=(0.626909,0.409048,0.409048); rgb(65pt)=(0.631745,0.412231,0.412231); rgb(66pt)=(0.636544,0.41539,0.41539); rgb(67pt)=(0.641307,0.418525,0.418525); rgb(68pt)=(0.646035,0.421637,0.421637); rgb(69pt)=(0.650729,0.424726,0.424726); rgb(70pt)=(0.655389,0.427793,0.427793); rgb(71pt)=(0.660016,0.430837,0.430837); rgb(72pt)=(0.664611,0.433861,0.433861); rgb(73pt)=(0.669174,0.436863,0.436863); rgb(74pt)=(0.673707,0.439845,0.439845); rgb(75pt)=(0.678209,0.442807,0.442807); rgb(76pt)=(0.682681,0.44575,0.44575); rgb(77pt)=(0.687125,0.448673,0.448673); rgb(78pt)=(0.69154,0.451577,0.451577); rgb(79pt)=(0.695927,0.454462,0.454462); rgb(80pt)=(0.700286,0.45733,0.45733); rgb(81pt)=(0.704618,0.460179,0.460179); rgb(82pt)=(0.708924,0.463011,0.463011); rgb(83pt)=(0.713204,0.465826,0.465826); rgb(84pt)=(0.717459,0.468623,0.468623); rgb(85pt)=(0.721688,0.471405,0.471405); rgb(86pt)=(0.725893,0.474169,0.474169); rgb(87pt)=(0.730073,0.476918,0.476918); rgb(88pt)=(0.73423,0.479651,0.479651); rgb(89pt)=(0.738363,0.482369,0.482369); rgb(90pt)=(0.742473,0.485071,0.485071); rgb(91pt)=(0.746561,0.487759,0.487759); rgb(92pt)=(0.750626,0.490431,0.490431); rgb(93pt)=(0.75467,0.493089,0.493089); rgb(94pt)=(0.758691,0.495733,0.495733); rgb(95pt)=(0.762692,0.498363,0.498363); rgb(96pt)=(0.764404,0.504433,0.500979); rgb(97pt)=(0.766112,0.51043,0.503582); rgb(98pt)=(0.767817,0.516358,0.506171); rgb(99pt)=(0.769517,0.522219,0.508747); rgb(100pt)=(0.771214,0.528014,0.51131); rgb(101pt)=(0.772907,0.533747,0.51386); rgb(102pt)=(0.774597,0.539418,0.516398); rgb(103pt)=(0.776282,0.545031,0.518923); rgb(104pt)=(0.777964,0.550586,0.521436); rgb(105pt)=(0.779643,0.556086,0.523937); rgb(106pt)=(0.781318,0.561532,0.526426); rgb(107pt)=(0.782989,0.566926,0.528903); rgb(108pt)=(0.784657,0.572269,0.531369); rgb(109pt)=(0.786321,0.577562,0.533823); rgb(110pt)=(0.787982,0.582808,0.536266); rgb(111pt)=(0.789639,0.588006,0.538699); rgb(112pt)=(0.791292,0.59316,0.54112); rgb(113pt)=(0.792943,0.598268,0.54353); rgb(114pt)=(0.79459,0.603334,0.54593); rgb(115pt)=(0.796233,0.608357,0.548319); rgb(116pt)=(0.797873,0.613339,0.550698); rgb(117pt)=(0.79951,0.618281,0.553066); rgb(118pt)=(0.801143,0.623184,0.555425); rgb(119pt)=(0.802773,0.628048,0.557773); rgb(120pt)=(0.8044,0.632875,0.560112); rgb(121pt)=(0.806023,0.637666,0.562441); rgb(122pt)=(0.807643,0.642421,0.56476); rgb(123pt)=(0.80926,0.647141,0.56707); rgb(124pt)=(0.810874,0.651826,0.569371); rgb(125pt)=(0.812484,0.656479,0.571662); rgb(126pt)=(0.814092,0.661098,0.573944); rgb(127pt)=(0.815696,0.665686,0.576217); rgb(128pt)=(0.817297,0.670242,0.578481); rgb(129pt)=(0.818895,0.674767,0.580737); rgb(130pt)=(0.820489,0.679262,0.582983); rgb(131pt)=(0.822081,0.683728,0.585221); rgb(132pt)=(0.823669,0.688164,0.58745); rgb(133pt)=(0.825255,0.692573,0.589671); rgb(134pt)=(0.826837,0.696953,0.591884); rgb(135pt)=(0.828417,0.701306,0.594089); rgb(136pt)=(0.829993,0.705632,0.596285); rgb(137pt)=(0.831567,0.709932,0.598473); rgb(138pt)=(0.833137,0.714206,0.600653); rgb(139pt)=(0.834705,0.718454,0.602826); rgb(140pt)=(0.836269,0.722678,0.60499); rgb(141pt)=(0.837831,0.726877,0.607147); rgb(142pt)=(0.83939,0.731051,0.609296); rgb(143pt)=(0.840946,0.735203,0.611438); rgb(144pt)=(0.842499,0.73933,0.613572); rgb(145pt)=(0.844049,0.743435,0.615699); rgb(146pt)=(0.845596,0.747518,0.617818); rgb(147pt)=(0.847141,0.751578,0.61993); rgb(148pt)=(0.848682,0.755616,0.622035); rgb(149pt)=(0.850221,0.759633,0.624133); rgb(150pt)=(0.851757,0.763629,0.626224); rgb(151pt)=(0.85329,0.767604,0.628308); rgb(152pt)=(0.854821,0.771558,0.630385); rgb(153pt)=(0.856349,0.775493,0.632456); rgb(154pt)=(0.857874,0.779407,0.634519); rgb(155pt)=(0.859396,0.783302,0.636576); rgb(156pt)=(0.860916,0.787178,0.638626); rgb(157pt)=(0.862433,0.791034,0.64067); rgb(158pt)=(0.863947,0.794872,0.642707); rgb(159pt)=(0.865459,0.798692,0.644737); rgb(160pt)=(0.866968,0.802493,0.646762); rgb(161pt)=(0.868475,0.806276,0.64878); rgb(162pt)=(0.869979,0.810042,0.650791); rgb(163pt)=(0.87148,0.81379,0.652797); rgb(164pt)=(0.872979,0.817522,0.654796); rgb(165pt)=(0.874475,0.821236,0.65679); rgb(166pt)=(0.875968,0.824933,0.658777); rgb(167pt)=(0.877459,0.828614,0.660758); rgb(168pt)=(0.878948,0.832279,0.662733); rgb(169pt)=(0.880434,0.835927,0.664703); rgb(170pt)=(0.881917,0.83956,0.666667); rgb(171pt)=(0.883398,0.843177,0.668625); rgb(172pt)=(0.884877,0.846779,0.670577); rgb(173pt)=(0.886353,0.850365,0.672523); rgb(174pt)=(0.887826,0.853936,0.674464); rgb(175pt)=(0.889297,0.857493,0.6764); rgb(176pt)=(0.890766,0.861035,0.678329); rgb(177pt)=(0.892232,0.864562,0.680254); rgb(178pt)=(0.893696,0.868075,0.682173); rgb(179pt)=(0.895158,0.871574,0.684086); rgb(180pt)=(0.896617,0.875058,0.685994); rgb(181pt)=(0.898073,0.878529,0.687897); rgb(182pt)=(0.899528,0.881987,0.689795); rgb(183pt)=(0.90098,0.88543,0.691687); rgb(184pt)=(0.90243,0.888861,0.693575); rgb(185pt)=(0.903877,0.892278,0.695457); rgb(186pt)=(0.905322,0.895682,0.697334); rgb(187pt)=(0.906765,0.899074,0.699206); rgb(188pt)=(0.908205,0.902452,0.701073); rgb(189pt)=(0.909643,0.905818,0.702935); rgb(190pt)=(0.911079,0.909172,0.704792); rgb(191pt)=(0.912513,0.912513,0.706644); rgb(192pt)=(0.913944,0.913944,0.712158); rgb(193pt)=(0.915373,0.915373,0.717629); rgb(194pt)=(0.9168,0.9168,0.723059); rgb(195pt)=(0.918225,0.918225,0.728449); rgb(196pt)=(0.919648,0.919648,0.733798); rgb(197pt)=(0.921068,0.921068,0.739109); rgb(198pt)=(0.922486,0.922486,0.744383); rgb(199pt)=(0.923902,0.923902,0.749619); rgb(200pt)=(0.925316,0.925316,0.754818); rgb(201pt)=(0.926727,0.926727,0.759983); rgb(202pt)=(0.928137,0.928137,0.765112); rgb(203pt)=(0.929544,0.929544,0.770207); rgb(204pt)=(0.930949,0.930949,0.775269); rgb(205pt)=(0.932352,0.932352,0.780298); rgb(206pt)=(0.933753,0.933753,0.785294); rgb(207pt)=(0.935152,0.935152,0.790259); rgb(208pt)=(0.936549,0.936549,0.795193); rgb(209pt)=(0.937944,0.937944,0.800097); rgb(210pt)=(0.939336,0.939336,0.804971); rgb(211pt)=(0.940727,0.940727,0.809815); rgb(212pt)=(0.942116,0.942116,0.814631); rgb(213pt)=(0.943502,0.943502,0.819418); rgb(214pt)=(0.944886,0.944886,0.824178); rgb(215pt)=(0.946269,0.946269,0.82891); rgb(216pt)=(0.947649,0.947649,0.833615); rgb(217pt)=(0.949028,0.949028,0.838294); rgb(218pt)=(0.950404,0.950404,0.842947); rgb(219pt)=(0.951779,0.951779,0.847574); rgb(220pt)=(0.953151,0.953151,0.852177); rgb(221pt)=(0.954521,0.954521,0.856754); rgb(222pt)=(0.95589,0.95589,0.861307); rgb(223pt)=(0.957256,0.957256,0.865837); rgb(224pt)=(0.958621,0.958621,0.870342); rgb(225pt)=(0.959984,0.959984,0.874825); rgb(226pt)=(0.961344,0.961344,0.879285); rgb(227pt)=(0.962703,0.962703,0.883722); rgb(228pt)=(0.96406,0.96406,0.888137); rgb(229pt)=(0.965415,0.965415,0.89253); rgb(230pt)=(0.966768,0.966768,0.896901); rgb(231pt)=(0.968119,0.968119,0.901252); rgb(232pt)=(0.969469,0.969469,0.905581); rgb(233pt)=(0.970816,0.970816,0.90989); rgb(234pt)=(0.972162,0.972162,0.914179); rgb(235pt)=(0.973505,0.973505,0.918447); rgb(236pt)=(0.974847,0.974847,0.922696); rgb(237pt)=(0.976187,0.976187,0.926926); rgb(238pt)=(0.977525,0.977525,0.931136); rgb(239pt)=(0.978862,0.978862,0.935327); rgb(240pt)=(0.980196,0.980196,0.9395); rgb(241pt)=(0.981529,0.981529,0.943654); rgb(242pt)=(0.98286,0.98286,0.947789); rgb(243pt)=(0.984189,0.984189,0.951907); rgb(244pt)=(0.985516,0.985516,0.956007); rgb(245pt)=(0.986842,0.986842,0.96009); rgb(246pt)=(0.988165,0.988165,0.964155); rgb(247pt)=(0.989487,0.989487,0.968204); rgb(248pt)=(0.990807,0.990807,0.972235); rgb(249pt)=(0.992126,0.992126,0.97625); rgb(250pt)=(0.993443,0.993443,0.980248); rgb(251pt)=(0.994757,0.994757,0.98423); rgb(252pt)=(0.996071,0.996071,0.988196); rgb(253pt)=(0.997382,0.997382,0.992146); rgb(254pt)=(0.998692,0.998692,0.996081); rgb(255pt)=(1,1,1)},
colorbar
]

\addplot[%
surf,
shader=flat corner, draw=black, colormap={mymap}{[1pt] rgb(0pt)=(0.0589256,0,0); rgb(1pt)=(0.0977692,0.051131,0.051131); rgb(2pt)=(0.125082,0.0723102,0.0723102); rgb(3pt)=(0.147418,0.0885615,0.0885615); rgb(4pt)=(0.166789,0.102262,0.102262); rgb(5pt)=(0.184134,0.114332,0.114332); rgb(6pt)=(0.19998,0.125245,0.125245); rgb(7pt)=(0.214659,0.13528,0.13528); rgb(8pt)=(0.228397,0.14462,0.14462); rgb(9pt)=(0.241354,0.153393,0.153393); rgb(10pt)=(0.25365,0.16169,0.16169); rgb(11pt)=(0.265377,0.169582,0.169582); rgb(12pt)=(0.276607,0.177123,0.177123); rgb(13pt)=(0.287399,0.184355,0.184355); rgb(14pt)=(0.2978,0.191315,0.191315); rgb(15pt)=(0.307849,0.19803,0.19803); rgb(16pt)=(0.317581,0.204524,0.204524); rgb(17pt)=(0.327024,0.210819,0.210819); rgb(18pt)=(0.336201,0.21693,0.21693); rgb(19pt)=(0.345134,0.222875,0.222875); rgb(20pt)=(0.353842,0.228665,0.228665); rgb(21pt)=(0.362341,0.234312,0.234312); rgb(22pt)=(0.370645,0.239826,0.239826); rgb(23pt)=(0.378766,0.245216,0.245216); rgb(24pt)=(0.386718,0.25049,0.25049); rgb(25pt)=(0.394509,0.255655,0.255655); rgb(26pt)=(0.402149,0.260718,0.260718); rgb(27pt)=(0.409647,0.265684,0.265684); rgb(28pt)=(0.41701,0.27056,0.27056); rgb(29pt)=(0.424245,0.275349,0.275349); rgb(30pt)=(0.431359,0.280056,0.280056); rgb(31pt)=(0.438357,0.284685,0.284685); rgb(32pt)=(0.445245,0.289241,0.289241); rgb(33pt)=(0.452029,0.293725,0.293725); rgb(34pt)=(0.458712,0.298142,0.298142); rgb(35pt)=(0.465299,0.302495,0.302495); rgb(36pt)=(0.471794,0.306786,0.306786); rgb(37pt)=(0.478201,0.311018,0.311018); rgb(38pt)=(0.484524,0.315193,0.315193); rgb(39pt)=(0.490764,0.319313,0.319313); rgb(40pt)=(0.496927,0.323381,0.323381); rgb(41pt)=(0.503014,0.327398,0.327398); rgb(42pt)=(0.509028,0.331367,0.331367); rgb(43pt)=(0.514972,0.335288,0.335288); rgb(44pt)=(0.520848,0.339165,0.339165); rgb(45pt)=(0.526659,0.342997,0.342997); rgb(46pt)=(0.532406,0.346787,0.346787); rgb(47pt)=(0.538092,0.350536,0.350536); rgb(48pt)=(0.543718,0.354246,0.354246); rgb(49pt)=(0.549287,0.357917,0.357917); rgb(50pt)=(0.554799,0.361551,0.361551); rgb(51pt)=(0.560258,0.365148,0.365148); rgb(52pt)=(0.565664,0.368711,0.368711); rgb(53pt)=(0.571018,0.372239,0.372239); rgb(54pt)=(0.576323,0.375735,0.375735); rgb(55pt)=(0.58158,0.379198,0.379198); rgb(56pt)=(0.586789,0.382629,0.382629); rgb(57pt)=(0.591953,0.386031,0.386031); rgb(58pt)=(0.597072,0.389402,0.389402); rgb(59pt)=(0.602148,0.392745,0.392745); rgb(60pt)=(0.607181,0.396059,0.396059); rgb(61pt)=(0.612172,0.399346,0.399346); rgb(62pt)=(0.617124,0.402606,0.402606); rgb(63pt)=(0.622035,0.40584,0.40584); rgb(64pt)=(0.626909,0.409048,0.409048); rgb(65pt)=(0.631745,0.412231,0.412231); rgb(66pt)=(0.636544,0.41539,0.41539); rgb(67pt)=(0.641307,0.418525,0.418525); rgb(68pt)=(0.646035,0.421637,0.421637); rgb(69pt)=(0.650729,0.424726,0.424726); rgb(70pt)=(0.655389,0.427793,0.427793); rgb(71pt)=(0.660016,0.430837,0.430837); rgb(72pt)=(0.664611,0.433861,0.433861); rgb(73pt)=(0.669174,0.436863,0.436863); rgb(74pt)=(0.673707,0.439845,0.439845); rgb(75pt)=(0.678209,0.442807,0.442807); rgb(76pt)=(0.682681,0.44575,0.44575); rgb(77pt)=(0.687125,0.448673,0.448673); rgb(78pt)=(0.69154,0.451577,0.451577); rgb(79pt)=(0.695927,0.454462,0.454462); rgb(80pt)=(0.700286,0.45733,0.45733); rgb(81pt)=(0.704618,0.460179,0.460179); rgb(82pt)=(0.708924,0.463011,0.463011); rgb(83pt)=(0.713204,0.465826,0.465826); rgb(84pt)=(0.717459,0.468623,0.468623); rgb(85pt)=(0.721688,0.471405,0.471405); rgb(86pt)=(0.725893,0.474169,0.474169); rgb(87pt)=(0.730073,0.476918,0.476918); rgb(88pt)=(0.73423,0.479651,0.479651); rgb(89pt)=(0.738363,0.482369,0.482369); rgb(90pt)=(0.742473,0.485071,0.485071); rgb(91pt)=(0.746561,0.487759,0.487759); rgb(92pt)=(0.750626,0.490431,0.490431); rgb(93pt)=(0.75467,0.493089,0.493089); rgb(94pt)=(0.758691,0.495733,0.495733); rgb(95pt)=(0.762692,0.498363,0.498363); rgb(96pt)=(0.764404,0.504433,0.500979); rgb(97pt)=(0.766112,0.51043,0.503582); rgb(98pt)=(0.767817,0.516358,0.506171); rgb(99pt)=(0.769517,0.522219,0.508747); rgb(100pt)=(0.771214,0.528014,0.51131); rgb(101pt)=(0.772907,0.533747,0.51386); rgb(102pt)=(0.774597,0.539418,0.516398); rgb(103pt)=(0.776282,0.545031,0.518923); rgb(104pt)=(0.777964,0.550586,0.521436); rgb(105pt)=(0.779643,0.556086,0.523937); rgb(106pt)=(0.781318,0.561532,0.526426); rgb(107pt)=(0.782989,0.566926,0.528903); rgb(108pt)=(0.784657,0.572269,0.531369); rgb(109pt)=(0.786321,0.577562,0.533823); rgb(110pt)=(0.787982,0.582808,0.536266); rgb(111pt)=(0.789639,0.588006,0.538699); rgb(112pt)=(0.791292,0.59316,0.54112); rgb(113pt)=(0.792943,0.598268,0.54353); rgb(114pt)=(0.79459,0.603334,0.54593); rgb(115pt)=(0.796233,0.608357,0.548319); rgb(116pt)=(0.797873,0.613339,0.550698); rgb(117pt)=(0.79951,0.618281,0.553066); rgb(118pt)=(0.801143,0.623184,0.555425); rgb(119pt)=(0.802773,0.628048,0.557773); rgb(120pt)=(0.8044,0.632875,0.560112); rgb(121pt)=(0.806023,0.637666,0.562441); rgb(122pt)=(0.807643,0.642421,0.56476); rgb(123pt)=(0.80926,0.647141,0.56707); rgb(124pt)=(0.810874,0.651826,0.569371); rgb(125pt)=(0.812484,0.656479,0.571662); rgb(126pt)=(0.814092,0.661098,0.573944); rgb(127pt)=(0.815696,0.665686,0.576217); rgb(128pt)=(0.817297,0.670242,0.578481); rgb(129pt)=(0.818895,0.674767,0.580737); rgb(130pt)=(0.820489,0.679262,0.582983); rgb(131pt)=(0.822081,0.683728,0.585221); rgb(132pt)=(0.823669,0.688164,0.58745); rgb(133pt)=(0.825255,0.692573,0.589671); rgb(134pt)=(0.826837,0.696953,0.591884); rgb(135pt)=(0.828417,0.701306,0.594089); rgb(136pt)=(0.829993,0.705632,0.596285); rgb(137pt)=(0.831567,0.709932,0.598473); rgb(138pt)=(0.833137,0.714206,0.600653); rgb(139pt)=(0.834705,0.718454,0.602826); rgb(140pt)=(0.836269,0.722678,0.60499); rgb(141pt)=(0.837831,0.726877,0.607147); rgb(142pt)=(0.83939,0.731051,0.609296); rgb(143pt)=(0.840946,0.735203,0.611438); rgb(144pt)=(0.842499,0.73933,0.613572); rgb(145pt)=(0.844049,0.743435,0.615699); rgb(146pt)=(0.845596,0.747518,0.617818); rgb(147pt)=(0.847141,0.751578,0.61993); rgb(148pt)=(0.848682,0.755616,0.622035); rgb(149pt)=(0.850221,0.759633,0.624133); rgb(150pt)=(0.851757,0.763629,0.626224); rgb(151pt)=(0.85329,0.767604,0.628308); rgb(152pt)=(0.854821,0.771558,0.630385); rgb(153pt)=(0.856349,0.775493,0.632456); rgb(154pt)=(0.857874,0.779407,0.634519); rgb(155pt)=(0.859396,0.783302,0.636576); rgb(156pt)=(0.860916,0.787178,0.638626); rgb(157pt)=(0.862433,0.791034,0.64067); rgb(158pt)=(0.863947,0.794872,0.642707); rgb(159pt)=(0.865459,0.798692,0.644737); rgb(160pt)=(0.866968,0.802493,0.646762); rgb(161pt)=(0.868475,0.806276,0.64878); rgb(162pt)=(0.869979,0.810042,0.650791); rgb(163pt)=(0.87148,0.81379,0.652797); rgb(164pt)=(0.872979,0.817522,0.654796); rgb(165pt)=(0.874475,0.821236,0.65679); rgb(166pt)=(0.875968,0.824933,0.658777); rgb(167pt)=(0.877459,0.828614,0.660758); rgb(168pt)=(0.878948,0.832279,0.662733); rgb(169pt)=(0.880434,0.835927,0.664703); rgb(170pt)=(0.881917,0.83956,0.666667); rgb(171pt)=(0.883398,0.843177,0.668625); rgb(172pt)=(0.884877,0.846779,0.670577); rgb(173pt)=(0.886353,0.850365,0.672523); rgb(174pt)=(0.887826,0.853936,0.674464); rgb(175pt)=(0.889297,0.857493,0.6764); rgb(176pt)=(0.890766,0.861035,0.678329); rgb(177pt)=(0.892232,0.864562,0.680254); rgb(178pt)=(0.893696,0.868075,0.682173); rgb(179pt)=(0.895158,0.871574,0.684086); rgb(180pt)=(0.896617,0.875058,0.685994); rgb(181pt)=(0.898073,0.878529,0.687897); rgb(182pt)=(0.899528,0.881987,0.689795); rgb(183pt)=(0.90098,0.88543,0.691687); rgb(184pt)=(0.90243,0.888861,0.693575); rgb(185pt)=(0.903877,0.892278,0.695457); rgb(186pt)=(0.905322,0.895682,0.697334); rgb(187pt)=(0.906765,0.899074,0.699206); rgb(188pt)=(0.908205,0.902452,0.701073); rgb(189pt)=(0.909643,0.905818,0.702935); rgb(190pt)=(0.911079,0.909172,0.704792); rgb(191pt)=(0.912513,0.912513,0.706644); rgb(192pt)=(0.913944,0.913944,0.712158); rgb(193pt)=(0.915373,0.915373,0.717629); rgb(194pt)=(0.9168,0.9168,0.723059); rgb(195pt)=(0.918225,0.918225,0.728449); rgb(196pt)=(0.919648,0.919648,0.733798); rgb(197pt)=(0.921068,0.921068,0.739109); rgb(198pt)=(0.922486,0.922486,0.744383); rgb(199pt)=(0.923902,0.923902,0.749619); rgb(200pt)=(0.925316,0.925316,0.754818); rgb(201pt)=(0.926727,0.926727,0.759983); rgb(202pt)=(0.928137,0.928137,0.765112); rgb(203pt)=(0.929544,0.929544,0.770207); rgb(204pt)=(0.930949,0.930949,0.775269); rgb(205pt)=(0.932352,0.932352,0.780298); rgb(206pt)=(0.933753,0.933753,0.785294); rgb(207pt)=(0.935152,0.935152,0.790259); rgb(208pt)=(0.936549,0.936549,0.795193); rgb(209pt)=(0.937944,0.937944,0.800097); rgb(210pt)=(0.939336,0.939336,0.804971); rgb(211pt)=(0.940727,0.940727,0.809815); rgb(212pt)=(0.942116,0.942116,0.814631); rgb(213pt)=(0.943502,0.943502,0.819418); rgb(214pt)=(0.944886,0.944886,0.824178); rgb(215pt)=(0.946269,0.946269,0.82891); rgb(216pt)=(0.947649,0.947649,0.833615); rgb(217pt)=(0.949028,0.949028,0.838294); rgb(218pt)=(0.950404,0.950404,0.842947); rgb(219pt)=(0.951779,0.951779,0.847574); rgb(220pt)=(0.953151,0.953151,0.852177); rgb(221pt)=(0.954521,0.954521,0.856754); rgb(222pt)=(0.95589,0.95589,0.861307); rgb(223pt)=(0.957256,0.957256,0.865837); rgb(224pt)=(0.958621,0.958621,0.870342); rgb(225pt)=(0.959984,0.959984,0.874825); rgb(226pt)=(0.961344,0.961344,0.879285); rgb(227pt)=(0.962703,0.962703,0.883722); rgb(228pt)=(0.96406,0.96406,0.888137); rgb(229pt)=(0.965415,0.965415,0.89253); rgb(230pt)=(0.966768,0.966768,0.896901); rgb(231pt)=(0.968119,0.968119,0.901252); rgb(232pt)=(0.969469,0.969469,0.905581); rgb(233pt)=(0.970816,0.970816,0.90989); rgb(234pt)=(0.972162,0.972162,0.914179); rgb(235pt)=(0.973505,0.973505,0.918447); rgb(236pt)=(0.974847,0.974847,0.922696); rgb(237pt)=(0.976187,0.976187,0.926926); rgb(238pt)=(0.977525,0.977525,0.931136); rgb(239pt)=(0.978862,0.978862,0.935327); rgb(240pt)=(0.980196,0.980196,0.9395); rgb(241pt)=(0.981529,0.981529,0.943654); rgb(242pt)=(0.98286,0.98286,0.947789); rgb(243pt)=(0.984189,0.984189,0.951907); rgb(244pt)=(0.985516,0.985516,0.956007); rgb(245pt)=(0.986842,0.986842,0.96009); rgb(246pt)=(0.988165,0.988165,0.964155); rgb(247pt)=(0.989487,0.989487,0.968204); rgb(248pt)=(0.990807,0.990807,0.972235); rgb(249pt)=(0.992126,0.992126,0.97625); rgb(250pt)=(0.993443,0.993443,0.980248); rgb(251pt)=(0.994757,0.994757,0.98423); rgb(252pt)=(0.996071,0.996071,0.988196); rgb(253pt)=(0.997382,0.997382,0.992146); rgb(254pt)=(0.998692,0.998692,0.996081); rgb(255pt)=(1,1,1)}, mesh/rows=16]
table[row sep=crcr, point meta=\thisrow{c}] {%
x	y	c\\
0	0	0.387524034075977\\
2	0	0.387524034075977\\
4	0	0.387524034075977\\
6	0	0.387524034075977\\
8	0	0.387524034075977\\
10	0	0.387524034075977\\
12	0	0.387524034075977\\
14	0	0.387524034075977\\
16	0	0.387524034075977\\
18	0	0.387524034075977\\
20	0	0.387524034075977\\
22	0	0.387524034075977\\
24	0	0.387524034075977\\
26	0	0.387524034075977\\
28	0	0.387524034075977\\
30	0	0.387524034075977\\
0	2	0.644795481578139\\
2	2	0.63913816048514\\
4	2	0.629217593882424\\
6	2	0.63807056417557\\
8	2	0.619250863708169\\
10	2	0.625949563361174\\
12	2	0.621176602432643\\
14	2	0.633559730605665\\
16	2	0.630558355654733\\
18	2	0.630148014628663\\
20	2	0.631624539791433\\
22	2	0.633610565807613\\
24	2	0.633345419523725\\
26	2	0.629780938659778\\
28	2	0.634365872594192\\
30	2	0.635408465931121\\
0	4	0.491506194565933\\
2	4	0.491279465495621\\
4	4	0.400954655590658\\
6	4	0.405022529620032\\
8	4	0.399510241819707\\
10	4	0.406537950128746\\
12	4	0.404121596200163\\
14	4	0.411830012739798\\
16	4	0.411859904086806\\
18	4	0.4122048374899\\
20	4	0.416000131152397\\
22	4	0.416460898434662\\
24	4	0.415711452974055\\
26	4	0.415539193947873\\
28	4	0.418342723980226\\
30	4	0.41811536508689\\
0	6	0.468625505069023\\
2	6	0.435985981612792\\
4	6	0.31722621283261\\
6	6	0.314353731832075\\
8	6	0.31385245058639\\
10	6	0.317124883548553\\
12	6	0.315155593356859\\
14	6	0.325749477989495\\
16	6	0.32400069130873\\
18	6	0.325189060844265\\
20	6	0.328735290865151\\
22	6	0.326422570366623\\
24	6	0.327341030595348\\
26	6	0.327894107357155\\
28	6	0.331287041809048\\
30	6	0.330003180236718\\
0	8	0.451850587741762\\
2	8	0.406479626480619\\
4	8	0.27321088221293\\
6	8	0.266299211268679\\
8	8	0.263802009371778\\
10	8	0.26798605223366\\
12	8	0.267639713023251\\
14	8	0.275626866412235\\
16	8	0.275384764100957\\
18	8	0.275543392171957\\
20	8	0.277608166095181\\
22	8	0.274652613852988\\
24	8	0.277245018593573\\
26	8	0.278283583887209\\
28	8	0.280196983207819\\
30	8	0.280719210273787\\
0	10	0.443186088968263\\
2	10	0.387019812177021\\
4	10	0.24656641278002\\
6	10	0.235078418208949\\
8	10	0.233551266746386\\
10	10	0.237329138945933\\
12	10	0.236891121043948\\
14	10	0.243379241065187\\
16	10	0.242480970932943\\
18	10	0.241555996588405\\
20	10	0.243085961727831\\
22	10	0.241765872131425\\
24	10	0.244140107158236\\
26	10	0.244954505222473\\
28	10	0.247218203248837\\
30	10	0.2467768012137\\
0	12	0.43713813038816\\
2	12	0.374760396677951\\
4	12	0.22687729708033\\
6	12	0.214715474566975\\
8	12	0.212284410634445\\
10	12	0.215636814435666\\
12	12	0.215095725646234\\
14	12	0.220461182738082\\
16	12	0.217930002977607\\
18	12	0.216878654117321\\
20	12	0.218887657906307\\
22	12	0.217495443994285\\
24	12	0.219922198425847\\
26	12	0.221332840677601\\
28	12	0.223118851077841\\
30	12	0.223375390988818\\
0	14	0.4335864348275\\
2	14	0.365388304738064\\
4	14	0.214203633333011\\
6	14	0.199800129051263\\
8	14	0.196918395174241\\
10	14	0.199598818384089\\
12	14	0.198515757154327\\
14	14	0.2017210747592\\
16	14	0.19959576674389\\
18	14	0.198859315984025\\
20	14	0.200090700871618\\
22	14	0.198827121692654\\
24	14	0.201635437006918\\
26	14	0.20337840876778\\
28	14	0.205459209319729\\
30	14	0.205500877812829\\
0	16	0.430639068248897\\
2	16	0.358242757652053\\
4	16	0.204377738916231\\
6	16	0.18878555068741\\
8	16	0.185289697731032\\
10	16	0.187296543657482\\
12	16	0.184671186083045\\
14	16	0.187340626758247\\
16	16	0.185712581706856\\
18	16	0.184736140481638\\
20	16	0.185454472023337\\
22	16	0.184800957110366\\
24	16	0.187418825849079\\
26	16	0.189467196399445\\
28	16	0.191442481322595\\
30	16	0.191632183876253\\
0	18	0.428478061417514\\
2	18	0.352678549061018\\
4	18	0.196794953094023\\
6	18	0.180233902816603\\
8	18	0.176041802976283\\
10	18	0.176485713379601\\
12	18	0.173828596091228\\
14	18	0.176318262765157\\
16	18	0.17435790268142\\
18	18	0.173118735739622\\
20	18	0.174229245288698\\
22	18	0.173356791945638\\
24	18	0.176327904443037\\
26	18	0.178273204942767\\
28	18	0.180122140560852\\
30	18	0.180681197443266\\
0	20	0.426697186917109\\
2	20	0.348236111458234\\
4	20	0.190829795789127\\
6	20	0.173280682653854\\
8	20	0.167956508621319\\
10	20	0.16796814956553\\
12	20	0.165103011381409\\
14	20	0.167235857107826\\
16	20	0.164922254327521\\
18	20	0.163999921676118\\
20	20	0.164785779056935\\
22	20	0.164539047107397\\
24	20	0.167204262785386\\
26	20	0.168999096489871\\
28	20	0.171124783935056\\
30	20	0.171413984178128\\
0	22	0.425231129076411\\
2	22	0.344416271061737\\
4	22	0.185940206360779\\
6	22	0.167118652109344\\
8	22	0.16157392704311\\
10	22	0.161072220776982\\
12	22	0.15781426550487\\
14	22	0.159453163373534\\
16	22	0.157409485778879\\
18	22	0.156452334818232\\
20	22	0.15726049798008\\
22	22	0.157071724694522\\
24	22	0.159395175122404\\
26	22	0.161603324216156\\
28	22	0.163375645202124\\
30	22	0.163480857569596\\
0	24	0.423988552586788\\
2	24	0.341168123377756\\
4	24	0.181635142711964\\
6	24	0.162185392204227\\
8	24	0.156111533907085\\
10	24	0.155312549508797\\
12	24	0.15155100499246\\
14	24	0.153169817964787\\
16	24	0.151235920676747\\
18	24	0.150323496967167\\
20	24	0.151012936552522\\
22	24	0.150495306472742\\
24	24	0.153223269429594\\
26	24	0.155258406748008\\
28	24	0.15671676083614\\
30	24	0.156595194835269\\
0	26	0.422943937134692\\
2	26	0.338306733467692\\
4	26	0.178062450617375\\
6	26	0.15819784743155\\
8	26	0.151551708284085\\
10	26	0.150182439705132\\
12	26	0.146449524203735\\
14	26	0.14800609698607\\
16	26	0.146130547324525\\
18	26	0.145014391734154\\
20	26	0.145441552553429\\
22	26	0.145452475546171\\
24	26	0.147840936155808\\
26	26	0.149711515985453\\
28	26	0.150887940118665\\
30	26	0.150589274591235\\
0	28	0.422002600987768\\
2	28	0.335867260002605\\
4	28	0.175107537621238\\
6	28	0.154645090741878\\
8	28	0.147554498696923\\
10	28	0.146101220251243\\
12	28	0.142233024156427\\
14	28	0.143787616665393\\
16	28	0.141694434799839\\
18	28	0.140442226942389\\
20	28	0.141103872283807\\
22	28	0.141038065049193\\
24	28	0.143266922847416\\
26	28	0.144704102206302\\
28	28	0.14578113146022\\
30	28	0.145296574905004\\
0	30	0.421172264539852\\
2	30	0.333780997247885\\
4	30	0.172494790894138\\
6	30	0.151539706513396\\
8	30	0.144366969936481\\
10	30	0.142674310399728\\
12	30	0.138836796624535\\
14	30	0.140054716276031\\
16	30	0.137840185075826\\
18	30	0.136754376033385\\
20	30	0.137291342739863\\
22	30	0.137134537935643\\
24	30	0.139051338300525\\
26	30	0.140290495132234\\
28	30	0.14114892487163\\
30	30	0.140430005643373\\
};
\end{axis}

\begin{axis}[%
width=3.633in,
height=2.425in,
at={(0.28in,0.25in)},
scale only axis,
point meta min=0,
point meta max=1,
xmin=0,
xmax=1,
ymin=0,
ymax=1,
axis line style={draw=none},
ticks=none,
axis x line*=bottom,
axis y line*=left
]
\node[below right, align=left, font=\color{white}]
at (rel axis cs:0.196,0.557) {Low\\ cross\\ correlation};
\node[below right, align=left, font=\color{white}]
at (rel axis cs:0.554,0.523) {Dominance\\ of\\ Noise};
\draw[line width=3.0pt, draw=mycolor2] (axis cs:0.192833836745348,0.256031882529797) rectangle (axis cs:0.337893360554872,0.948254104752022);
\draw[line width=3.0pt, draw=mycolor1] (axis cs:0.512830562236641,0.254993068185632) rectangle (axis cs:0.797994252712832,0.948747711887088);
\end{axis}
\end{tikzpicture}%

%% file: TIFS_figures_tikz/Nokia_snr20_mp.tex
%
%
\definecolor{mycolor1}{rgb}{0.63529,0.07843,0.18431}%
\begin{tikzpicture}

\begin{axis}[%
width=2.5in,
height=2.000in,
at={(0.666in,0.569in)},
scale only axis,
point meta min=0.141548192733656,
point meta max=0.673381198915271,
xmin=0,
xmax=30,
xlabel style={font=\color{white!15!black}},
xlabel={Dimensions to omit},
ymin=2,
ymax=30,
ylabel style={font=\color{white!15!black}},
ylabel={$\tilde{D}$ - Dimensions to keep from $\tilde{D}_1$ to $\tilde{D}_2$},
axis background/.style={fill=white},
axis x line*=bottom,
axis y line*=left,
colormap={mymap}{[1pt] rgb(0pt)=(0,0.5,0.4); rgb(255pt)=(1,1,0.4)},
colorbar
]

\addplot[%
surf,
shader=flat corner, draw=black, colormap={mymap}{[1pt] rgb(0pt)=(0,0.5,0.4); rgb(255pt)=(1,1,0.4)}, mesh/rows=16]
table[row sep=crcr, point meta=\thisrow{c}] {%
x	y	c\\
0	0	0.129145089285714\\
2	0	0.129145089285714\\
4	0	0.129145089285714\\
6	0	0.129145089285714\\
8	0	0.129145089285714\\
10	0	0.129145089285714\\
12	0	0.129145089285714\\
14	0	0.129145089285714\\
16	0	0.129145089285714\\
18	0	0.129145089285714\\
20	0	0.129145089285714\\
22	0	0.129145089285714\\
24	0	0.129145089285714\\
26	0	0.129145089285714\\
28	0	0.129145089285714\\
30	0	0.129145089285714\\
0	2	0.111623883928571\\
2	2	0.113535714285714\\
4	2	0.107547991071429\\
6	2	0.129816964285714\\
8	2	0.1523828125\\
10	2	0.171254464285714\\
12	2	0.229127232142857\\
14	2	0.273083705357143\\
16	2	0.339623883928571\\
18	2	0.3793125\\
20	2	0.363707589285714\\
22	2	0.379775669642857\\
24	2	0.387266741071429\\
26	2	0.4183046875\\
28	2	0.429936383928571\\
30	2	0.442833705357143\\
0	4	0.0909821428571429\\
2	4	0.095546875\\
4	4	0.119274553571429\\
6	4	0.140533482142857\\
8	4	0.158334821428571\\
10	4	0.2053203125\\
12	4	0.215508928571429\\
14	4	0.302998883928571\\
16	4	0.286944196428571\\
18	4	0.358782366071429\\
20	4	0.343834821428571\\
22	4	0.365949776785714\\
24	4	0.378266741071429\\
26	4	0.400385044642857\\
28	4	0.416627232142857\\
30	4	0.4187265625\\
0	6	0.115688616071429\\
2	6	0.102216517857143\\
4	6	0.126709821428571\\
6	6	0.136417410714286\\
8	6	0.189996651785714\\
10	6	0.211787946428571\\
12	6	0.246715401785714\\
14	6	0.261284598214286\\
16	6	0.294518973214286\\
18	6	0.338231026785714\\
20	6	0.352180803571429\\
22	6	0.361362723214286\\
24	6	0.376065848214286\\
26	6	0.393143973214286\\
28	6	0.397506696428572\\
30	6	0.406108258928571\\
0	8	0.110189732142857\\
2	8	0.104832589285714\\
4	8	0.131089285714286\\
6	8	0.155147321428571\\
8	8	0.197045758928571\\
10	8	0.232572544642857\\
12	8	0.237733258928571\\
14	8	0.269131696428571\\
16	8	0.278322544642857\\
18	8	0.349521205357143\\
20	8	0.337270089285714\\
22	8	0.362191964285714\\
24	8	0.366510044642857\\
26	8	0.387579241071429\\
28	8	0.392127232142857\\
30	8	0.413373883928571\\
0	10	0.0813560267857143\\
2	10	0.0884654017857143\\
4	10	0.153483258928571\\
6	10	0.153123883928571\\
8	10	0.210001116071429\\
10	10	0.215888392857143\\
12	10	0.246684151785714\\
14	10	0.278177455357143\\
16	10	0.293232142857143\\
18	10	0.348258928571429\\
20	10	0.3362734375\\
22	10	0.359339285714286\\
24	10	0.364126116071429\\
26	10	0.378535714285714\\
28	10	0.398106026785714\\
30	10	0.419080357142857\\
0	12	0.147324776785714\\
2	12	0.126572544642857\\
4	12	0.150122767857143\\
6	12	0.181229910714286\\
8	12	0.200428571428571\\
10	12	0.227859375\\
12	12	0.23728125\\
14	12	0.283974330357143\\
16	12	0.292868303571429\\
18	12	0.342764508928571\\
20	12	0.330066964285714\\
22	12	0.352002232142857\\
24	12	0.361393973214286\\
26	12	0.386950892857143\\
28	12	0.400782366071429\\
30	12	0.413675223214286\\
0	14	0.1062734375\\
2	14	0.134693080357143\\
4	14	0.163234375\\
6	14	0.186758928571429\\
8	14	0.2023671875\\
10	14	0.222220982142857\\
12	14	0.249675223214286\\
14	14	0.275809151785714\\
16	14	0.296732142857143\\
18	14	0.339260044642857\\
20	14	0.341082589285714\\
22	14	0.357204241071429\\
24	14	0.377674107142857\\
26	14	0.393375\\
28	14	0.396502232142857\\
30	14	0.424174107142857\\
0	16	0.1133671875\\
2	16	0.101252232142857\\
4	16	0.160625\\
6	16	0.184360491071429\\
8	16	0.203465401785714\\
10	16	0.2386640625\\
12	16	0.248265625\\
14	16	0.279892857142857\\
16	16	0.301795758928571\\
18	16	0.340631696428571\\
20	16	0.33553125\\
22	16	0.366214285714286\\
24	16	0.369162946428571\\
26	16	0.393613839285714\\
28	16	0.411828125\\
30	16	0.421581473214286\\
0	18	0.100582589285714\\
2	18	0.124831473214286\\
4	18	0.173508928571429\\
6	18	0.2070546875\\
8	18	0.228043526785714\\
10	18	0.223987723214286\\
12	18	0.269940848214286\\
14	18	0.296801339285714\\
16	18	0.297904017857143\\
18	18	0.332889508928571\\
20	18	0.341522321428571\\
22	18	0.361379464285714\\
24	18	0.383064732142857\\
26	18	0.394556919642857\\
28	18	0.415872767857143\\
30	18	0.427801339285714\\
0	20	0.0988738839285714\\
2	20	0.102508928571429\\
4	20	0.166717633928571\\
6	20	0.191841517857143\\
8	20	0.213469866071429\\
10	20	0.234935267857143\\
12	20	0.269404017857143\\
14	20	0.285642857142857\\
16	20	0.305116071428571\\
18	20	0.345357142857143\\
20	20	0.345633928571429\\
22	20	0.3679921875\\
24	20	0.385293526785714\\
26	20	0.398877232142857\\
28	20	0.414522321428571\\
30	20	0.434123883928571\\
0	22	0.127321428571429\\
2	22	0.12359375\\
4	22	0.173616071428571\\
6	22	0.197021205357143\\
8	22	0.224994419642857\\
10	22	0.243731026785714\\
12	22	0.254988839285714\\
14	22	0.293612723214286\\
16	22	0.324597098214286\\
18	22	0.336641741071429\\
20	22	0.353393973214286\\
22	22	0.370044642857143\\
24	22	0.392325892857143\\
26	22	0.409672991071429\\
28	22	0.422900669642857\\
30	22	0.440959821428571\\
0	24	0.0947332589285714\\
2	24	0.125034598214286\\
4	24	0.172387276785714\\
6	24	0.196368303571429\\
8	24	0.228079241071429\\
10	24	0.2518203125\\
12	24	0.275410714285714\\
14	24	0.307607142857143\\
16	24	0.320185267857143\\
18	24	0.347483258928571\\
20	24	0.352100446428571\\
22	24	0.379808035714286\\
24	24	0.391456473214286\\
26	24	0.421214285714286\\
28	24	0.429025669642857\\
30	24	0.448376116071429\\
0	26	0.125262276785714\\
2	26	0.1156953125\\
4	26	0.182736607142857\\
6	26	0.206376116071429\\
8	26	0.238955357142857\\
10	26	0.242770089285714\\
12	26	0.278758928571429\\
14	26	0.302922991071429\\
16	26	0.313511160714286\\
18	26	0.345813616071429\\
20	26	0.365588169642857\\
22	26	0.3892265625\\
24	26	0.4032265625\\
26	26	0.417883928571429\\
28	26	0.428716517857143\\
30	26	0.449404017857143\\
0	28	0.103552455357143\\
2	28	0.131425223214286\\
4	28	0.188020089285714\\
6	28	0.207943080357143\\
8	28	0.227011160714286\\
10	28	0.262479910714286\\
12	28	0.277372767857143\\
14	28	0.305396205357143\\
16	28	0.328496651785714\\
18	28	0.365299107142857\\
20	28	0.3681484375\\
22	28	0.391127232142857\\
24	28	0.406947544642857\\
26	28	0.419770089285714\\
28	28	0.437544642857143\\
30	28	0.455702008928571\\
0	30	0.154169642857143\\
2	30	0.148186383928571\\
4	30	0.1984375\\
6	30	0.214934151785714\\
8	30	0.255013392857143\\
10	30	0.256643973214286\\
12	30	0.284745535714286\\
14	30	0.3256171875\\
16	30	0.327308035714286\\
18	30	0.363345982142857\\
20	30	0.3781875\\
22	30	0.398418526785714\\
24	30	0.407272321428571\\
26	30	0.436189732142857\\
28	30	0.440508928571429\\
30	30	0.455991071428571\\
};
\end{axis}

\begin{axis}[%
width=3.633in,
height=2.425in,
at={(0.28in,0.25in)},
scale only axis,
point meta min=0,
point meta max=1,
xmin=0,
xmax=1,
ymin=0,
ymax=1,
axis line style={draw=none},
ticks=none,
axis x line*=bottom,
axis y line*=left
]
\node[below right, align=left, font=\color{white}]
at (rel axis cs:0.196,0.557) {Low\\ mismatch\\ probability};
\node[below right, align=left, font=\color{purple}]
at (rel axis cs:0.554,0.523) {High\\ mismatch\\ probability};
\draw[line width=3.0pt, draw=black] (axis cs:0.192833836745348,0.256031882529797) rectangle (axis cs:0.337893360554872,0.948254104752022);
\draw[line width=3.0pt, draw=mycolor1] (axis cs:0.512830562236641,0.254993068185632) rectangle (axis cs:0.797994252712832,0.948747711887088);
\end{axis}
\end{tikzpicture}%

%% file: TIFS_figures_tikz/Nokia_snr5_corr.tex
%
%
\definecolor{mycolor1}{rgb}{0.85098,0.32549,0.09804}%
\begin{tikzpicture}

\begin{axis}[%
width=2.5in,
height=2.000in,
at={(0.666in,0.569in)},
scale only axis,
point meta min=0.141548192733656,
point meta max=0.673381198915271,
xmin=0,
xmax=30,
xlabel style={font=\color{white!15!black}},
xlabel={Dimensions to omit},
ymin=2,
ymax=30,
ylabel style={font=\color{white!15!black}},
ylabel={$\tilde{D}$ - Dimensions to keep from $\tilde{D}_1$ to $\tilde{D}_2$},
axis background/.style={fill=white},
axis x line*=bottom,
axis y line*=left,
colormap={mymap}{[1pt] rgb(0pt)=(0.0589256,0,0); rgb(1pt)=(0.0977692,0.051131,0.051131); rgb(2pt)=(0.125082,0.0723102,0.0723102); rgb(3pt)=(0.147418,0.0885615,0.0885615); rgb(4pt)=(0.166789,0.102262,0.102262); rgb(5pt)=(0.184134,0.114332,0.114332); rgb(6pt)=(0.19998,0.125245,0.125245); rgb(7pt)=(0.214659,0.13528,0.13528); rgb(8pt)=(0.228397,0.14462,0.14462); rgb(9pt)=(0.241354,0.153393,0.153393); rgb(10pt)=(0.25365,0.16169,0.16169); rgb(11pt)=(0.265377,0.169582,0.169582); rgb(12pt)=(0.276607,0.177123,0.177123); rgb(13pt)=(0.287399,0.184355,0.184355); rgb(14pt)=(0.2978,0.191315,0.191315); rgb(15pt)=(0.307849,0.19803,0.19803); rgb(16pt)=(0.317581,0.204524,0.204524); rgb(17pt)=(0.327024,0.210819,0.210819); rgb(18pt)=(0.336201,0.21693,0.21693); rgb(19pt)=(0.345134,0.222875,0.222875); rgb(20pt)=(0.353842,0.228665,0.228665); rgb(21pt)=(0.362341,0.234312,0.234312); rgb(22pt)=(0.370645,0.239826,0.239826); rgb(23pt)=(0.378766,0.245216,0.245216); rgb(24pt)=(0.386718,0.25049,0.25049); rgb(25pt)=(0.394509,0.255655,0.255655); rgb(26pt)=(0.402149,0.260718,0.260718); rgb(27pt)=(0.409647,0.265684,0.265684); rgb(28pt)=(0.41701,0.27056,0.27056); rgb(29pt)=(0.424245,0.275349,0.275349); rgb(30pt)=(0.431359,0.280056,0.280056); rgb(31pt)=(0.438357,0.284685,0.284685); rgb(32pt)=(0.445245,0.289241,0.289241); rgb(33pt)=(0.452029,0.293725,0.293725); rgb(34pt)=(0.458712,0.298142,0.298142); rgb(35pt)=(0.465299,0.302495,0.302495); rgb(36pt)=(0.471794,0.306786,0.306786); rgb(37pt)=(0.478201,0.311018,0.311018); rgb(38pt)=(0.484524,0.315193,0.315193); rgb(39pt)=(0.490764,0.319313,0.319313); rgb(40pt)=(0.496927,0.323381,0.323381); rgb(41pt)=(0.503014,0.327398,0.327398); rgb(42pt)=(0.509028,0.331367,0.331367); rgb(43pt)=(0.514972,0.335288,0.335288); rgb(44pt)=(0.520848,0.339165,0.339165); rgb(45pt)=(0.526659,0.342997,0.342997); rgb(46pt)=(0.532406,0.346787,0.346787); rgb(47pt)=(0.538092,0.350536,0.350536); rgb(48pt)=(0.543718,0.354246,0.354246); rgb(49pt)=(0.549287,0.357917,0.357917); rgb(50pt)=(0.554799,0.361551,0.361551); rgb(51pt)=(0.560258,0.365148,0.365148); rgb(52pt)=(0.565664,0.368711,0.368711); rgb(53pt)=(0.571018,0.372239,0.372239); rgb(54pt)=(0.576323,0.375735,0.375735); rgb(55pt)=(0.58158,0.379198,0.379198); rgb(56pt)=(0.586789,0.382629,0.382629); rgb(57pt)=(0.591953,0.386031,0.386031); rgb(58pt)=(0.597072,0.389402,0.389402); rgb(59pt)=(0.602148,0.392745,0.392745); rgb(60pt)=(0.607181,0.396059,0.396059); rgb(61pt)=(0.612172,0.399346,0.399346); rgb(62pt)=(0.617124,0.402606,0.402606); rgb(63pt)=(0.622035,0.40584,0.40584); rgb(64pt)=(0.626909,0.409048,0.409048); rgb(65pt)=(0.631745,0.412231,0.412231); rgb(66pt)=(0.636544,0.41539,0.41539); rgb(67pt)=(0.641307,0.418525,0.418525); rgb(68pt)=(0.646035,0.421637,0.421637); rgb(69pt)=(0.650729,0.424726,0.424726); rgb(70pt)=(0.655389,0.427793,0.427793); rgb(71pt)=(0.660016,0.430837,0.430837); rgb(72pt)=(0.664611,0.433861,0.433861); rgb(73pt)=(0.669174,0.436863,0.436863); rgb(74pt)=(0.673707,0.439845,0.439845); rgb(75pt)=(0.678209,0.442807,0.442807); rgb(76pt)=(0.682681,0.44575,0.44575); rgb(77pt)=(0.687125,0.448673,0.448673); rgb(78pt)=(0.69154,0.451577,0.451577); rgb(79pt)=(0.695927,0.454462,0.454462); rgb(80pt)=(0.700286,0.45733,0.45733); rgb(81pt)=(0.704618,0.460179,0.460179); rgb(82pt)=(0.708924,0.463011,0.463011); rgb(83pt)=(0.713204,0.465826,0.465826); rgb(84pt)=(0.717459,0.468623,0.468623); rgb(85pt)=(0.721688,0.471405,0.471405); rgb(86pt)=(0.725893,0.474169,0.474169); rgb(87pt)=(0.730073,0.476918,0.476918); rgb(88pt)=(0.73423,0.479651,0.479651); rgb(89pt)=(0.738363,0.482369,0.482369); rgb(90pt)=(0.742473,0.485071,0.485071); rgb(91pt)=(0.746561,0.487759,0.487759); rgb(92pt)=(0.750626,0.490431,0.490431); rgb(93pt)=(0.75467,0.493089,0.493089); rgb(94pt)=(0.758691,0.495733,0.495733); rgb(95pt)=(0.762692,0.498363,0.498363); rgb(96pt)=(0.764404,0.504433,0.500979); rgb(97pt)=(0.766112,0.51043,0.503582); rgb(98pt)=(0.767817,0.516358,0.506171); rgb(99pt)=(0.769517,0.522219,0.508747); rgb(100pt)=(0.771214,0.528014,0.51131); rgb(101pt)=(0.772907,0.533747,0.51386); rgb(102pt)=(0.774597,0.539418,0.516398); rgb(103pt)=(0.776282,0.545031,0.518923); rgb(104pt)=(0.777964,0.550586,0.521436); rgb(105pt)=(0.779643,0.556086,0.523937); rgb(106pt)=(0.781318,0.561532,0.526426); rgb(107pt)=(0.782989,0.566926,0.528903); rgb(108pt)=(0.784657,0.572269,0.531369); rgb(109pt)=(0.786321,0.577562,0.533823); rgb(110pt)=(0.787982,0.582808,0.536266); rgb(111pt)=(0.789639,0.588006,0.538699); rgb(112pt)=(0.791292,0.59316,0.54112); rgb(113pt)=(0.792943,0.598268,0.54353); rgb(114pt)=(0.79459,0.603334,0.54593); rgb(115pt)=(0.796233,0.608357,0.548319); rgb(116pt)=(0.797873,0.613339,0.550698); rgb(117pt)=(0.79951,0.618281,0.553066); rgb(118pt)=(0.801143,0.623184,0.555425); rgb(119pt)=(0.802773,0.628048,0.557773); rgb(120pt)=(0.8044,0.632875,0.560112); rgb(121pt)=(0.806023,0.637666,0.562441); rgb(122pt)=(0.807643,0.642421,0.56476); rgb(123pt)=(0.80926,0.647141,0.56707); rgb(124pt)=(0.810874,0.651826,0.569371); rgb(125pt)=(0.812484,0.656479,0.571662); rgb(126pt)=(0.814092,0.661098,0.573944); rgb(127pt)=(0.815696,0.665686,0.576217); rgb(128pt)=(0.817297,0.670242,0.578481); rgb(129pt)=(0.818895,0.674767,0.580737); rgb(130pt)=(0.820489,0.679262,0.582983); rgb(131pt)=(0.822081,0.683728,0.585221); rgb(132pt)=(0.823669,0.688164,0.58745); rgb(133pt)=(0.825255,0.692573,0.589671); rgb(134pt)=(0.826837,0.696953,0.591884); rgb(135pt)=(0.828417,0.701306,0.594089); rgb(136pt)=(0.829993,0.705632,0.596285); rgb(137pt)=(0.831567,0.709932,0.598473); rgb(138pt)=(0.833137,0.714206,0.600653); rgb(139pt)=(0.834705,0.718454,0.602826); rgb(140pt)=(0.836269,0.722678,0.60499); rgb(141pt)=(0.837831,0.726877,0.607147); rgb(142pt)=(0.83939,0.731051,0.609296); rgb(143pt)=(0.840946,0.735203,0.611438); rgb(144pt)=(0.842499,0.73933,0.613572); rgb(145pt)=(0.844049,0.743435,0.615699); rgb(146pt)=(0.845596,0.747518,0.617818); rgb(147pt)=(0.847141,0.751578,0.61993); rgb(148pt)=(0.848682,0.755616,0.622035); rgb(149pt)=(0.850221,0.759633,0.624133); rgb(150pt)=(0.851757,0.763629,0.626224); rgb(151pt)=(0.85329,0.767604,0.628308); rgb(152pt)=(0.854821,0.771558,0.630385); rgb(153pt)=(0.856349,0.775493,0.632456); rgb(154pt)=(0.857874,0.779407,0.634519); rgb(155pt)=(0.859396,0.783302,0.636576); rgb(156pt)=(0.860916,0.787178,0.638626); rgb(157pt)=(0.862433,0.791034,0.64067); rgb(158pt)=(0.863947,0.794872,0.642707); rgb(159pt)=(0.865459,0.798692,0.644737); rgb(160pt)=(0.866968,0.802493,0.646762); rgb(161pt)=(0.868475,0.806276,0.64878); rgb(162pt)=(0.869979,0.810042,0.650791); rgb(163pt)=(0.87148,0.81379,0.652797); rgb(164pt)=(0.872979,0.817522,0.654796); rgb(165pt)=(0.874475,0.821236,0.65679); rgb(166pt)=(0.875968,0.824933,0.658777); rgb(167pt)=(0.877459,0.828614,0.660758); rgb(168pt)=(0.878948,0.832279,0.662733); rgb(169pt)=(0.880434,0.835927,0.664703); rgb(170pt)=(0.881917,0.83956,0.666667); rgb(171pt)=(0.883398,0.843177,0.668625); rgb(172pt)=(0.884877,0.846779,0.670577); rgb(173pt)=(0.886353,0.850365,0.672523); rgb(174pt)=(0.887826,0.853936,0.674464); rgb(175pt)=(0.889297,0.857493,0.6764); rgb(176pt)=(0.890766,0.861035,0.678329); rgb(177pt)=(0.892232,0.864562,0.680254); rgb(178pt)=(0.893696,0.868075,0.682173); rgb(179pt)=(0.895158,0.871574,0.684086); rgb(180pt)=(0.896617,0.875058,0.685994); rgb(181pt)=(0.898073,0.878529,0.687897); rgb(182pt)=(0.899528,0.881987,0.689795); rgb(183pt)=(0.90098,0.88543,0.691687); rgb(184pt)=(0.90243,0.888861,0.693575); rgb(185pt)=(0.903877,0.892278,0.695457); rgb(186pt)=(0.905322,0.895682,0.697334); rgb(187pt)=(0.906765,0.899074,0.699206); rgb(188pt)=(0.908205,0.902452,0.701073); rgb(189pt)=(0.909643,0.905818,0.702935); rgb(190pt)=(0.911079,0.909172,0.704792); rgb(191pt)=(0.912513,0.912513,0.706644); rgb(192pt)=(0.913944,0.913944,0.712158); rgb(193pt)=(0.915373,0.915373,0.717629); rgb(194pt)=(0.9168,0.9168,0.723059); rgb(195pt)=(0.918225,0.918225,0.728449); rgb(196pt)=(0.919648,0.919648,0.733798); rgb(197pt)=(0.921068,0.921068,0.739109); rgb(198pt)=(0.922486,0.922486,0.744383); rgb(199pt)=(0.923902,0.923902,0.749619); rgb(200pt)=(0.925316,0.925316,0.754818); rgb(201pt)=(0.926727,0.926727,0.759983); rgb(202pt)=(0.928137,0.928137,0.765112); rgb(203pt)=(0.929544,0.929544,0.770207); rgb(204pt)=(0.930949,0.930949,0.775269); rgb(205pt)=(0.932352,0.932352,0.780298); rgb(206pt)=(0.933753,0.933753,0.785294); rgb(207pt)=(0.935152,0.935152,0.790259); rgb(208pt)=(0.936549,0.936549,0.795193); rgb(209pt)=(0.937944,0.937944,0.800097); rgb(210pt)=(0.939336,0.939336,0.804971); rgb(211pt)=(0.940727,0.940727,0.809815); rgb(212pt)=(0.942116,0.942116,0.814631); rgb(213pt)=(0.943502,0.943502,0.819418); rgb(214pt)=(0.944886,0.944886,0.824178); rgb(215pt)=(0.946269,0.946269,0.82891); rgb(216pt)=(0.947649,0.947649,0.833615); rgb(217pt)=(0.949028,0.949028,0.838294); rgb(218pt)=(0.950404,0.950404,0.842947); rgb(219pt)=(0.951779,0.951779,0.847574); rgb(220pt)=(0.953151,0.953151,0.852177); rgb(221pt)=(0.954521,0.954521,0.856754); rgb(222pt)=(0.95589,0.95589,0.861307); rgb(223pt)=(0.957256,0.957256,0.865837); rgb(224pt)=(0.958621,0.958621,0.870342); rgb(225pt)=(0.959984,0.959984,0.874825); rgb(226pt)=(0.961344,0.961344,0.879285); rgb(227pt)=(0.962703,0.962703,0.883722); rgb(228pt)=(0.96406,0.96406,0.888137); rgb(229pt)=(0.965415,0.965415,0.89253); rgb(230pt)=(0.966768,0.966768,0.896901); rgb(231pt)=(0.968119,0.968119,0.901252); rgb(232pt)=(0.969469,0.969469,0.905581); rgb(233pt)=(0.970816,0.970816,0.90989); rgb(234pt)=(0.972162,0.972162,0.914179); rgb(235pt)=(0.973505,0.973505,0.918447); rgb(236pt)=(0.974847,0.974847,0.922696); rgb(237pt)=(0.976187,0.976187,0.926926); rgb(238pt)=(0.977525,0.977525,0.931136); rgb(239pt)=(0.978862,0.978862,0.935327); rgb(240pt)=(0.980196,0.980196,0.9395); rgb(241pt)=(0.981529,0.981529,0.943654); rgb(242pt)=(0.98286,0.98286,0.947789); rgb(243pt)=(0.984189,0.984189,0.951907); rgb(244pt)=(0.985516,0.985516,0.956007); rgb(245pt)=(0.986842,0.986842,0.96009); rgb(246pt)=(0.988165,0.988165,0.964155); rgb(247pt)=(0.989487,0.989487,0.968204); rgb(248pt)=(0.990807,0.990807,0.972235); rgb(249pt)=(0.992126,0.992126,0.97625); rgb(250pt)=(0.993443,0.993443,0.980248); rgb(251pt)=(0.994757,0.994757,0.98423); rgb(252pt)=(0.996071,0.996071,0.988196); rgb(253pt)=(0.997382,0.997382,0.992146); rgb(254pt)=(0.998692,0.998692,0.996081); rgb(255pt)=(1,1,1)},
colorbar
]

\addplot[%
surf,
shader=flat corner, draw=black, colormap={mymap}{[1pt] rgb(0pt)=(0.0589256,0,0); rgb(1pt)=(0.0977692,0.051131,0.051131); rgb(2pt)=(0.125082,0.0723102,0.0723102); rgb(3pt)=(0.147418,0.0885615,0.0885615); rgb(4pt)=(0.166789,0.102262,0.102262); rgb(5pt)=(0.184134,0.114332,0.114332); rgb(6pt)=(0.19998,0.125245,0.125245); rgb(7pt)=(0.214659,0.13528,0.13528); rgb(8pt)=(0.228397,0.14462,0.14462); rgb(9pt)=(0.241354,0.153393,0.153393); rgb(10pt)=(0.25365,0.16169,0.16169); rgb(11pt)=(0.265377,0.169582,0.169582); rgb(12pt)=(0.276607,0.177123,0.177123); rgb(13pt)=(0.287399,0.184355,0.184355); rgb(14pt)=(0.2978,0.191315,0.191315); rgb(15pt)=(0.307849,0.19803,0.19803); rgb(16pt)=(0.317581,0.204524,0.204524); rgb(17pt)=(0.327024,0.210819,0.210819); rgb(18pt)=(0.336201,0.21693,0.21693); rgb(19pt)=(0.345134,0.222875,0.222875); rgb(20pt)=(0.353842,0.228665,0.228665); rgb(21pt)=(0.362341,0.234312,0.234312); rgb(22pt)=(0.370645,0.239826,0.239826); rgb(23pt)=(0.378766,0.245216,0.245216); rgb(24pt)=(0.386718,0.25049,0.25049); rgb(25pt)=(0.394509,0.255655,0.255655); rgb(26pt)=(0.402149,0.260718,0.260718); rgb(27pt)=(0.409647,0.265684,0.265684); rgb(28pt)=(0.41701,0.27056,0.27056); rgb(29pt)=(0.424245,0.275349,0.275349); rgb(30pt)=(0.431359,0.280056,0.280056); rgb(31pt)=(0.438357,0.284685,0.284685); rgb(32pt)=(0.445245,0.289241,0.289241); rgb(33pt)=(0.452029,0.293725,0.293725); rgb(34pt)=(0.458712,0.298142,0.298142); rgb(35pt)=(0.465299,0.302495,0.302495); rgb(36pt)=(0.471794,0.306786,0.306786); rgb(37pt)=(0.478201,0.311018,0.311018); rgb(38pt)=(0.484524,0.315193,0.315193); rgb(39pt)=(0.490764,0.319313,0.319313); rgb(40pt)=(0.496927,0.323381,0.323381); rgb(41pt)=(0.503014,0.327398,0.327398); rgb(42pt)=(0.509028,0.331367,0.331367); rgb(43pt)=(0.514972,0.335288,0.335288); rgb(44pt)=(0.520848,0.339165,0.339165); rgb(45pt)=(0.526659,0.342997,0.342997); rgb(46pt)=(0.532406,0.346787,0.346787); rgb(47pt)=(0.538092,0.350536,0.350536); rgb(48pt)=(0.543718,0.354246,0.354246); rgb(49pt)=(0.549287,0.357917,0.357917); rgb(50pt)=(0.554799,0.361551,0.361551); rgb(51pt)=(0.560258,0.365148,0.365148); rgb(52pt)=(0.565664,0.368711,0.368711); rgb(53pt)=(0.571018,0.372239,0.372239); rgb(54pt)=(0.576323,0.375735,0.375735); rgb(55pt)=(0.58158,0.379198,0.379198); rgb(56pt)=(0.586789,0.382629,0.382629); rgb(57pt)=(0.591953,0.386031,0.386031); rgb(58pt)=(0.597072,0.389402,0.389402); rgb(59pt)=(0.602148,0.392745,0.392745); rgb(60pt)=(0.607181,0.396059,0.396059); rgb(61pt)=(0.612172,0.399346,0.399346); rgb(62pt)=(0.617124,0.402606,0.402606); rgb(63pt)=(0.622035,0.40584,0.40584); rgb(64pt)=(0.626909,0.409048,0.409048); rgb(65pt)=(0.631745,0.412231,0.412231); rgb(66pt)=(0.636544,0.41539,0.41539); rgb(67pt)=(0.641307,0.418525,0.418525); rgb(68pt)=(0.646035,0.421637,0.421637); rgb(69pt)=(0.650729,0.424726,0.424726); rgb(70pt)=(0.655389,0.427793,0.427793); rgb(71pt)=(0.660016,0.430837,0.430837); rgb(72pt)=(0.664611,0.433861,0.433861); rgb(73pt)=(0.669174,0.436863,0.436863); rgb(74pt)=(0.673707,0.439845,0.439845); rgb(75pt)=(0.678209,0.442807,0.442807); rgb(76pt)=(0.682681,0.44575,0.44575); rgb(77pt)=(0.687125,0.448673,0.448673); rgb(78pt)=(0.69154,0.451577,0.451577); rgb(79pt)=(0.695927,0.454462,0.454462); rgb(80pt)=(0.700286,0.45733,0.45733); rgb(81pt)=(0.704618,0.460179,0.460179); rgb(82pt)=(0.708924,0.463011,0.463011); rgb(83pt)=(0.713204,0.465826,0.465826); rgb(84pt)=(0.717459,0.468623,0.468623); rgb(85pt)=(0.721688,0.471405,0.471405); rgb(86pt)=(0.725893,0.474169,0.474169); rgb(87pt)=(0.730073,0.476918,0.476918); rgb(88pt)=(0.73423,0.479651,0.479651); rgb(89pt)=(0.738363,0.482369,0.482369); rgb(90pt)=(0.742473,0.485071,0.485071); rgb(91pt)=(0.746561,0.487759,0.487759); rgb(92pt)=(0.750626,0.490431,0.490431); rgb(93pt)=(0.75467,0.493089,0.493089); rgb(94pt)=(0.758691,0.495733,0.495733); rgb(95pt)=(0.762692,0.498363,0.498363); rgb(96pt)=(0.764404,0.504433,0.500979); rgb(97pt)=(0.766112,0.51043,0.503582); rgb(98pt)=(0.767817,0.516358,0.506171); rgb(99pt)=(0.769517,0.522219,0.508747); rgb(100pt)=(0.771214,0.528014,0.51131); rgb(101pt)=(0.772907,0.533747,0.51386); rgb(102pt)=(0.774597,0.539418,0.516398); rgb(103pt)=(0.776282,0.545031,0.518923); rgb(104pt)=(0.777964,0.550586,0.521436); rgb(105pt)=(0.779643,0.556086,0.523937); rgb(106pt)=(0.781318,0.561532,0.526426); rgb(107pt)=(0.782989,0.566926,0.528903); rgb(108pt)=(0.784657,0.572269,0.531369); rgb(109pt)=(0.786321,0.577562,0.533823); rgb(110pt)=(0.787982,0.582808,0.536266); rgb(111pt)=(0.789639,0.588006,0.538699); rgb(112pt)=(0.791292,0.59316,0.54112); rgb(113pt)=(0.792943,0.598268,0.54353); rgb(114pt)=(0.79459,0.603334,0.54593); rgb(115pt)=(0.796233,0.608357,0.548319); rgb(116pt)=(0.797873,0.613339,0.550698); rgb(117pt)=(0.79951,0.618281,0.553066); rgb(118pt)=(0.801143,0.623184,0.555425); rgb(119pt)=(0.802773,0.628048,0.557773); rgb(120pt)=(0.8044,0.632875,0.560112); rgb(121pt)=(0.806023,0.637666,0.562441); rgb(122pt)=(0.807643,0.642421,0.56476); rgb(123pt)=(0.80926,0.647141,0.56707); rgb(124pt)=(0.810874,0.651826,0.569371); rgb(125pt)=(0.812484,0.656479,0.571662); rgb(126pt)=(0.814092,0.661098,0.573944); rgb(127pt)=(0.815696,0.665686,0.576217); rgb(128pt)=(0.817297,0.670242,0.578481); rgb(129pt)=(0.818895,0.674767,0.580737); rgb(130pt)=(0.820489,0.679262,0.582983); rgb(131pt)=(0.822081,0.683728,0.585221); rgb(132pt)=(0.823669,0.688164,0.58745); rgb(133pt)=(0.825255,0.692573,0.589671); rgb(134pt)=(0.826837,0.696953,0.591884); rgb(135pt)=(0.828417,0.701306,0.594089); rgb(136pt)=(0.829993,0.705632,0.596285); rgb(137pt)=(0.831567,0.709932,0.598473); rgb(138pt)=(0.833137,0.714206,0.600653); rgb(139pt)=(0.834705,0.718454,0.602826); rgb(140pt)=(0.836269,0.722678,0.60499); rgb(141pt)=(0.837831,0.726877,0.607147); rgb(142pt)=(0.83939,0.731051,0.609296); rgb(143pt)=(0.840946,0.735203,0.611438); rgb(144pt)=(0.842499,0.73933,0.613572); rgb(145pt)=(0.844049,0.743435,0.615699); rgb(146pt)=(0.845596,0.747518,0.617818); rgb(147pt)=(0.847141,0.751578,0.61993); rgb(148pt)=(0.848682,0.755616,0.622035); rgb(149pt)=(0.850221,0.759633,0.624133); rgb(150pt)=(0.851757,0.763629,0.626224); rgb(151pt)=(0.85329,0.767604,0.628308); rgb(152pt)=(0.854821,0.771558,0.630385); rgb(153pt)=(0.856349,0.775493,0.632456); rgb(154pt)=(0.857874,0.779407,0.634519); rgb(155pt)=(0.859396,0.783302,0.636576); rgb(156pt)=(0.860916,0.787178,0.638626); rgb(157pt)=(0.862433,0.791034,0.64067); rgb(158pt)=(0.863947,0.794872,0.642707); rgb(159pt)=(0.865459,0.798692,0.644737); rgb(160pt)=(0.866968,0.802493,0.646762); rgb(161pt)=(0.868475,0.806276,0.64878); rgb(162pt)=(0.869979,0.810042,0.650791); rgb(163pt)=(0.87148,0.81379,0.652797); rgb(164pt)=(0.872979,0.817522,0.654796); rgb(165pt)=(0.874475,0.821236,0.65679); rgb(166pt)=(0.875968,0.824933,0.658777); rgb(167pt)=(0.877459,0.828614,0.660758); rgb(168pt)=(0.878948,0.832279,0.662733); rgb(169pt)=(0.880434,0.835927,0.664703); rgb(170pt)=(0.881917,0.83956,0.666667); rgb(171pt)=(0.883398,0.843177,0.668625); rgb(172pt)=(0.884877,0.846779,0.670577); rgb(173pt)=(0.886353,0.850365,0.672523); rgb(174pt)=(0.887826,0.853936,0.674464); rgb(175pt)=(0.889297,0.857493,0.6764); rgb(176pt)=(0.890766,0.861035,0.678329); rgb(177pt)=(0.892232,0.864562,0.680254); rgb(178pt)=(0.893696,0.868075,0.682173); rgb(179pt)=(0.895158,0.871574,0.684086); rgb(180pt)=(0.896617,0.875058,0.685994); rgb(181pt)=(0.898073,0.878529,0.687897); rgb(182pt)=(0.899528,0.881987,0.689795); rgb(183pt)=(0.90098,0.88543,0.691687); rgb(184pt)=(0.90243,0.888861,0.693575); rgb(185pt)=(0.903877,0.892278,0.695457); rgb(186pt)=(0.905322,0.895682,0.697334); rgb(187pt)=(0.906765,0.899074,0.699206); rgb(188pt)=(0.908205,0.902452,0.701073); rgb(189pt)=(0.909643,0.905818,0.702935); rgb(190pt)=(0.911079,0.909172,0.704792); rgb(191pt)=(0.912513,0.912513,0.706644); rgb(192pt)=(0.913944,0.913944,0.712158); rgb(193pt)=(0.915373,0.915373,0.717629); rgb(194pt)=(0.9168,0.9168,0.723059); rgb(195pt)=(0.918225,0.918225,0.728449); rgb(196pt)=(0.919648,0.919648,0.733798); rgb(197pt)=(0.921068,0.921068,0.739109); rgb(198pt)=(0.922486,0.922486,0.744383); rgb(199pt)=(0.923902,0.923902,0.749619); rgb(200pt)=(0.925316,0.925316,0.754818); rgb(201pt)=(0.926727,0.926727,0.759983); rgb(202pt)=(0.928137,0.928137,0.765112); rgb(203pt)=(0.929544,0.929544,0.770207); rgb(204pt)=(0.930949,0.930949,0.775269); rgb(205pt)=(0.932352,0.932352,0.780298); rgb(206pt)=(0.933753,0.933753,0.785294); rgb(207pt)=(0.935152,0.935152,0.790259); rgb(208pt)=(0.936549,0.936549,0.795193); rgb(209pt)=(0.937944,0.937944,0.800097); rgb(210pt)=(0.939336,0.939336,0.804971); rgb(211pt)=(0.940727,0.940727,0.809815); rgb(212pt)=(0.942116,0.942116,0.814631); rgb(213pt)=(0.943502,0.943502,0.819418); rgb(214pt)=(0.944886,0.944886,0.824178); rgb(215pt)=(0.946269,0.946269,0.82891); rgb(216pt)=(0.947649,0.947649,0.833615); rgb(217pt)=(0.949028,0.949028,0.838294); rgb(218pt)=(0.950404,0.950404,0.842947); rgb(219pt)=(0.951779,0.951779,0.847574); rgb(220pt)=(0.953151,0.953151,0.852177); rgb(221pt)=(0.954521,0.954521,0.856754); rgb(222pt)=(0.95589,0.95589,0.861307); rgb(223pt)=(0.957256,0.957256,0.865837); rgb(224pt)=(0.958621,0.958621,0.870342); rgb(225pt)=(0.959984,0.959984,0.874825); rgb(226pt)=(0.961344,0.961344,0.879285); rgb(227pt)=(0.962703,0.962703,0.883722); rgb(228pt)=(0.96406,0.96406,0.888137); rgb(229pt)=(0.965415,0.965415,0.89253); rgb(230pt)=(0.966768,0.966768,0.896901); rgb(231pt)=(0.968119,0.968119,0.901252); rgb(232pt)=(0.969469,0.969469,0.905581); rgb(233pt)=(0.970816,0.970816,0.90989); rgb(234pt)=(0.972162,0.972162,0.914179); rgb(235pt)=(0.973505,0.973505,0.918447); rgb(236pt)=(0.974847,0.974847,0.922696); rgb(237pt)=(0.976187,0.976187,0.926926); rgb(238pt)=(0.977525,0.977525,0.931136); rgb(239pt)=(0.978862,0.978862,0.935327); rgb(240pt)=(0.980196,0.980196,0.9395); rgb(241pt)=(0.981529,0.981529,0.943654); rgb(242pt)=(0.98286,0.98286,0.947789); rgb(243pt)=(0.984189,0.984189,0.951907); rgb(244pt)=(0.985516,0.985516,0.956007); rgb(245pt)=(0.986842,0.986842,0.96009); rgb(246pt)=(0.988165,0.988165,0.964155); rgb(247pt)=(0.989487,0.989487,0.968204); rgb(248pt)=(0.990807,0.990807,0.972235); rgb(249pt)=(0.992126,0.992126,0.97625); rgb(250pt)=(0.993443,0.993443,0.980248); rgb(251pt)=(0.994757,0.994757,0.98423); rgb(252pt)=(0.996071,0.996071,0.988196); rgb(253pt)=(0.997382,0.997382,0.992146); rgb(254pt)=(0.998692,0.998692,0.996081); rgb(255pt)=(1,1,1)}, mesh/rows=16]
table[row sep=crcr, point meta=\thisrow{c}] {%
x	y	c\\
0	0	0.249455445444342\\
2	0	0.249455445444342\\
4	0	0.249455445444342\\
6	0	0.249455445444342\\
8	0	0.249455445444342\\
10	0	0.249455445444342\\
12	0	0.249455445444342\\
14	0	0.249455445444342\\
16	0	0.249455445444342\\
18	0	0.249455445444342\\
20	0	0.249455445444342\\
22	0	0.249455445444342\\
24	0	0.249455445444342\\
26	0	0.249455445444342\\
28	0	0.249455445444342\\
30	0	0.249455445444342\\
0	2	0.643836295152719\\
2	2	0.63813926683101\\
4	2	0.633699410142932\\
6	2	0.634632065031485\\
8	2	0.631052348337706\\
10	2	0.63405320056796\\
12	2	0.633074584475339\\
14	2	0.634222319330724\\
16	2	0.632772502800351\\
18	2	0.633774383556843\\
20	2	0.634891443281733\\
22	2	0.633910654985491\\
24	2	0.634580876833406\\
26	2	0.635171688211518\\
28	2	0.634896216866427\\
30	2	0.634957182881855\\
0	4	0.485804635998666\\
2	4	0.474590378368936\\
4	4	0.398084746937799\\
6	4	0.409462695204076\\
8	4	0.414238296678353\\
10	4	0.416585551859744\\
12	4	0.416994728766081\\
14	4	0.419722136766783\\
16	4	0.418333894447392\\
18	4	0.420981807117471\\
20	4	0.421267353714825\\
22	4	0.420719812356853\\
24	4	0.421139659766049\\
26	4	0.420988164869607\\
28	4	0.421149738508838\\
30	4	0.42009460148965\\
0	6	0.45619652728581\\
2	6	0.409718572814161\\
4	6	0.312058577782373\\
6	6	0.319768853672469\\
8	6	0.326353442945197\\
10	6	0.329862643088888\\
12	6	0.332201868804292\\
14	6	0.333289356972217\\
16	6	0.332828315089321\\
18	6	0.334848224234169\\
20	6	0.334943807414304\\
22	6	0.334921252570986\\
24	6	0.334437658959106\\
26	6	0.334310364646034\\
28	6	0.334043562560572\\
30	6	0.333332634041307\\
0	8	0.435429780403676\\
2	8	0.371455312416388\\
4	8	0.266020452797288\\
6	8	0.270868911425899\\
8	8	0.277375540326611\\
10	8	0.280633555292215\\
12	8	0.28223397229468\\
14	8	0.283720307863433\\
16	8	0.28380306872923\\
18	8	0.285409852257254\\
20	8	0.284984559444815\\
22	8	0.284593992489699\\
24	8	0.284444217586492\\
26	8	0.284587163585029\\
28	8	0.284547406563187\\
30	8	0.284672898539941\\
0	10	0.422122878120421\\
2	10	0.343662936539619\\
4	10	0.23601274277971\\
6	10	0.239681473167806\\
8	10	0.245461168742016\\
10	10	0.247342882148833\\
12	10	0.249527406267772\\
14	10	0.250922955962494\\
16	10	0.251557942383256\\
18	10	0.251788477763165\\
20	10	0.251560495265171\\
22	10	0.251002895124947\\
24	10	0.251020121301901\\
26	10	0.251253980423487\\
28	10	0.251692050497089\\
30	10	0.251333064711755\\
0	12	0.411787467486152\\
2	12	0.322442456797226\\
4	12	0.214824395258812\\
6	12	0.217435944148617\\
8	12	0.22118758881439\\
10	12	0.223322122491934\\
12	12	0.225771580264358\\
14	12	0.226559863612867\\
16	12	0.226978599078186\\
18	12	0.227492475042189\\
20	12	0.226629621217111\\
22	12	0.226369365741258\\
24	12	0.226823041443344\\
26	12	0.227312203574852\\
28	12	0.226968503401088\\
30	12	0.22707848546989\\
0	14	0.403664441373596\\
2	14	0.305596784051396\\
4	14	0.198646886398281\\
6	14	0.199870606756586\\
8	14	0.203019868912523\\
10	14	0.205326480563533\\
12	14	0.207229287770093\\
14	14	0.207496421032792\\
16	14	0.208245149514805\\
18	14	0.2080828857944\\
20	14	0.207980098162326\\
22	14	0.207851449056662\\
24	14	0.207999355086473\\
26	14	0.208224497811321\\
28	14	0.208191970985406\\
30	14	0.207908519585619\\
0	16	0.396576513299264\\
2	16	0.291378724297192\\
4	16	0.185392413962856\\
6	16	0.185963701589108\\
8	16	0.188846326493614\\
10	16	0.190494395014703\\
12	16	0.191811762086206\\
14	16	0.192413167211522\\
16	16	0.192618767843614\\
18	16	0.193074140526465\\
20	16	0.192933209583795\\
22	16	0.192596112164279\\
24	16	0.192850466569213\\
26	16	0.193031956642483\\
28	16	0.192909454205396\\
30	16	0.192933988019993\\
0	18	0.39003578164959\\
2	18	0.278832159761154\\
4	18	0.174742860136576\\
6	18	0.174457420266671\\
8	18	0.176641511227898\\
10	18	0.17815126837139\\
12	18	0.179433148122459\\
14	18	0.179757855470351\\
16	18	0.18017631654313\\
18	18	0.180624724350746\\
20	18	0.180160968533095\\
22	18	0.180189605096696\\
24	18	0.180358614718367\\
26	18	0.180370557950806\\
28	18	0.180438510262928\\
30	18	0.180502964221619\\
0	20	0.383924921538426\\
2	20	0.268000362428748\\
4	20	0.165615950677939\\
6	20	0.164496367050766\\
8	20	0.166506300870365\\
10	20	0.167647442206255\\
12	20	0.168833430207853\\
14	20	0.169298962766455\\
16	20	0.169665197113495\\
18	20	0.169615671070738\\
20	20	0.169594382881357\\
22	20	0.169554320314827\\
24	20	0.16966336591584\\
26	20	0.169802397313901\\
28	20	0.169866123172561\\
30	20	0.169898336217211\\
0	22	0.37835366079719\\
2	22	0.258154488147243\\
4	22	0.157357850372629\\
6	22	0.155940087051318\\
8	22	0.157740327962257\\
10	22	0.15872509134486\\
12	22	0.159870198467823\\
14	22	0.160282240955992\\
16	22	0.160158360267489\\
18	22	0.160502175323718\\
20	22	0.160332093831355\\
22	22	0.16044097147027\\
24	22	0.160643222391043\\
26	22	0.160583952889793\\
28	22	0.160595528009414\\
30	22	0.160690513404864\\
0	24	0.37327434281365\\
2	24	0.249371780738949\\
4	24	0.150249661757337\\
6	24	0.148453817491605\\
8	24	0.149976627480192\\
10	24	0.151054809014697\\
12	24	0.152158037749566\\
14	24	0.151972159798262\\
16	24	0.152261533662654\\
18	24	0.152295552548643\\
20	24	0.152194220342639\\
22	24	0.152626987075311\\
24	24	0.152696540058622\\
26	24	0.152280286677132\\
28	24	0.152507349116233\\
30	24	0.152613228339569\\
0	26	0.368484724446029\\
2	26	0.241320070747239\\
4	26	0.143873682033198\\
6	26	0.141808623794218\\
8	26	0.143336728956651\\
10	26	0.144198744720723\\
12	26	0.144886242358099\\
14	26	0.145006585182262\\
16	26	0.145073199150723\\
18	26	0.145156748003806\\
20	26	0.145236643677351\\
22	26	0.145512056987313\\
24	26	0.145356044562678\\
26	26	0.145127416602759\\
28	26	0.145404685059174\\
30	26	0.145307533661401\\
0	28	0.363950130581249\\
2	28	0.233952424148237\\
4	28	0.138068913126475\\
6	28	0.136091392777546\\
8	28	0.137341145455708\\
10	28	0.137911020739187\\
12	28	0.138707864026265\\
14	28	0.138545372914715\\
16	28	0.13867489096864\\
18	28	0.138881590358444\\
20	28	0.138962021550251\\
22	28	0.139019385420756\\
24	28	0.138885794988285\\
26	28	0.138761383959775\\
28	28	0.138839807705806\\
30	28	0.138733217429635\\
0	30	0.359642511762689\\
2	30	0.227197305976694\\
4	30	0.133152734908774\\
6	30	0.13086633376493\\
8	30	0.131764722320313\\
10	30	0.132285077741197\\
12	30	0.132909072663071\\
14	30	0.132813496607057\\
16	30	0.132968474144853\\
18	30	0.13333903320575\\
20	30	0.133099085198492\\
22	30	0.133192808546625\\
24	30	0.133112143780472\\
26	30	0.132932259048207\\
28	30	0.132896129086133\\
30	30	0.132793502954325\\
};
\end{axis}

\begin{axis}[%
width=3.633in,
height=2.425in,
at={(0.25in,0.25in)},
scale only axis,
point meta min=0,
point meta max=1,
xmin=0,
xmax=1,
ymin=0,
ymax=1,
axis line style={draw=none},
ticks=none,
axis x line*=bottom,
axis y line*=left
]
\node[below right, align=left, font=\color{white}]
at (rel axis cs:0.506,0.597) {Dominance\\of\\Noise};
\draw[line width=3.0pt, draw=mycolor1] (axis cs:0.301566170712609,0.256853408029879) rectangle (axis cs:0.79863686223373,0.959813258636795);
\end{axis}
\end{tikzpicture}%

%% file: TIFS_figures_tikz/Nokia_snr5_mp.tex
%
%
\definecolor{mycolor1}{rgb}{0.63529,0.07843,0.18431}%
\begin{tikzpicture}

\begin{axis}[%
width=2.5in,
height=2.000in,
at={(0.666in,0.569in)},
scale only axis,
point meta min=0.141548192733656,
point meta max=0.673381198915271,
xmin=0,
xmax=30,
xlabel style={font=\color{white!15!black}},
xlabel={Dimensions to omit},
ymin=2,
ymax=30,
ylabel style={font=\color{white!15!black}},
ylabel={$\tilde{D}$ - Dimensions to keep from $\tilde{D}_1$ to $\tilde{D}_2$},
axis background/.style={fill=white},
axis x line*=bottom,
axis y line*=left,
colormap={mymap}{[1pt] rgb(0pt)=(0,0.5,0.4); rgb(255pt)=(1,1,0.4)},
colorbar
]

\addplot[%
surf,
shader=flat corner, draw=black, colormap={mymap}{[1pt] rgb(0pt)=(0,0.5,0.4); rgb(255pt)=(1,1,0.4)}, mesh/rows=16]
table[row sep=crcr, point meta=\thisrow{c}] {%
x	y	c\\
0	0	0.331725446428571\\
2	0	0.331725446428571\\
4	0	0.331725446428571\\
6	0	0.331725446428571\\
8	0	0.331725446428571\\
10	0	0.331725446428571\\
12	0	0.331725446428571\\
14	0	0.331725446428571\\
16	0	0.331725446428571\\
18	0	0.331725446428571\\
20	0	0.331725446428571\\
22	0	0.331725446428571\\
24	0	0.331725446428571\\
26	0	0.331725446428571\\
28	0	0.331725446428571\\
30	0	0.331725446428571\\
0	2	0.1558828125\\
2	2	0.169433035714286\\
4	2	0.3328671875\\
6	2	0.422639508928571\\
8	2	0.462705357142857\\
10	2	0.486372767857143\\
12	2	0.497296875\\
14	2	0.503707589285714\\
16	2	0.498444196428571\\
18	2	0.503931919642857\\
20	2	0.496902901785714\\
22	2	0.500411830357143\\
24	2	0.499265625\\
26	2	0.500055803571429\\
28	2	0.503896205357143\\
30	2	0.502541294642857\\
0	4	0.127526785714286\\
2	4	0.219886160714286\\
4	4	0.382267857142857\\
6	4	0.440074776785714\\
8	4	0.471127232142857\\
10	4	0.492301339285714\\
12	4	0.494533482142857\\
14	4	0.500176339285714\\
16	4	0.5043203125\\
18	4	0.502715401785714\\
20	4	0.500147321428571\\
22	4	0.496869419642857\\
24	4	0.497022321428571\\
26	4	0.499955357142857\\
28	4	0.502104910714286\\
30	4	0.502348214285714\\
0	6	0.126775669642857\\
2	6	0.2490625\\
4	6	0.413776785714286\\
6	6	0.459040178571429\\
8	6	0.479002232142857\\
10	6	0.488590401785714\\
12	6	0.491315848214286\\
14	6	0.500042410714286\\
16	6	0.497941964285714\\
18	6	0.501558035714286\\
20	6	0.5035078125\\
22	6	0.500030133928571\\
24	6	0.501696428571429\\
26	6	0.502371651785714\\
28	6	0.503147321428571\\
30	6	0.502154017857143\\
0	8	0.149591517857143\\
2	8	0.303089285714286\\
4	8	0.418883928571429\\
6	8	0.464547991071429\\
8	8	0.48384375\\
10	8	0.494400669642857\\
12	8	0.497293526785714\\
14	8	0.494496651785714\\
16	8	0.498625\\
18	8	0.501717633928571\\
20	8	0.4998359375\\
22	8	0.501544642857143\\
24	8	0.498380580357143\\
26	8	0.496642857142857\\
28	8	0.5026875\\
30	8	0.497180803571429\\
0	10	0.152159598214286\\
2	10	0.317751116071429\\
4	10	0.444845982142857\\
6	10	0.474051339285714\\
8	10	0.483402901785714\\
10	10	0.493944196428571\\
12	10	0.501609375\\
14	10	0.498869419642857\\
16	10	0.497549107142857\\
18	10	0.498172991071429\\
20	10	0.495983258928571\\
22	10	0.501849330357143\\
24	10	0.496935267857143\\
26	10	0.494957589285714\\
28	10	0.500987723214286\\
30	10	0.500113839285714\\
0	12	0.158310267857143\\
2	12	0.332818080357143\\
4	12	0.444819196428571\\
6	12	0.475356026785714\\
8	12	0.489672991071429\\
10	12	0.491131696428571\\
12	12	0.502402901785714\\
14	12	0.502071428571429\\
16	12	0.495214285714286\\
18	12	0.504283482142857\\
20	12	0.504868303571429\\
22	12	0.508196428571429\\
24	12	0.498245535714286\\
26	12	0.498871651785714\\
28	12	0.498675223214286\\
30	12	0.503981026785714\\
0	14	0.174080357142857\\
2	14	0.343267857142857\\
4	14	0.461130580357143\\
6	14	0.478928571428571\\
8	14	0.491831473214286\\
10	14	0.492100446428571\\
12	14	0.501260044642857\\
14	14	0.502286830357143\\
16	14	0.502629464285714\\
18	14	0.4956875\\
20	14	0.499646205357143\\
22	14	0.503764508928571\\
24	14	0.497185267857143\\
26	14	0.497629464285714\\
28	14	0.501081473214286\\
30	14	0.502909598214286\\
0	16	0.197813616071429\\
2	16	0.360790178571429\\
4	16	0.461784598214286\\
6	16	0.482302455357143\\
8	16	0.491943080357143\\
10	16	0.49484375\\
12	16	0.497472098214286\\
14	16	0.499584821428571\\
16	16	0.48634375\\
18	16	0.497316964285714\\
20	16	0.504386160714286\\
22	16	0.4970546875\\
24	16	0.499069196428572\\
26	16	0.503129464285714\\
28	16	0.500542410714286\\
30	16	0.496613839285714\\
0	18	0.198380580357143\\
2	18	0.378334821428571\\
4	18	0.468013392857143\\
6	18	0.486477678571429\\
8	18	0.4889140625\\
10	18	0.498426339285714\\
12	18	0.498456473214286\\
14	18	0.497381696428571\\
16	18	0.50165625\\
18	18	0.501809151785714\\
20	18	0.494073660714286\\
22	18	0.495236607142857\\
24	18	0.5066015625\\
26	18	0.501108258928571\\
28	18	0.502879464285714\\
30	18	0.498578125\\
0	20	0.2136875\\
2	20	0.382645089285714\\
4	20	0.466763392857143\\
6	20	0.487577008928571\\
8	20	0.493908482142857\\
10	20	0.500920758928571\\
12	20	0.497716517857143\\
14	20	0.496111607142857\\
16	20	0.498427455357143\\
18	20	0.499235491071429\\
20	20	0.50140625\\
22	20	0.491552455357143\\
24	20	0.496296875\\
26	20	0.501147321428571\\
28	20	0.497662946428571\\
30	20	0.497915178571429\\
0	22	0.211631696428571\\
2	22	0.395642857142857\\
4	22	0.473724330357143\\
6	22	0.484661830357143\\
8	22	0.494152901785714\\
10	22	0.493035714285714\\
12	22	0.491619419642857\\
14	22	0.506619419642857\\
16	22	0.498354910714286\\
18	22	0.4996171875\\
20	22	0.495706473214286\\
22	22	0.502806919642857\\
24	22	0.501133928571429\\
26	22	0.493290178571429\\
28	22	0.496613839285714\\
30	22	0.5020859375\\
0	24	0.216794642857143\\
2	24	0.388725446428571\\
4	24	0.477220982142857\\
6	24	0.486704241071429\\
8	24	0.495267857142857\\
10	24	0.494830357142857\\
12	24	0.491411830357143\\
14	24	0.496049107142857\\
16	24	0.494456473214286\\
18	24	0.4967421875\\
20	24	0.502598214285714\\
22	24	0.501280133928571\\
24	24	0.503431919642857\\
26	24	0.498425223214286\\
28	24	0.499362723214286\\
30	24	0.502342633928571\\
0	26	0.224321428571429\\
2	26	0.399053571428571\\
4	26	0.476989955357143\\
6	26	0.492993303571429\\
8	26	0.497466517857143\\
10	26	0.498924107142857\\
12	26	0.499319196428571\\
14	26	0.506879464285714\\
16	26	0.504254464285714\\
18	26	0.49628125\\
20	26	0.50278125\\
22	26	0.501094866071429\\
24	26	0.497348214285714\\
26	26	0.491947544642857\\
28	26	0.500435267857143\\
30	26	0.498421875\\
0	28	0.214267857142857\\
2	28	0.402943080357143\\
4	28	0.480301339285714\\
6	28	0.493452008928571\\
8	28	0.496521205357143\\
10	28	0.492029017857143\\
12	28	0.505296875\\
14	28	0.499244419642857\\
16	28	0.50246875\\
18	28	0.497137276785714\\
20	28	0.503050223214286\\
22	28	0.498032366071429\\
24	28	0.497366071428571\\
26	28	0.503518973214286\\
28	28	0.499133928571429\\
30	28	0.501756696428571\\
0	30	0.238302455357143\\
2	30	0.414354910714286\\
4	30	0.478104910714286\\
6	30	0.499729910714286\\
8	30	0.491042410714286\\
10	30	0.495411830357143\\
12	30	0.497248883928572\\
14	30	0.498858258928571\\
16	30	0.498300223214286\\
18	30	0.502925223214286\\
20	30	0.502724330357143\\
22	30	0.497111607142857\\
24	30	0.5002109375\\
26	30	0.497991071428571\\
28	30	0.4940625\\
30	30	0.501050223214286\\
};
\end{axis}

\begin{axis}[%
width=3.633in,
height=2.425in,
at={(0.25in,0.25in)},
scale only axis,
point meta min=0,
point meta max=1,
xmin=0,
xmax=1,
ymin=0,
ymax=1,
axis line style={draw=none},
ticks=none,
axis x line*=bottom,
axis y line*=left
]
\node[below right, align=left, font=\color{purple}]
at (rel axis cs:0.506,0.597) {High\\mismatch\\probability};
\draw[line width=3.0pt, draw=mycolor1] (axis cs:0.301566170712609,0.256853408029879) rectangle (axis cs:0.79863686223373,0.959813258636795);
\end{axis}
\end{tikzpicture}%

%% file: TIFS_figures_tikz/nokia_hsic_new.tex
%
%
\definecolor{mycolor1}{rgb}{0.00000,0.44700,0.74100}%
\begin{tikzpicture}

\begin{axis}[%
width=2.421in,
height=2.154in,
at={(0.758in,0.692in)},
scale only axis,
xmin=0,
xmax=30,
xlabel style={font=\color{white!15!black}},
xlabel={$\tilde{D}_1$},
ymin=0,
ymax=35,
ylabel style={font=\color{white!15!black}},
ylabel={Average $\overline {\Delta}$},
axis background/.style={fill=white},
xmajorgrids,
ymajorgrids,
legend style={legend cell align=left, align=left, draw=white!15!black}
]
\addplot [color=mycolor1, line width=2.0pt, mark=asterisk, mark options={solid, mycolor1}]
  table[row sep=crcr]{%
0	32.0060456519476\\
2	12.9662967314702\\
4	4.3560860482005\\
6	3.81114163805788\\
8	3.42330645643008\\
10	3.32984047471319\\
12	2.88951283309871\\
14	2.76782841114154\\
16	2.57770174856211\\
18	1.96604299049197\\
20	1.88016329979986\\
22	1.62335558613577\\
24	1.32096845455231\\
26	2.00266136443472\\
28	2.15298386401992\\
30	1.85221805784506\\
};
\addlegendentry{SNR=20dB}

\addplot [color=red, line width=2.0pt, mark=star, mark options={solid, red}]
  table[row sep=crcr]{%
0	10.2879267509317\\
2	5.37819938600513\\
4	2.03766247026712\\
6	1.65534556343991\\
8	1.68631270034292\\
10	1.40241814529334\\
12	1.35807396532966\\
14	1.3702350407188\\
16	1.65322328513776\\
18	2.31979871050903\\
20	2.23071150777273\\
22	2.5515634398569\\
24	2.34536907658978\\
26	2.57018915095238\\
28	2.77445633886758\\
30	2.63948167401454\\
};
\addlegendentry{SNR=5dB}

\end{axis}
\end{tikzpicture}%